\documentclass[linenumbers]{aastex631}

\usepackage[normalem]{ulem}

\begin{document}
\nolinenumbers
\title{Case studies on pre-eruptive X-class flares using R-value in the lower solar atmosphere \footnote{Submitted on 18/04/2024}}

\author[0000-0002-1367-8326]{Shreeyesh Biswal}
\affiliation{Solar Physics \& Space Plasma Research Center (SP2RC), School of Mathematics and Statistics, University of Sheffield, Hounsfield Road, Sheffield, S3 7RH, UK}
\affiliation{Department of Physics - Section Astrogeophysics, University of Ioannina, Ioannina - 45110, Greece}

\author[0000-0002-0049-4798]{Marianna B. Kors\'os}
\affiliation{University of Sheffield, Department of Automatic Control and Systems Engineering, Amy Johnson Building, Portabello Street, Sheffield, S1 3JD, UK}
\affiliation{Department of Astronomy, E\"otv\"os Lor\'and University, P\'azm\'any P\'eter s\'et\'any 1/A, Budapest H-1117, Hungary}
\affiliation{Gyula Bay Zoltan Solar Observatory (GSO), Hungarian Solar Physics Foundation (HSPF), Pet\H{o}fi t\'er 3., Gyula H-5700, Hungary}

\author[0000-0001-6913-1330]
{Manolis K. Georgoulis}
\affiliation{Johns Hopkins University Applied Physics Laboratory Laurel, MD 20375, USA}
\affiliation{Research Center for Astronomy and Applied Mathematics of the Academy of Athens, 11527 Athens, Greece}

\author[0000-0003-0475-2886]{Alexander Nindos}
\affiliation{Department of Physics - Section Astrogeophysics, University of Ioannina, Ioannina - 45110, Greece}

\author[0000-0003-3345-9697]
{Spiros Patsourakos}
\affiliation{Department of Physics - Section Astrogeophysics, University of Ioannina, Ioannina - 45110, Greece}

\author[0000-0003-3439-4127]{Robertus Erdélyi}
\affiliation{Solar Physics \& Space Plasma Research Center (SP2RC), School of Mathematics and Statistics, University of Sheffield, Hounsfield Road, Sheffield, S3 7RH, UK}
\affiliation{Department of Astronomy, E\"otv\"os Lor\'and University, P\'azm\'any P\'eter s\'et\'any 1/A, Budapest H-1117, Hungary}
\affiliation{Gyula Bay Zoltan Solar Observatory (GSO), Hungarian Solar Physics Foundation (HSPF), Pet\H{o}fi t\'er 3., Gyula H-5700, Hungary}


\phantom{...}

\begin{abstract}
\nolinenumbers
\phantom{...}

The R-value is a measure of the strength of photospheric magnetic Polarity Inversion Lines (PILs) in Active Regions (ARs). This work investigates the possibility of a relation between R-value variations and the occurrence of X-class flares in ARs, not in the solar photosphere, as usual, but above it in regions, closer to where flares occur. The modus operandi is to extrapolate the Solar Dynamic Observatory's (SDO) Helioseismic and Magnetic Imager (HMI) magnetogram data up to a height of 3.24 Mm above the photosphere and then compute the R-value based on the extrapolated magnetic field. Recent studies have shown that certain flare-predictive parameters such as the horizontal gradient of the vertical magnetic field  and magnetic helicity may improve flare prediction lead times significantly if studied at a specific height range above the photosphere, called the Optimal Height Range (OHR). Here we define the OHR as a collection of heights where a sudden but sustained increase in R-value is found. For the eight case studies discussed in this paper, our results indicate that it is possible for OHRs to exist in the low solar atmosphere (between 0.36 - 3.24 Mm), where R-value spikes occur 48-68 hrs before the first X-class flare of an emerging AR. The temporal evolution of R-value before the first X-class flare for an emerging AR is also found to be  distinct from that of non-flaring ARs. For X-class flares associated with non-emerging ARs, an OHR could not be found.

\end{abstract}


\section{Introduction}
\label{sec:intro}

\phantom{...}

A solar flare is an intense burst of electromagnetic radiation from the Sun. It is caused due to magnetic reconnection in the solar atmosphere \citep{Kopp1976}. Early models that addressed the relation between solar flares and magnetic reconnection were two dimensional (2D) in nature, with one of the most popular being the CSHKP "standard flare" model (\citealp{Carmichael1964}; \citealp{Sturrock1966}; \citealp{Hirayama1974}; \citealp{Kopp1976}). Thanks to advancements in space research and technology, solar energetic  phenomena, including the standard flare model, are now studied in three dimensions (3D) via magnetohydrodynamic (MHD) models (\citealp{Janvier2017}; \citealp{Korsos2018}; \citealp{Pontin2022}). Solar flares are often associated with eruptive phenomena called Coronal Mass Ejections (CMEs). A CME is an ejection of a sizable coronal magnetic structure, thought to be a helical magnetic flux rope, into the heliosphere \citep{Low1994,Dere1999}. When this flux rope (also known as a magnetic cloud in the interplanetary space) is directed towards Earth, it has the potential to interact with the terrestrial, geomagnetic field. This interaction may induce a geomagnetic storm and cause damage to our technosphere from space all the way to Earth's surface for major disturbances. 

\phantom{...}
 
Solar flares and CMEs are two distinct manifestations of a common underlying mechanism of magnetic energy release (\citealp{Gosling1990}; \citealp{Low1994}; \citealp{Harrison1995}; \citealp{Gopalswamy2016}). Solar flare X-ray intensity fluxes are indeed well correlated to their corresponding CME energies \citep{Youssef2012}. Other studies show the relation between flares and CMEs as the synchronization  of flare emissions (HXR and temporal derivative of SXRs) and CME acceleration (\citealp{Zhang2004}; \citealp{Temmer_2010}). Since it is known that stronger flares, especially X-class events (x-ray intensity flux greater than 10$^{-4}$ Wm$^{-2}$), have a high probability of CME association \citep{Yashiro2005}, predicting X-class flares is a problem of particular importance in Space Weather research. \\

The key physical process leading to the manifestation of ARs is the emergence of toroidal magnetic flux tubes in the photosphere due to buoyancy in the convection zone (\citealp{Parker1955}; \citealp{Parker1979}). MHD models have successfully simulated and accounted for the inception of flux in the photosphere and its subsequent transport to the corona in 2D (\citealp{Shibata1989}; \citealp{Shibata1990}; \citealp{Kaisig1990}) and 3D (\citealp{Fan2001}; \citealp{Achrontis2004}).  Physical processes such as the evolution of an unstable flux rope (\citealp{Aulanier2010}; \citealp{Aulanier2012}; \citealp{Kusano2012}) and the evolution of the current layer and magnetic reconnection in 3D have also been studied extensively (\citealp{Kliem2013}; \citealp{Janvier2013}; \citealp{Janvier2017}). One way to predict solar flares is to track the changes occurring in the magnetic flux patterns of flare-producing ARs and assess how they differ from ARs that do not produce flares. \citet{Toriumi2019} give an overview of processes and features associated with the formation of flare-producing ARs. Significant processes linked to the production of flare-producing ARs include the formation of $\delta$-sunspots (\citealp{Kunzel1959}; \citealp{Sammis2000}; \citealp{Tian2002}) and the appearance of high-gradient magnetic PILs in the photosphere (\citealp{Falconer2002}; \citealp{Falconer2003}; \citealp{Schrijver2007}). \citet{Sammis2000} further showed that more complex sunspots, especially the ones identified to be $\delta$, $\beta\delta$ and $\beta\gamma\delta$ produce stronger flares and that flare strength (in terms of peak x-ray irradiance) is positively correlated with sunspot area. \\

Detailed tracking of PIL evolution is also useful to predict flares. PILs are interfaces between flux patches of opposite magnetic polarity, where the vertical magnetic field component $B_z$ "neutralizes" (i.e., becomes zero) along them. The presence of high-gradient PILs, where the vertical magnetic field component enhances dramatically just off the PIL,  is a characteristic eruptive flare source pattern and such PILs are often an outcome of shearing of the photospheric magnetic field and convergence of opposite-polarity flux patches \citep{Georgoulis2019}. Morphological parameters such as $G_M$ and $WG_M$, that take the horizontal gradient of the vertical magnetic flux into account, have also been previously introduced \citep{Korsos2014, Korsos2015}. For the calculation of $G_M$, two areas having the maximum positive and the maximum negative magnetic polarities are identified and the difference between their fluxes is divided by the distance between their area weighted centroids. $WG_M$ is a more generalised form of $G_M$ where the calculation includes not two, but several regions of opposite polarities. With regards to the PILs, there exist several other morphological parameters like unsigned flux, PIL gradients (\citealp{Falconer2002}; \citealp{Falconer2003}), R-value \citep{Schrijver2007}, effective connected magnetic field strength \citep{Georgoulis2007} and length (total and maximum) of PILs \citep{Mason2010} that could be used to address solar flare prediction probability quantitatively. \citet{Schrijver2007} found that when the peak R-value computed in the photosphere reaches about $2 \times 10^{21} $ Mx, then the probability of occurrence of a major flare in 24 hrs is close to unity. PILs can also be studied from the perspective of electric current density directly, instead of studying proxies of magnetic non-potentiality, but this requires the full magnetic field vector. In fact, strong PILs are the only photospheric structures that support non-neutralised electric currents, as in a nonzero volume current in coronal flux tubes \citep{Georgoulis2012}. From a study on the temporal evolution of non-neutralised currents, \citet{Kontogiannis2017} established a correspondence between them and key physical processes like the appearance of PILs and flux rope formation. Having provided a brief outline of parameters (or predictors) that mathematically incorporate several key features directly linked to solar flare productivity, it is noteworthy to mention that a total of 209 such 'predictors' have been identified by the European Union FLARECAST (Flare Likelihood and Region Eruption predicting) project \citep{Georgoulis2021}. The FLARECAST project (2015-18)\footnote{http://flarecast.eu/} conclusively showed that, due to stochasticity in flare occurrence, flare prediction is an inherently probabilistic problem \citep{Campi2019}.\\ 

Recent developments in solar flare prediction have suggested that the prediction of flare onset can be improved by several hours, if key predictors are studied above the photosphere in the lower solar atmosphere (LSA). For example, from a study of 13 flare producing ARs of Solar Cycle 24 (SC24), \cite{Korsos2020} showed that it may be possible to improve prediction lead time by 2-8 hrs by tracking the temporal evolution of the $WG_M$ morphological parameter at a height range of 1000-1800 km in the LSA instead of carrying out the same exercise in the photosphere. This LSA height range of 1000-1800 km serves, then, as an Optimal Height Range (OHR). Motivated by these promising results, the objective of this paper is to systematically explore the concept of OHR with the 'R-value' parameter. An application of this concept is core to the Solar Activity Magnetic Monitor Network (SAMNet\footnote{http://hspf.eu/samnet.html}) that aims to achieve, in practice by their proposed ground-based sentinel network, an improved flare forecasting by determining the OHR \citep{Erdelyi2022}. \\

This paper is mainly centered around the R-value parameter, which in basic terms is a quantification of the unsigned flux near high-gradient PILs. The outline of this paper is as follows: Section \ref{sec:methods} focuses on the analysis procedure and it contains a detailed description of the algorithm used for the computation of R-value. The essence of the algorithm, which is to detect high-gradient PILs in a given magnetic field map, is described within the context of the second step, presented in Section {\ref{sec:methods}}. Section \ref{sec:ardataset} contains information on the AR dataset and the criteria used for the selection of ARs. The results and discussion are presented in Section \ref{sec:rd}, followed by the summary and conclusions in Section \ref{sec:cs}. Additional pertinent information is described in the Appendix. \\

\section{ANALYSIS METHOD}
\label{sec:methods}

\phantom{...}

The overall analysis procedure can be described in three main steps: \\

\begin{figure}[h]
\centering
\includegraphics[width=.85\linewidth]{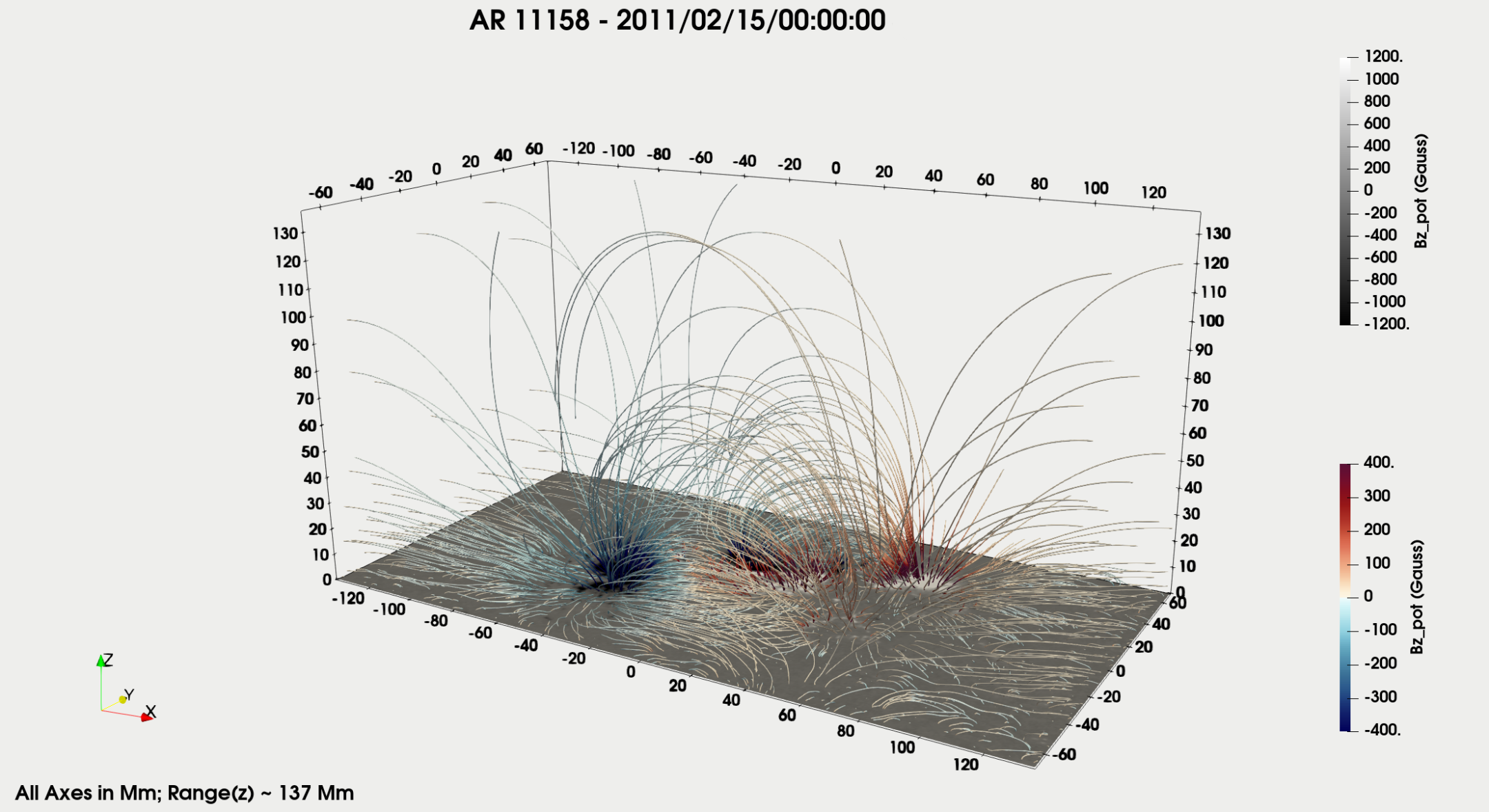}
\caption{A 3D visualisation of the PF-extrapolated magnetic field for AR 11158 at 00:00 UTC, 15 Feb 2011, created using Paraview (\href{https://www.paraview.org/}{https://www.paraview.org/}); the colourbar on top right denotes the $B_z$ values at the photosphere (map at the bottom of the grid, also reproduced in Figure 2a); the colourbar on bottom right denotes extrapolated $B_z$ values. The colourbars have been saturated to $\pm$ 1200 G on the photosphere and to $\pm$ 400 G above the photosphere. The data has been taken from an ISEE open-source database (\href{https://hinode.isee.nagoya-u.ac.jp/}{https://hinode.isee.nagoya-u.ac.jp/}), courtesy of \cite{Kusano2020}.}
\label{fig:extpf}
\end{figure}

First, for each AR, a Potential-Field (PF) extrapolation of the radial component of the magnetic field is performed using as input the SDO/HMI 2D Spaceweather HMI Active Region Patch (SHARP) vector magnetogram  corresponding to the AR  (\citealp{Pesnell2012}; \citealp{Scherrer2012}). If the area corresponding to a given SHARP includes more than one AR, it is suitably cropped to isolate the AR under focus. This is the case for NOAA ARs 11520 and 12017. The extrapolation is performed up to a height of 3.24 Mm in the LSA using 10 discrete planes, each spaced 0.36 Mm apart from its neighboring planes. The cadence of choice is 1 hr. We use a Linear Force Free Field (LFFF) extrapolation technique that relies on the Fast Fourier Transform approach. Naturally, a PF extrapolation is achieved by setting the force-free parameter $\alpha$ to zero \citep{alissandrakis81,Gary1989}. Although PF extrapolation has its limitations, it is precise enough for a first-hand estimate, mathematically simpler and quicker to operate compared to more sophisticated extrapolations \citep[e.g.,][]{Korsos2024}. By using the PF extrapolation, we are looking at the photospheric morphology over long time scales, without the intention of modelling dynamical features in the lower solar atmosphere (for reference, see \citealp{Wiegelmann2012}; for applications see \citealp{Korsos2020} and \citealp{Korsos2022}). The output of this step is a 3D data-grid of the vertical magnetic field component $B_z$ - we do not use the horizontal potential field. A sample visualisation of the 3D data-grid is shown in Figure \ref{fig:extpf}. \\

Second, starting with the 3D data-grid, the unsigned flux $\phi$ near PILs and R-value are computed in the 0-3.24 Mm height range at 1 hr cadence for the time-windows specified in Table \ref{table:dataset} (Section \ref{sec:ardataset}). The codes used to compute the unsigned magnetic flux near PILs and R-value are adapted from the FLARECAST Bitbucket Project Repository\footnote{https://dev.flarecast.eu/stash/projects/FE/repos}. The algorithm used for computing the R-value is an adaptation of the one described in \cite{Schrijver2007}. The difference is that we use the radial field component from vector magnetograms, while \cite{Schrijver2007} used the Line of Sight (LoS) component. Estimating the R-value relies on two input parameters: magnetic field threshold $B_{th}$ and separation distance $D_{sep}$, which control the identification of high-gradient PILs. The threshold $B_{th}$ is used to compute bitmaps corresponding to positive and negative flux. In the positive polarity bitmap, the elements are assigned the value '1' where $B_z > +B_{th}$ and '0' otherwise. Similarly, in the negative-polarity bitmap, the elements are assigned the value '1' where $B_z < -B_{th}$ and '0' otherwise. These bitmaps are then dilated and their product yields a map $M$ where high-field regions can be identified from non-zero values. The map '$M$', indicating high-polarity regions, is then convolved with an area-normalized Gaussian $G$ (characterized by a FWHM = $D_{sep}$), resulting in a weight map $W$ that assigns more weight to regions closer to high-gradient PILs as opposed to regions that are further apart (see Equation \ref{eq:1}).

\begin{equation}
    W = M(B_{th}) \ast G(D_{sep})
    \label{eq:1}
\end{equation}

\begin{figure}[h]
\centering

\begin{subfigure}
\centering
\includegraphics[width=.32\linewidth]{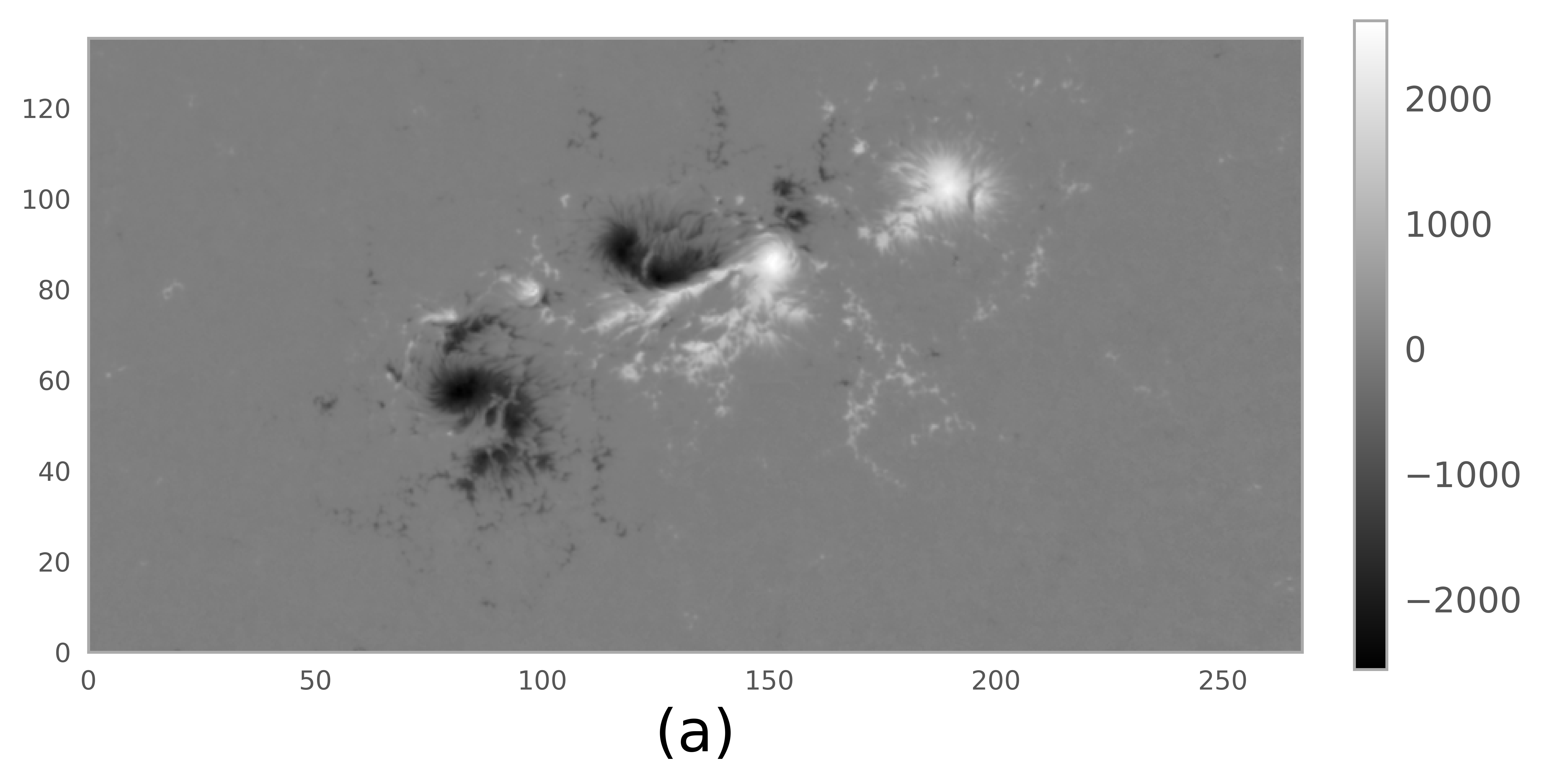}  
\end{subfigure}
\begin{subfigure}
\centering
\includegraphics[width=.32\linewidth]{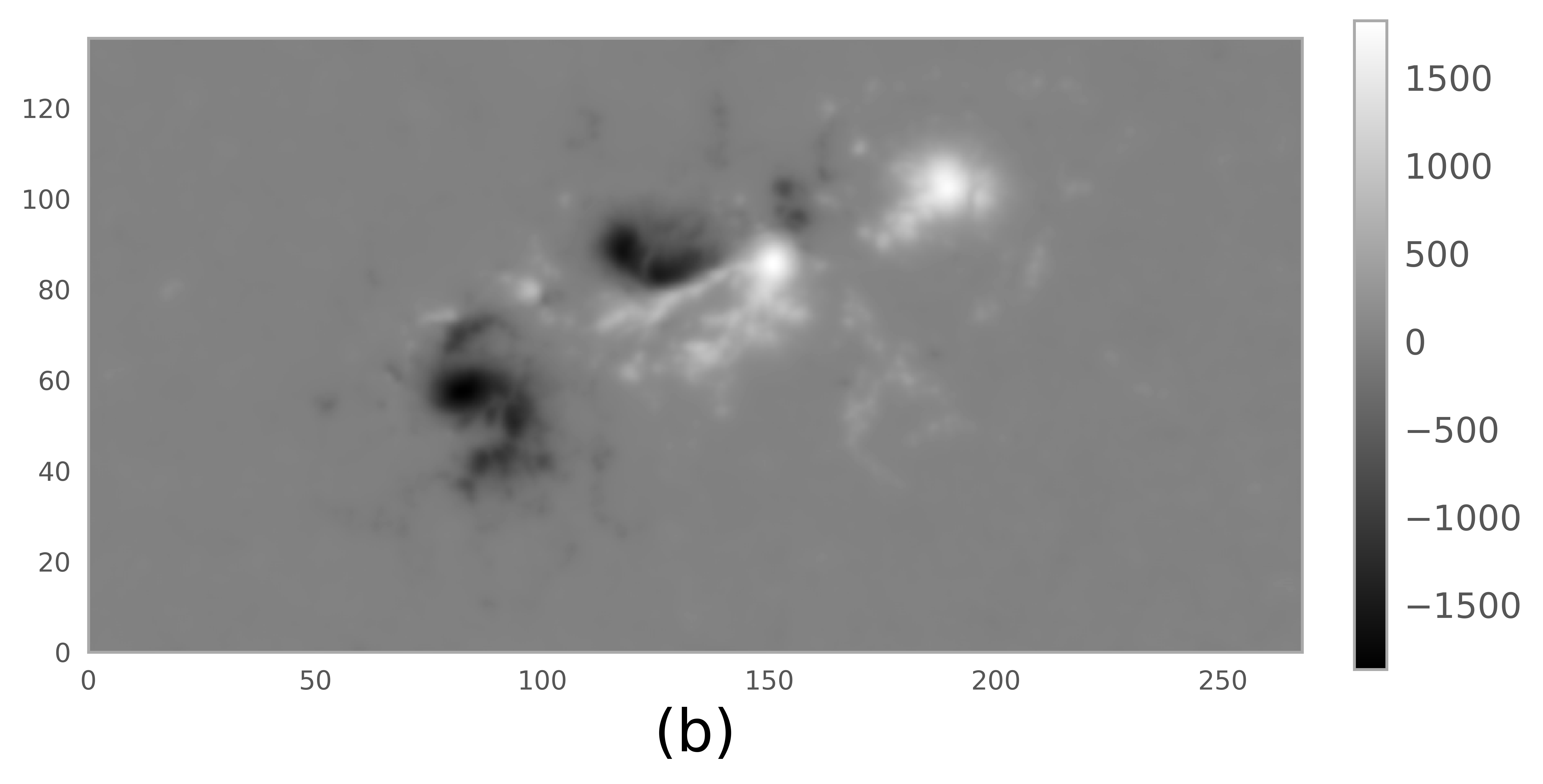}
\end{subfigure}
\begin{subfigure}
\centering
\includegraphics[width=.32\linewidth]{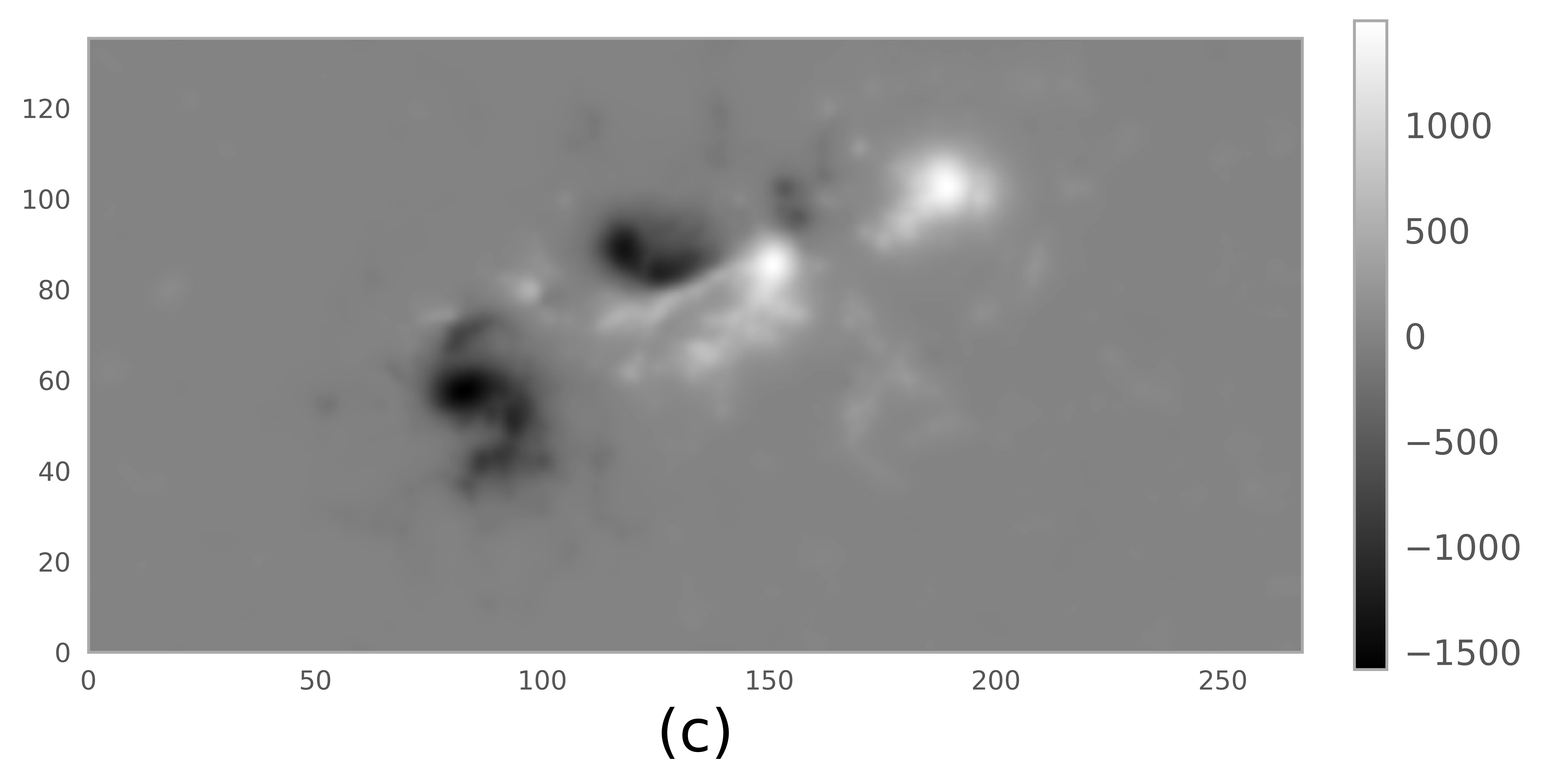}
\end{subfigure}


\begin{subfigure}
\centering
\includegraphics[width=.32\linewidth]{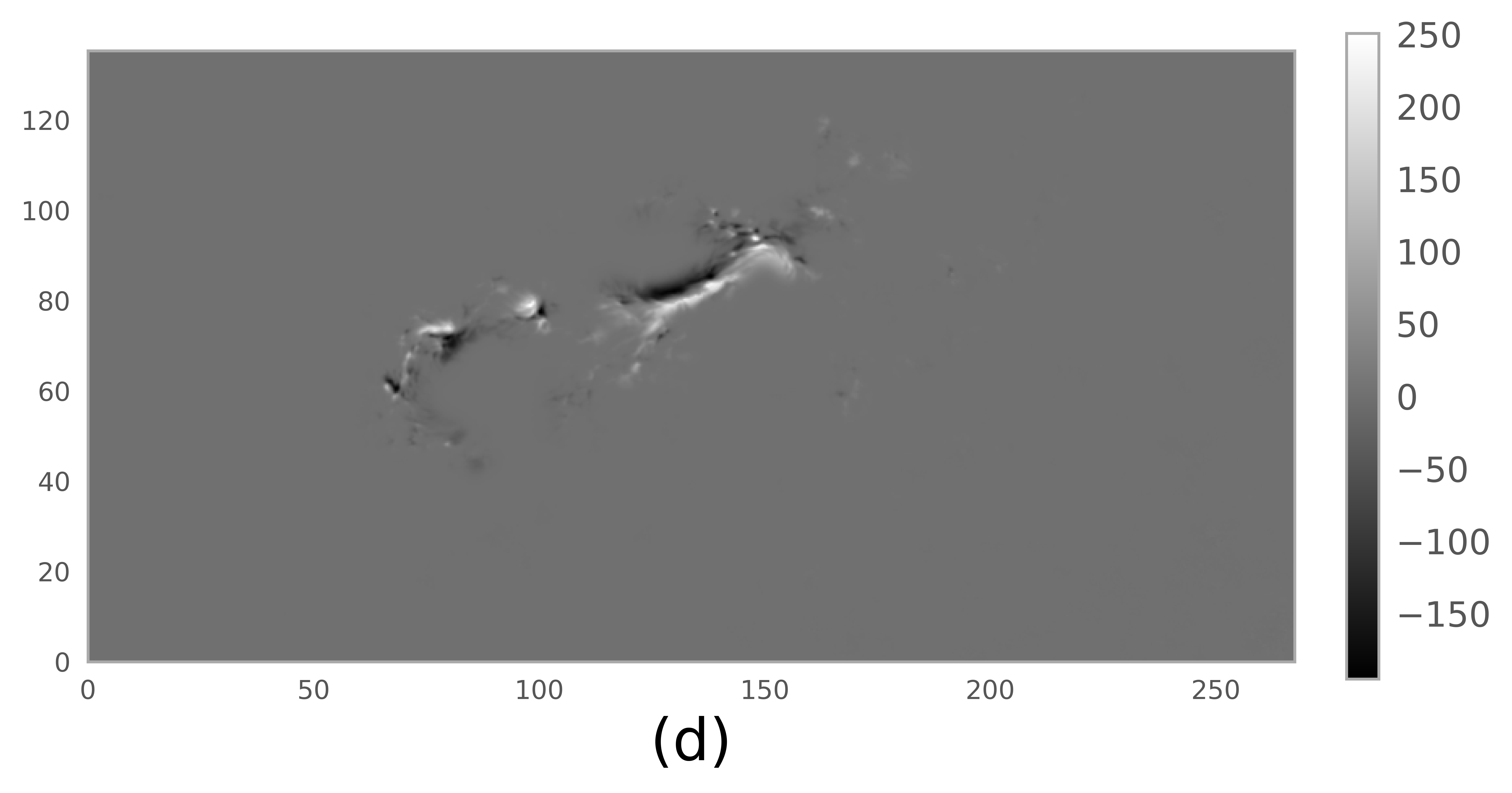}  
\end{subfigure}
\begin{subfigure}
\centering
\includegraphics[width=.32\linewidth]{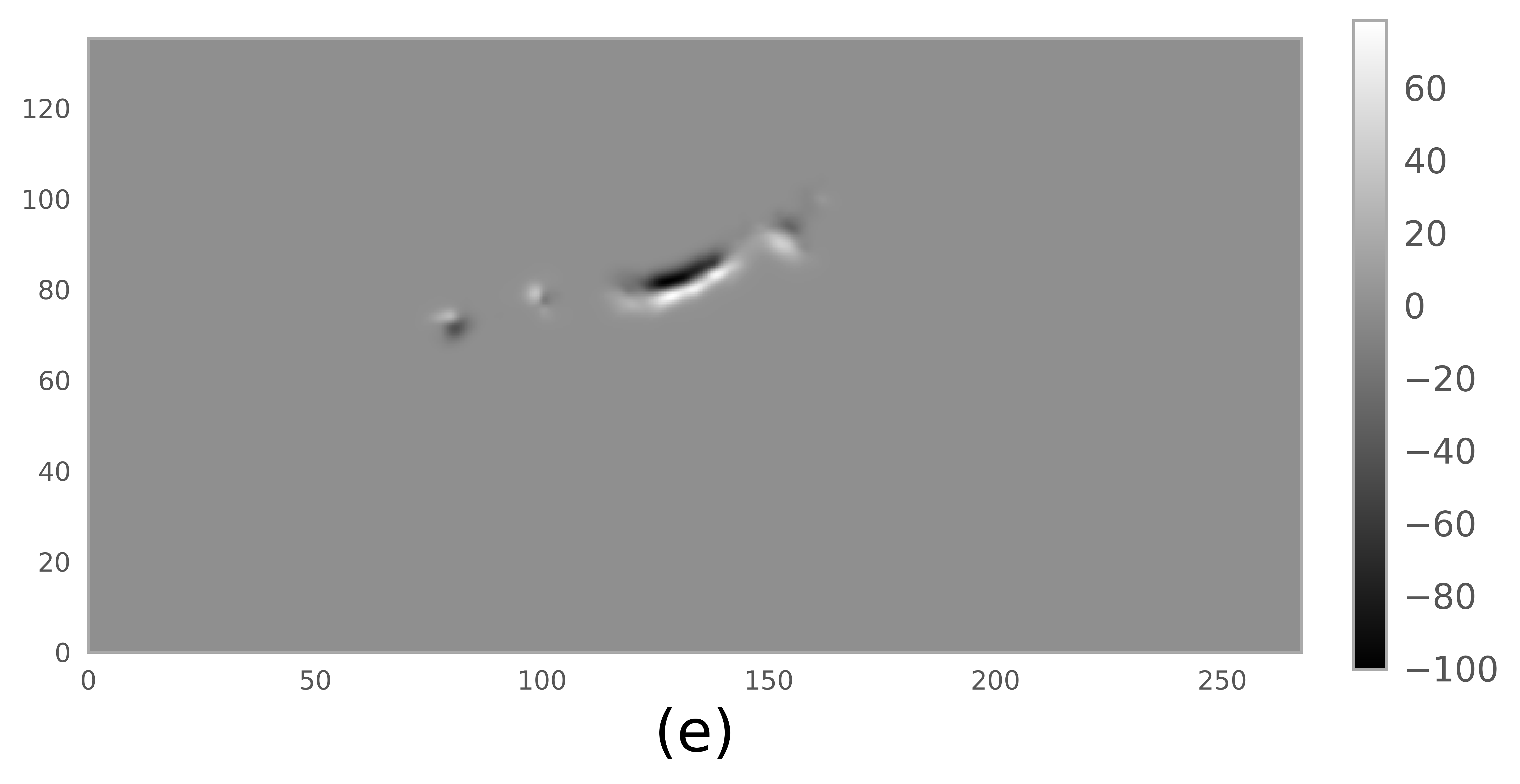}
\end{subfigure}
\begin{subfigure}
\centering
\includegraphics[width=.32\linewidth]{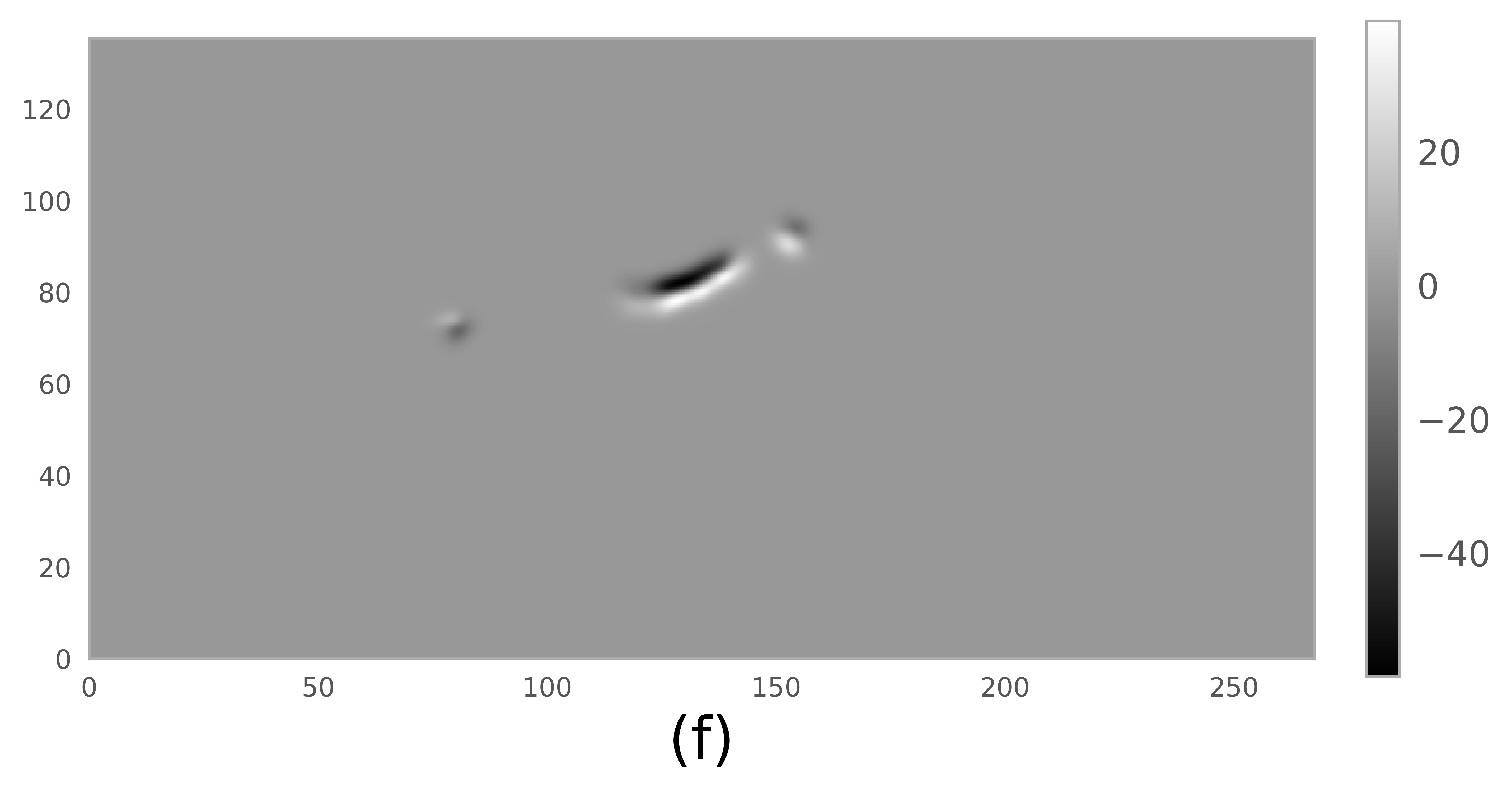}
\end{subfigure}


\begin{subfigure}
\centering
\includegraphics[width=.32\linewidth]{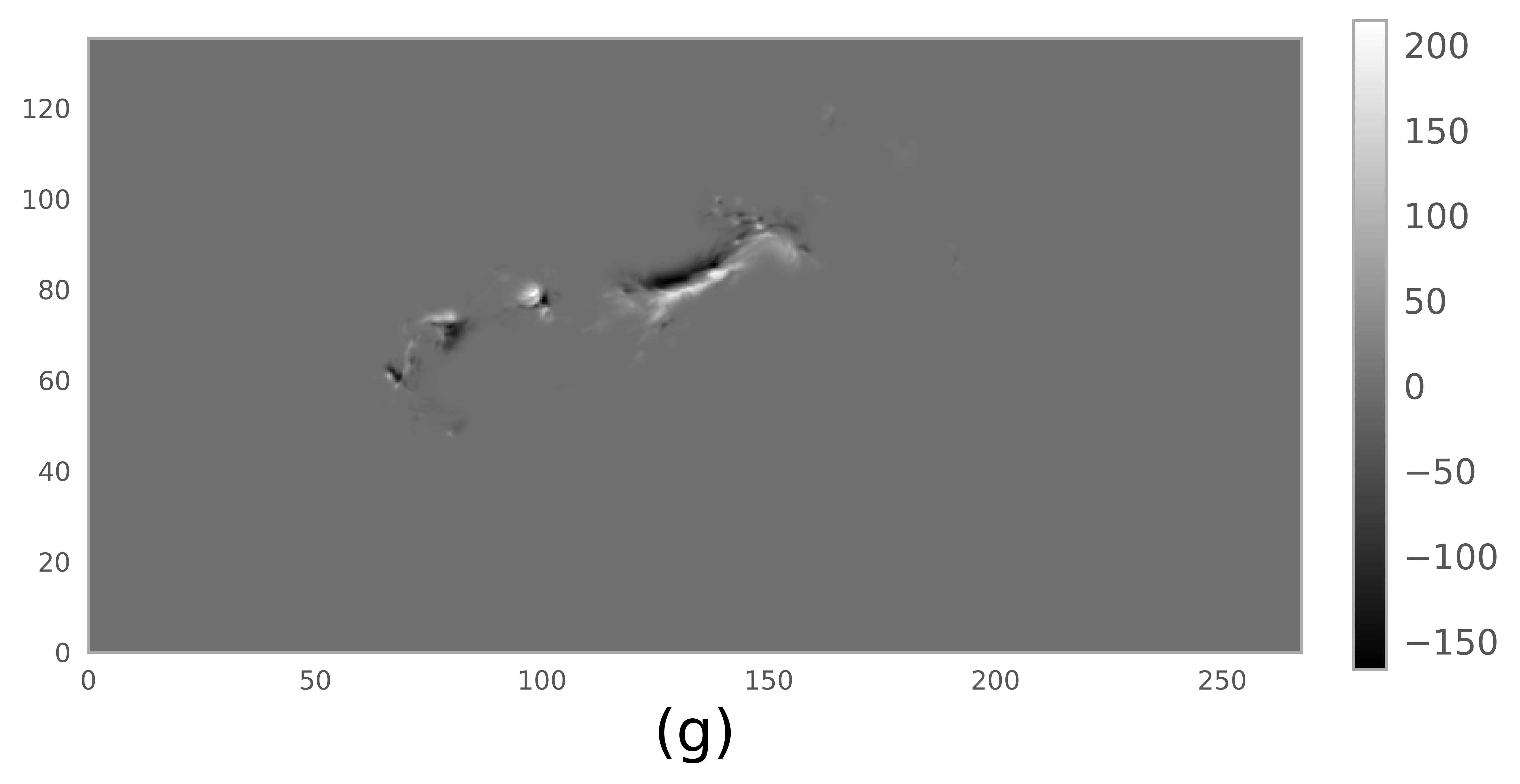} 
\end{subfigure}
\begin{subfigure}
\centering
\includegraphics[width=.32\linewidth]{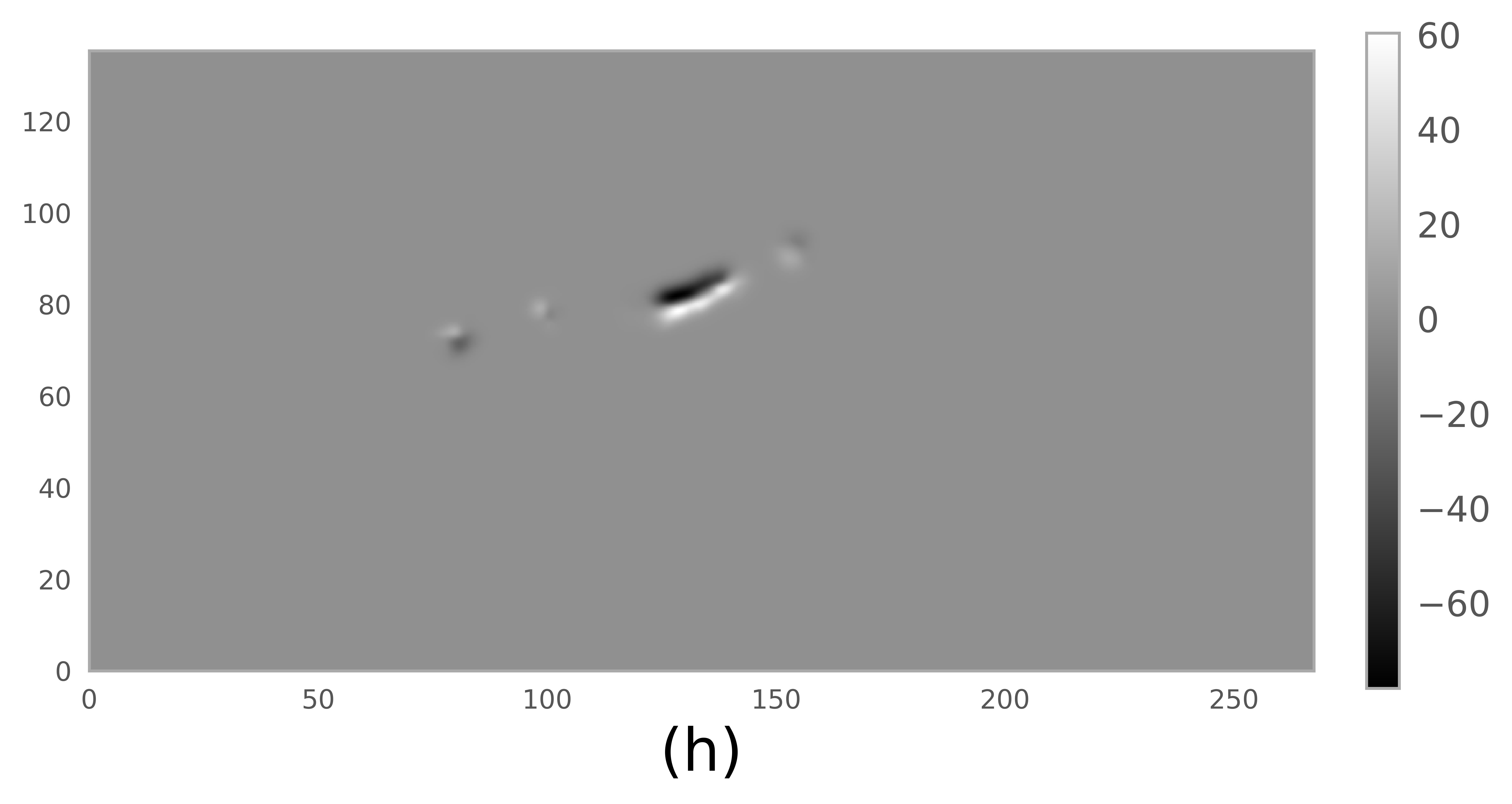}
\end{subfigure}
\begin{subfigure}
\centering
\includegraphics[width=.32\linewidth]{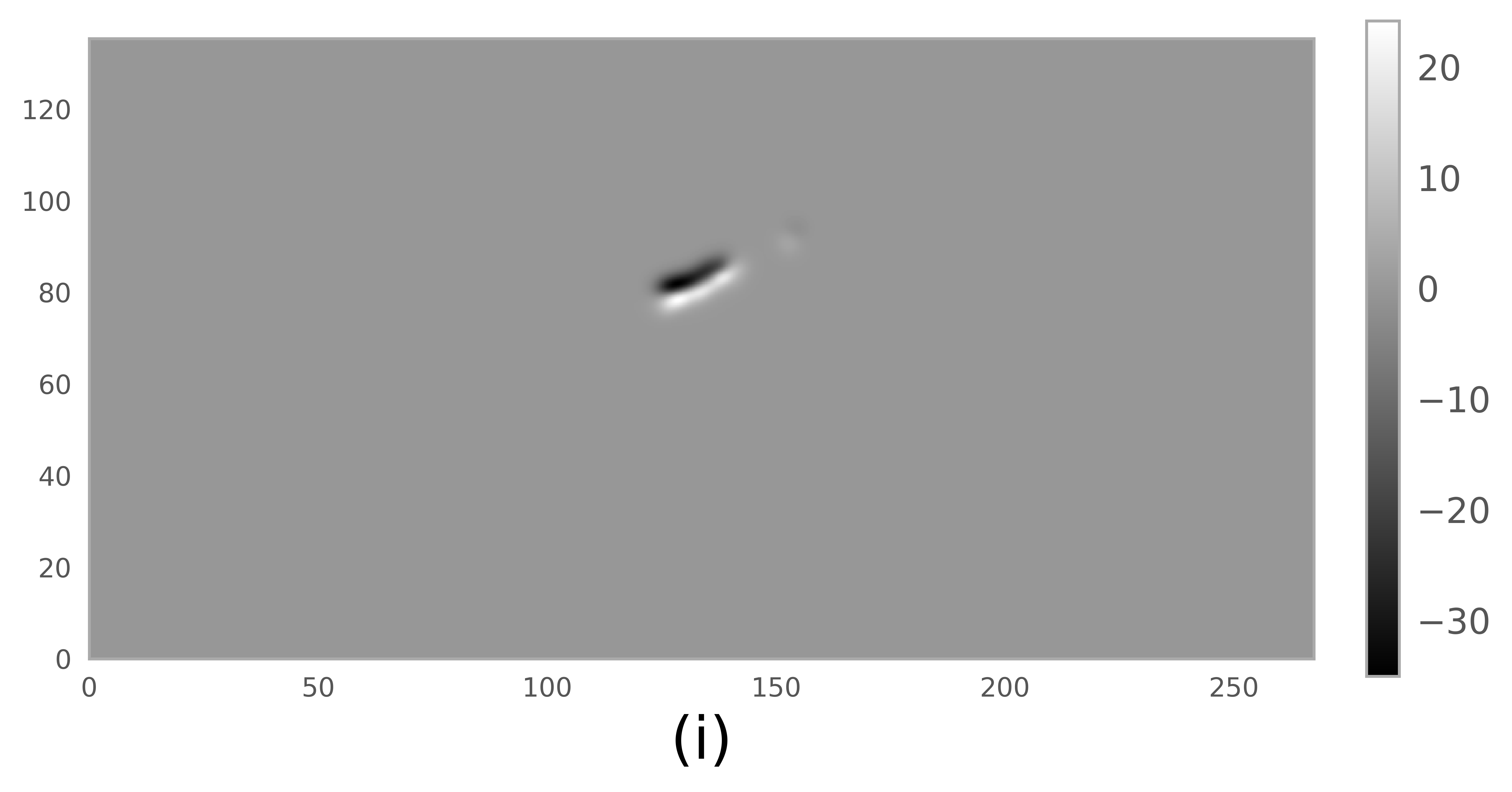}
\end{subfigure}


\begin{subfigure}
\centering
\includegraphics[width=.32\linewidth]{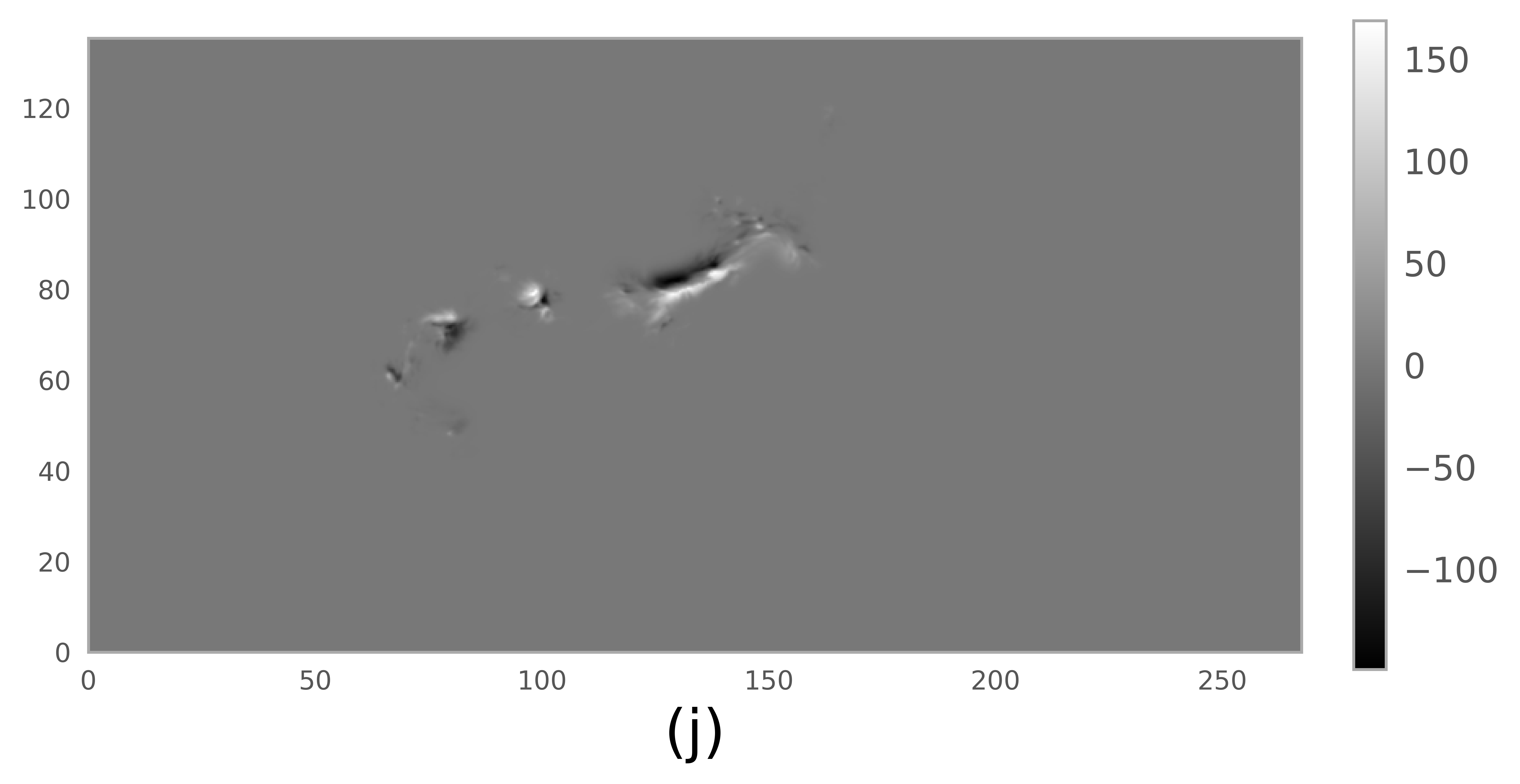} 
\end{subfigure}
\begin{subfigure}
\centering
\includegraphics[width=.32\linewidth]{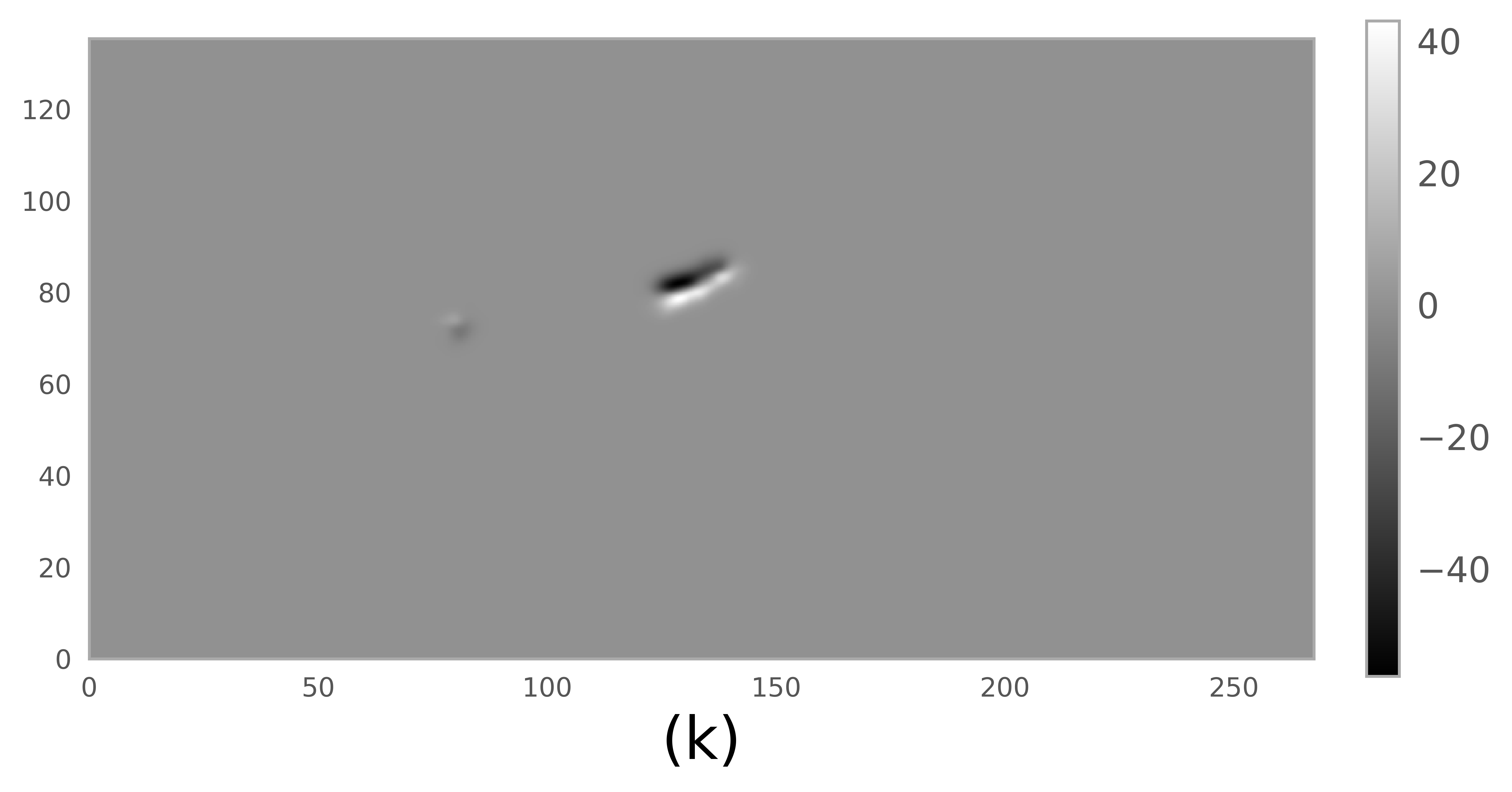}
\end{subfigure}
\begin{subfigure}
\centering
\includegraphics[width=.32\linewidth]{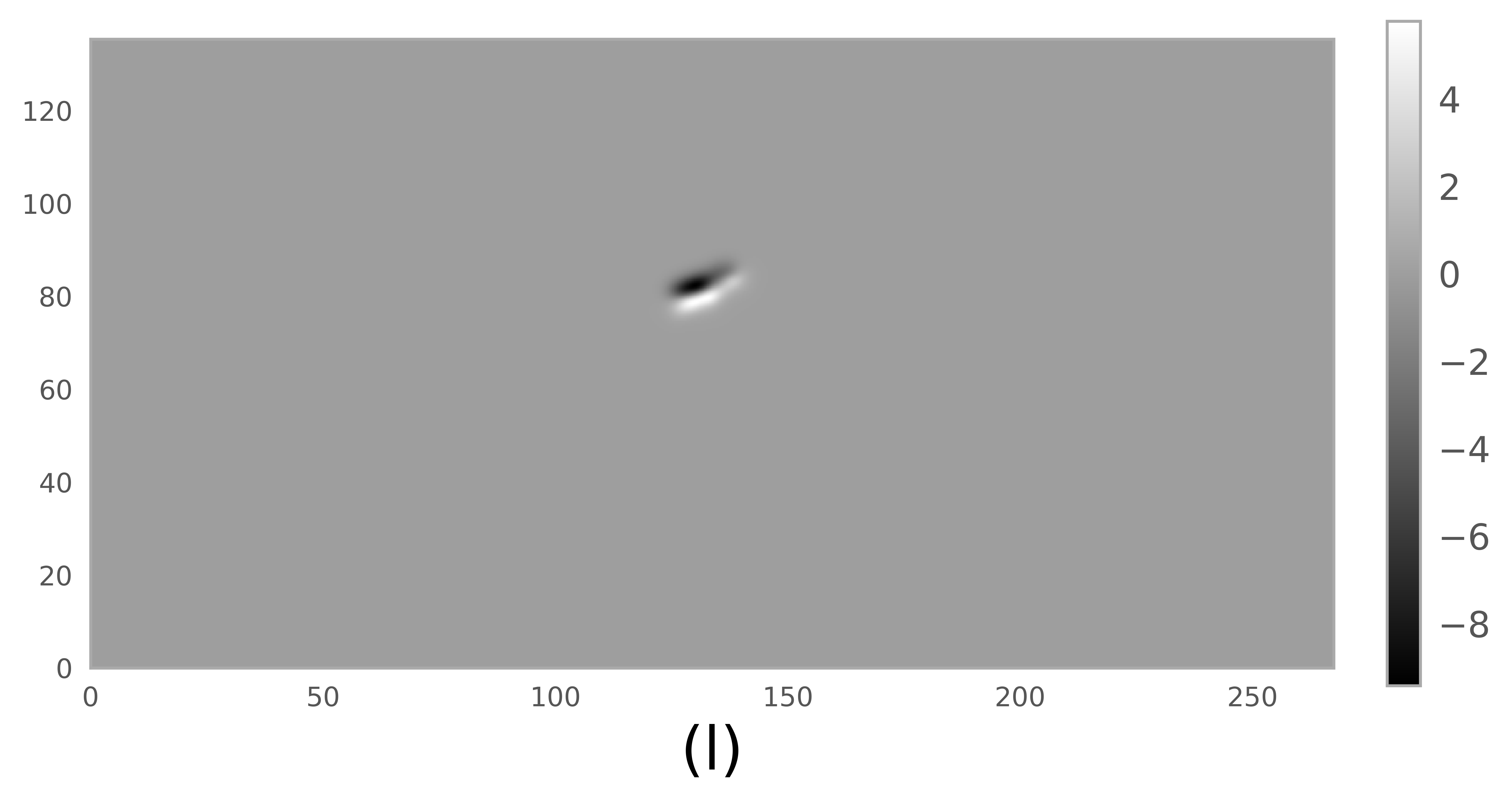}
\end{subfigure}

\caption{Plots corresponding to AR 11158 on 15/02/2011 00:00 UTC; a-c: $B_z$ maps at the photosphere, 1.08 Mm and 1.80 Mm, respectively; d-f: $R_{(50,15)}$ maps corresponding to magnetic field maps a-c, respectively; g-i: $R_{(100,15)}$ maps corresponding to magnetic field maps a-c, respectively; j-l: $R_{(150,15)}$ maps corresponding to magnetic field maps a-c respectively. The colourbars in magnetograms a-c denote $B_z$ (G), while in all other maps they represent $B_z$ (G) after the application of filtering. It is important to note how at higher altitudes (maps f, i and l), due to weakening of the field, the high-gradient regions tend to become more localised.}
\label{fig:3dextpf_sample1}
\end{figure}

This weight map is then multiplied with the original magnetogram data (or magnetic field map) $B_{map}$. Examples of the resulting output maps are shown in Figure \ref{fig:3dextpf_sample1}. The sum of absolute values of all elements multiplied with an area element $A$ gives the R-value (see Equation \ref{eq:2}). $A$ ($\sim 0.5\arcsec$) is approximately 1.3141 $\times$ $10^{15}$ cm$^{2}$ in CGS units.

\begin{equation}
    R_{val} = A \sum_{ij} |B_{map}|_{ij}.W_{ij}
    \label{eq:2}
\end{equation}

Third, following the R-value calculation in the photosphere and above, the data are visualised with the help of stack plots varying as a function of time (see the GitHub project repository\footnote{\href{https://github.com/shreeyesh-biswal/Rvalue_3D}{https://github.com/shreeyesh-biswal/Rvalue$\textunderscore$3D}} for all stack plots and codes). Since the exact dependence  of R-value on $B_{th}$ and $D_{sep}$ is not known, the R-value is computed for different combinations of $B_{th}$ and $D_{sep}$ (see Table \ref{table:modelset}). \citet{Schrijver2007} argued that, statistically, a threshold $B_{th}$ = 150 G could be used. Since the extrapolated fields are weaker than photospheric fields, and since our choice of $B_{th}$ does not vary with height, using a lower threshold for $B_{th}$ is helpful to identify high-gradient PILs at higher altitudes. The noise level associated with the photospheric data is $\sim$ 10 G \citep{Liu2012}. Therefore, it is not helpful to reduce $B_{th}$ below 50 G. On $D_{sep}$, \citet{Schrijver2007} found that two thirds of the values of the distribution for $D$ (i.e. the minimum distance between a PIL and the brightest point in the EUV images) were less than 15 Mm. Hence \citet{Schrijver2007} took $D_{sep}$ = 15 Mm for the computation of R-value on the photosphere. 

{\begin{table}[ht]
\begin{center}
\hskip-1.8cm
\begin{tabular}{| c | c | c | c |}
 \hline
 \multicolumn{4}{|c|}{} \\
 [-0.9 em]
 \multicolumn{4}{|c|}{Experimental Models for R-value} \\[0.4 em]
 \hline
 \phantom{x} & \phantom{x} & \phantom{x} & \phantom{x} \\ [-1.2em]
 No. & $B_{th}$ & $D_{sep}$ & Model Notation \\[0.2 em]
\hline
 01 & 150 G & 15 Mm & $R_{(150,15)}$ \\
 02 & 150 G & 10 Mm & $R_{(150,10)}$ \\
 03 & 100 G & 15 Mm & $R_{(100,15)}$ \\
 04 & 50 G & 15 Mm & $R_{(50,15)}$ \\
 05 & 50 G & 10 Mm & $R_{(50,10)}$ \\
\hline
\end{tabular}
\end{center}
\caption{Model specifications and notations for the calculation of the R-value; we alternate between values of 10 and 15 Mm for $D_{sep}$ and we use three different values for $B_{th}$, namely 50, 100, and 150 G.}
\label{table:modelset}
\end{table}}

\clearpage
\newpage

\section{ACTIVE REGION DATASET} 
\label{sec:ardataset}

\phantom{...}

We selected 8 ARs that hosted 11 X-class solar flares within a certain time interval of interest in each case (see Table \ref{table:dataset}). We chose these ARs and their corresponding temporal windows for two main reasons. First, in order to ensure that the magnetic field data is reliable for extrapolation. Since magnetic field observations have severe projection effects beyond $60^{\circ}$ from the solar central meridian \citep{Bobra2014}, all these ARs were located within $60^{\circ}$ EW throughout their corresponding time windows. Second, at any time within these time intervals of interest, each of the listed ARs hosted a $\delta$-sunspot, which is a feature indicating the presence of high-gradient PILs. All ARs have a minimum interval of 48 hrs between the observational start time and the corresponding flare onset. The objective is to examine whether knowledge of an imminent X-class flare is possible at least 24 hours in advance. Furthermore, we require that the SDO/HMI SHARP magnetogram data are continuously available for at least 120 hrs. The HMI data product used is the Lambert cylindrical equal-area projection of the photospheric vector magnetic field, identifiable by the following extension: hmi.sharp$\textunderscore$cea$\textunderscore$720s. \\

{\begin{table}[ht!]
\centering
\begin{tabular}{| c | c | c | c | c | c | c | c |}
 \hline
 \multicolumn{8}{|c|}{} \\
 [-0.9 em]
 \multicolumn{8}{|c|}{Observed Data} \\[0.4 em]
 \hline
 \phantom{x} & \phantom{x} & \phantom{x} & \phantom{x} & \phantom{x} & \phantom{x} & \phantom{x} & \phantom{x} \\ [-0.9em]
No. & AR & Class & $T_{start}$ &  $T_{end}$ & $T_{flare{\phantom{.}}onset}$ 
 & $T_{flare{\phantom{.}}onset}$ - $T_{start}$ (hrs) &  SHARP Size (Mm$^2$) \\
[0.3em]
\hline
 01 & 11158 & X2.2 & 2011/02/11 00:00 & 2011/02/17 00:00 & 2011/02/15 01:44 & 97.73 & 267.48 x 135.36 \\
 02 & 11166 & X1.5 & 2011/03/05 00:00 & 2011/03/10 00:00 & 2011/03/09 23:13 & 119.22 & 263.52 x 137.16 \\
 03 & 11283 & X2.1 & 2011/09/01 00:00 & 2011/09/09 00:00 & 2011/09/06 22:12 & 142.2 & 351.00 x 182.16 \\
 & 11283 & X1.8 & 2011/09/01 00:00 & 2011/09/09 00:00 & 2011/09/07 22:32 & 166.53 & 351.00 x 182.16 \\
 04 & 11520 & X1.4 & 2012/07/08 00:00 & 2012/07/15 00:00 & 2012/07/12 15:37 & 111.62 & 234.00 x 180.00 \\
 05 & 12017 & X1.0 & 2014/03/23 00:00 & 2014/04/01 00:00 & 2014/03/29 17:35 & 161.58 & 230.76 x 87.48 \\
 06 & 12158 & X1.6 & 2014/09/07 00:00 & 2014/09/14 00:00 & 2014/09/10 17:21 & 89.35 & 202.68 x 191.16 \\
 07 & 12297 & X2.1 & 2015/03/09 00:00 & 2015/03/15 00:00 & 
 2015/03/11 16:11 & 64.18 & 358.20 x 215.64 \\
 08 & 12673 & X2.2 & 2017/09/03 00:00 & 2017/09/08 00:00 & 2017/09/06 08:57 & 80.95 & 247.32 x 160.92 \\
 & 12673 & X9.3 & 2017/09/03 00:00 & 2017/09/08 00:00 & 
 2017/09/06 11:53 & 83.88 & 247.32 x 160.92 \\
 & 12673 & X1.3 & 2017/09/03 00:00 & 2017/09/08 00:00 & 2017/09/07 14:20 & 110.33 & 247.32 x 160.92 \\
 \hline
\end{tabular}

\phantom{...}
\caption{Table listing the details of the studied ARs, GOES flare classes, 
observation intervals (between $T_{start}$ and $T_{end}$), flare onset time ($T_{flare{\phantom{.}}onset}$), time difference between flare onset and start of the observing interval (in hours), and SHARP linear dimensions (in Mm$^2$).}
\label{table:dataset}
\end{table}}

\section{RESULTS AND DISCUSSION}
\label{sec:rd}

\phantom{...}

After a set of numerical sensitivity tests on several ARs, it was found that changing $D_{sep}$ from 15 to 10 Mm did not impact R-value as much as changing $B_{th}$ from 150 to 100 or 50 G. We took into account the statistical findings of \citealp{Schrijver2007} while selecting the numerical values for $B_{th}$ and $D_{sep}$ prior to the sensitivity tests. For the sake of brevity, one representative example is presented in Appendix A. For more examples, please visit the GitHub project repository. For simplicity, we keep $D_{sep}$ fixed at 15 Mm, considering only R-value results obtained from the following pairs of thresholds $R_{(B_{th},D_{sep})}$: $R_{(150,15)}$, $R_{(100,15)}$ and $R_{(50,15)}$. The similarity of patterns in (i) $R_{(50,15)}$ and $R_{(50,10)}$ and (ii) $R_{(150,15)}$ and $R_{(150,10)}$ can be seen in Figure \ref{fig:11166} (Appendix A). The pre-flare R-value trends are classified into two categories based on whether the unsigned flux increases or decreases near PILs, before the occurrence of first X-class flare for each AR. However, there was also the case of AR 11283 that fell into neither of these categories and it has been discussed in detail in Appendix B. In Section \ref{ssec:rd_nf}, the variation of R-value for non-flaring cases has been presented to help the reader understand how it differs from the cases immediately before an X-class flare (refer to Sections \ref{ssec:rd_inc} and \ref{ssec:rd_dec}). Assuming that pre-flare conditions prior to the occurrence of X-class flares in ARs are radically different from quiescent ARs, we have also explored the R-value in height and time for a few intermediate cases. In Appendix C, 4 ARs have been discussed where the R-value is studied prior to the occurrence of the first M-class flare in each case. A similar study, but for an AR with a $\delta$ sunspot hosting only C-class flares, is also presented in Appendix C. A detailed discussion on the lead time is presented in Section \ref{ssec:inf}. Finally, Section \ref{ssec:swx} presents a preliminary idea of how we envision the R-value to be relevant towards the development of a novel flare-prediction method.

\begin{figure}[h]
\centering
\textbf{Sample plots for Unsigned Flux near PILs (in $10^{20}$ Mx)}\par\medskip

\begin{subfigure}
  \centering
  \includegraphics[width=.32\linewidth]{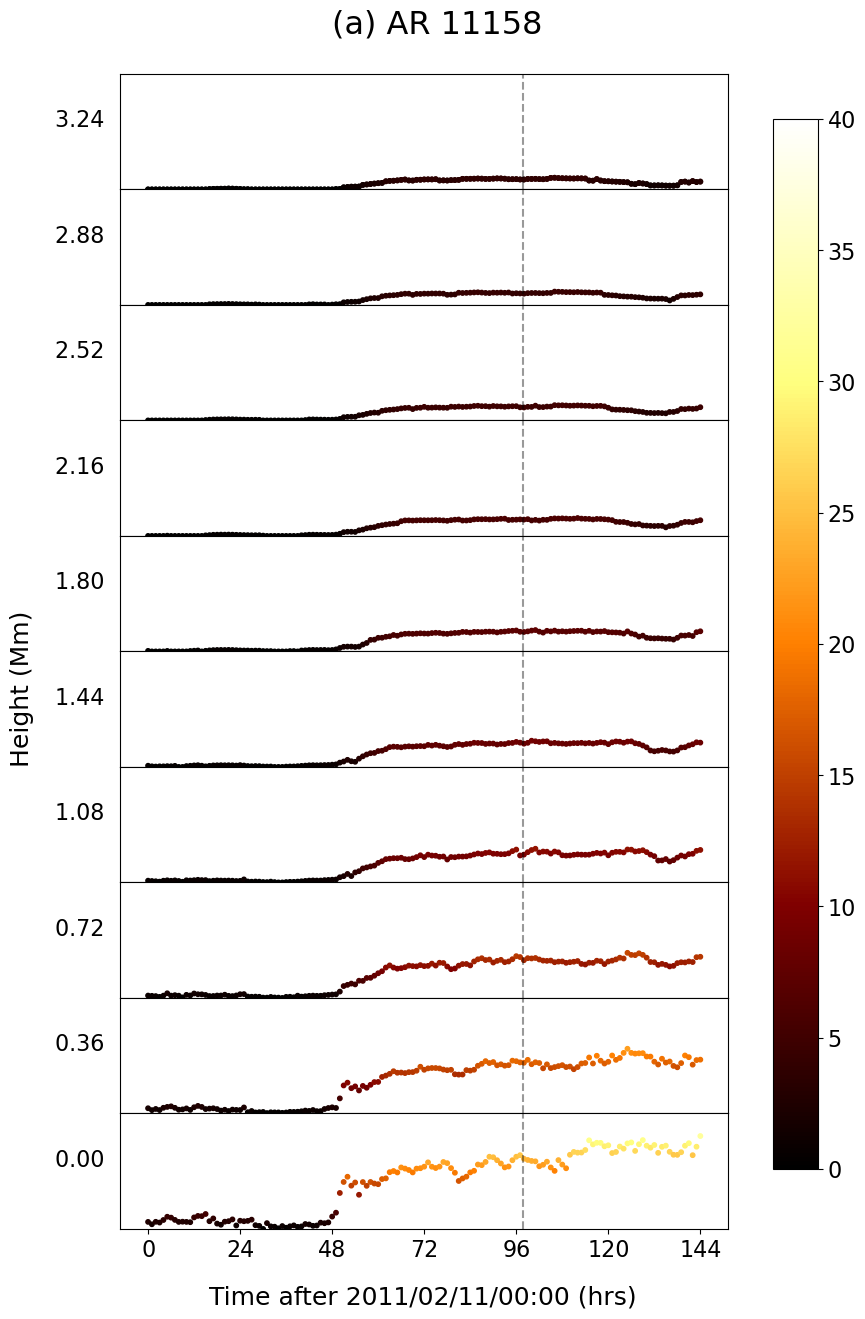}  
\end{subfigure}
\begin{subfigure}
  \centering
  \includegraphics[width=.32\linewidth]{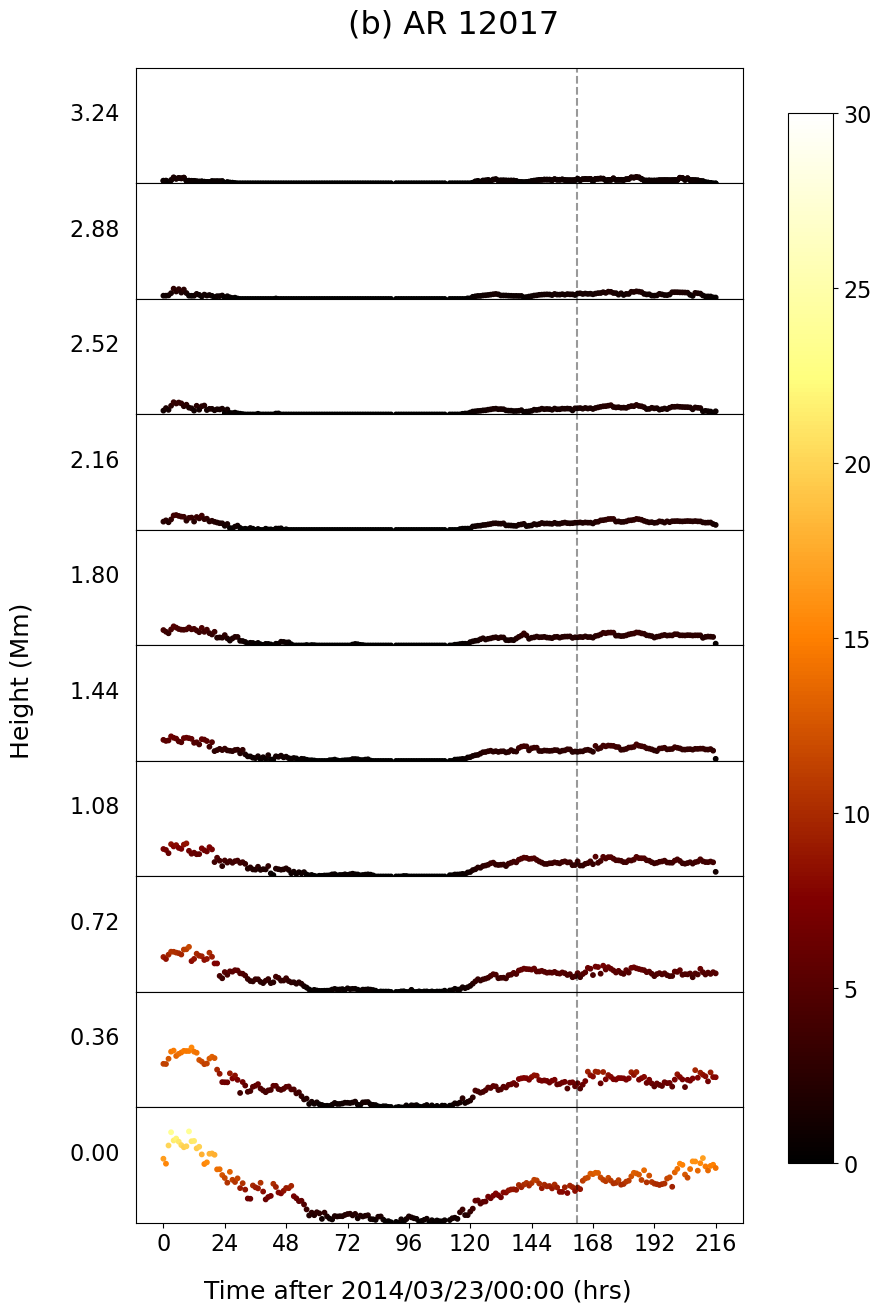}
\end{subfigure}
\begin{subfigure}
  \centering
  \includegraphics[width=.32\linewidth]{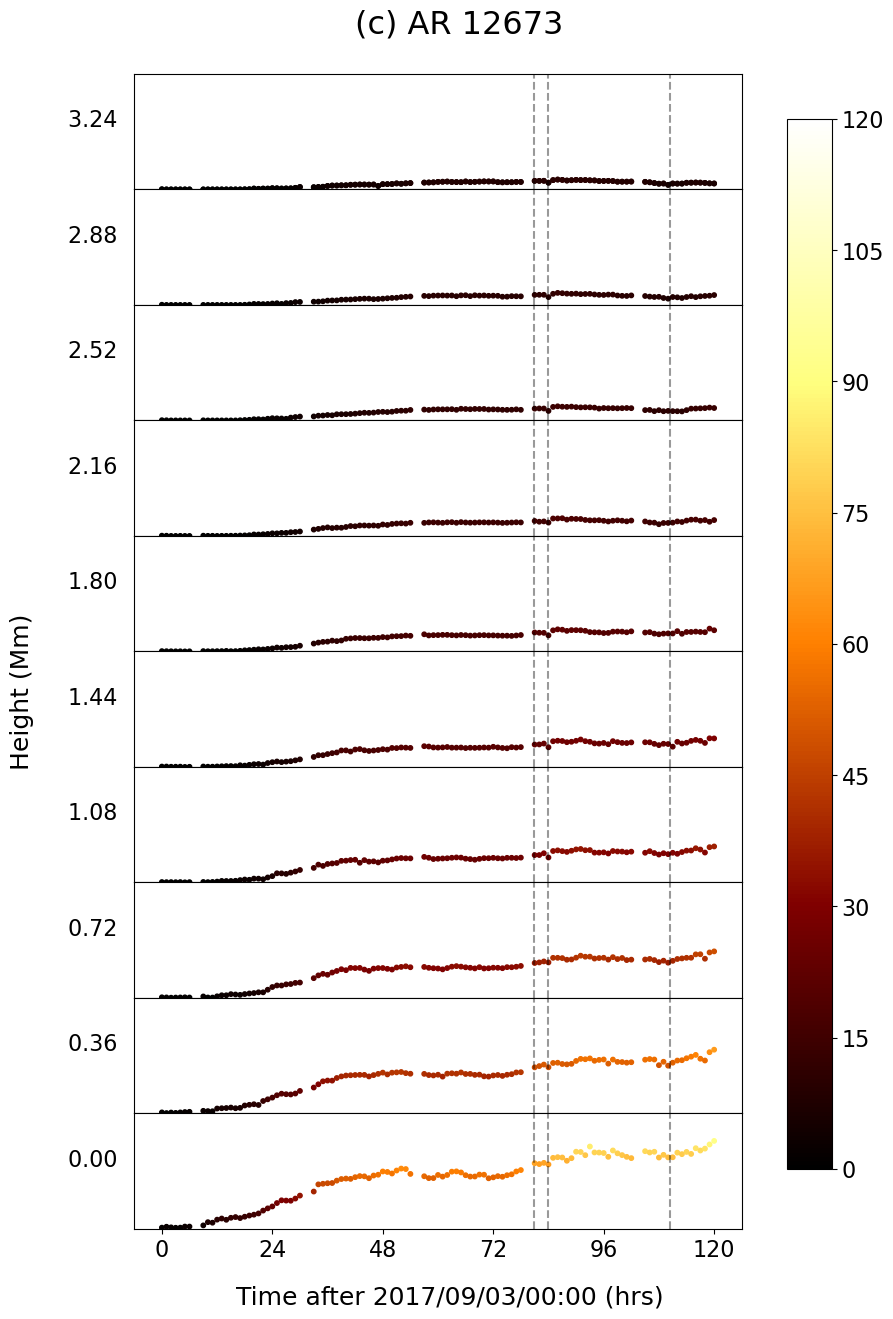}
\end{subfigure}

\caption{Multiple-height stack plots for unsigned flux near PILs for (a) AR 11158, (b) AR 12017 and (c) AR 12673; colourbar indicates the unsigned flux (in $10^{20}$ Mx); Flux emergence is observed prior to the occurrence of X-class flare in each case (marked by vertical dashed lines in each plot).}
\label{fig:fluxclubbed}
\end{figure}

\subsection{ARs associated with a prior increase in the unsigned flux near PILs}
\label{ssec:rd_inc}

\phantom{...}

\begin{figure}[!ht]
\vspace*{-0.58cm}
\centering
\textbf{R-value plots for AR 11158}\par\medskip

\begin{subfigure}
  \centering
  \includegraphics[width=.32\linewidth]{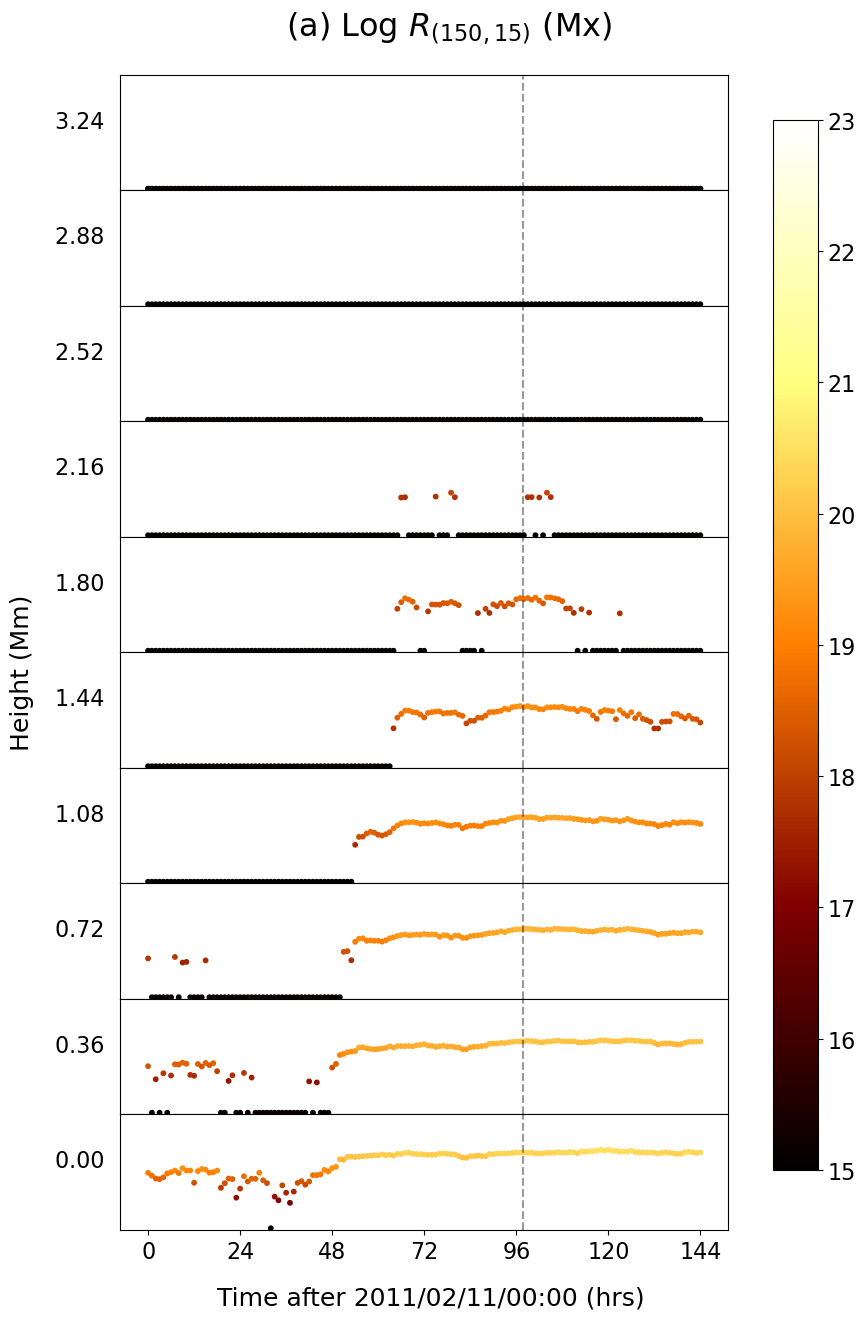}  
\end{subfigure}
\begin{subfigure}
  \centering
  \includegraphics[width=.32\linewidth]{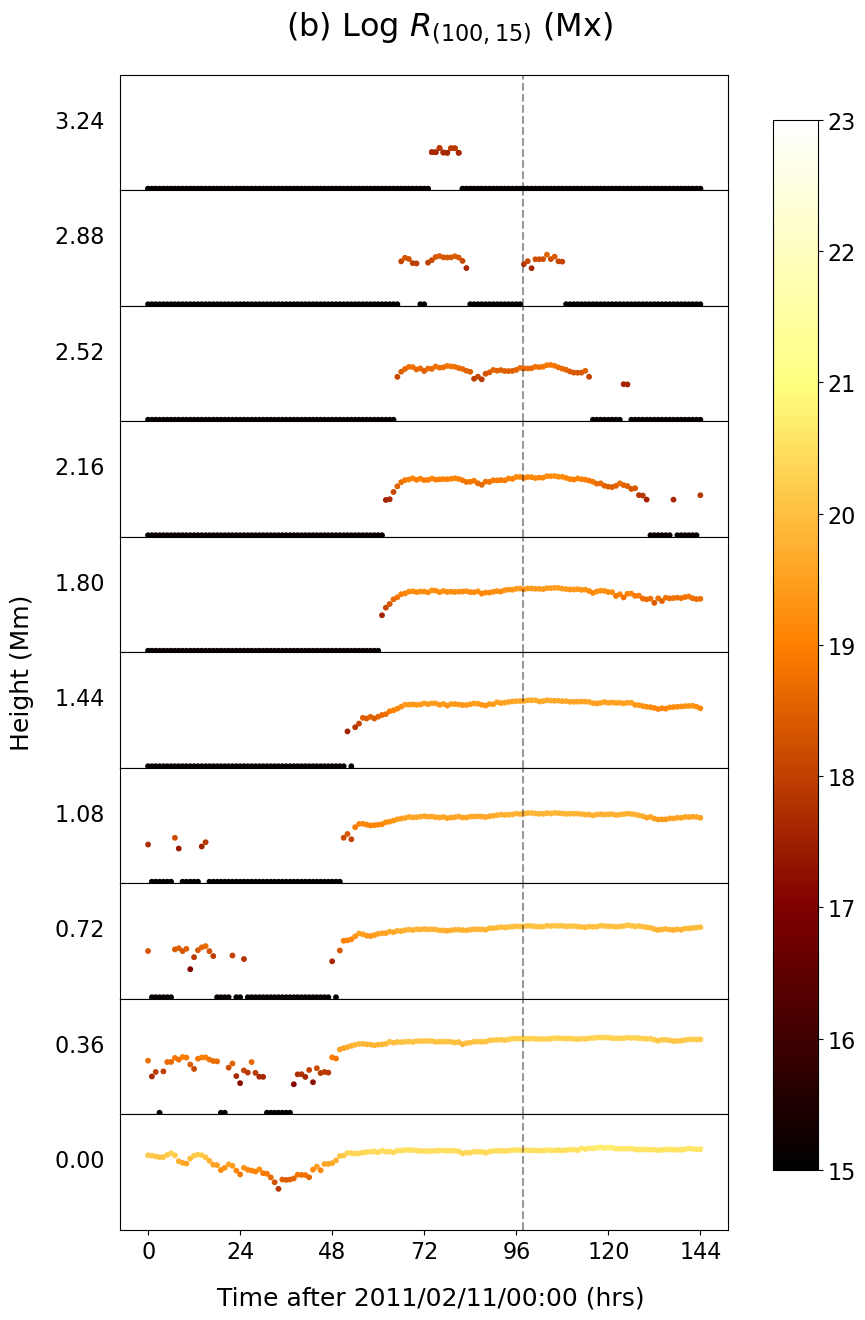}
\end{subfigure}
\begin{subfigure}
  \centering
  \includegraphics[width=.32\linewidth]{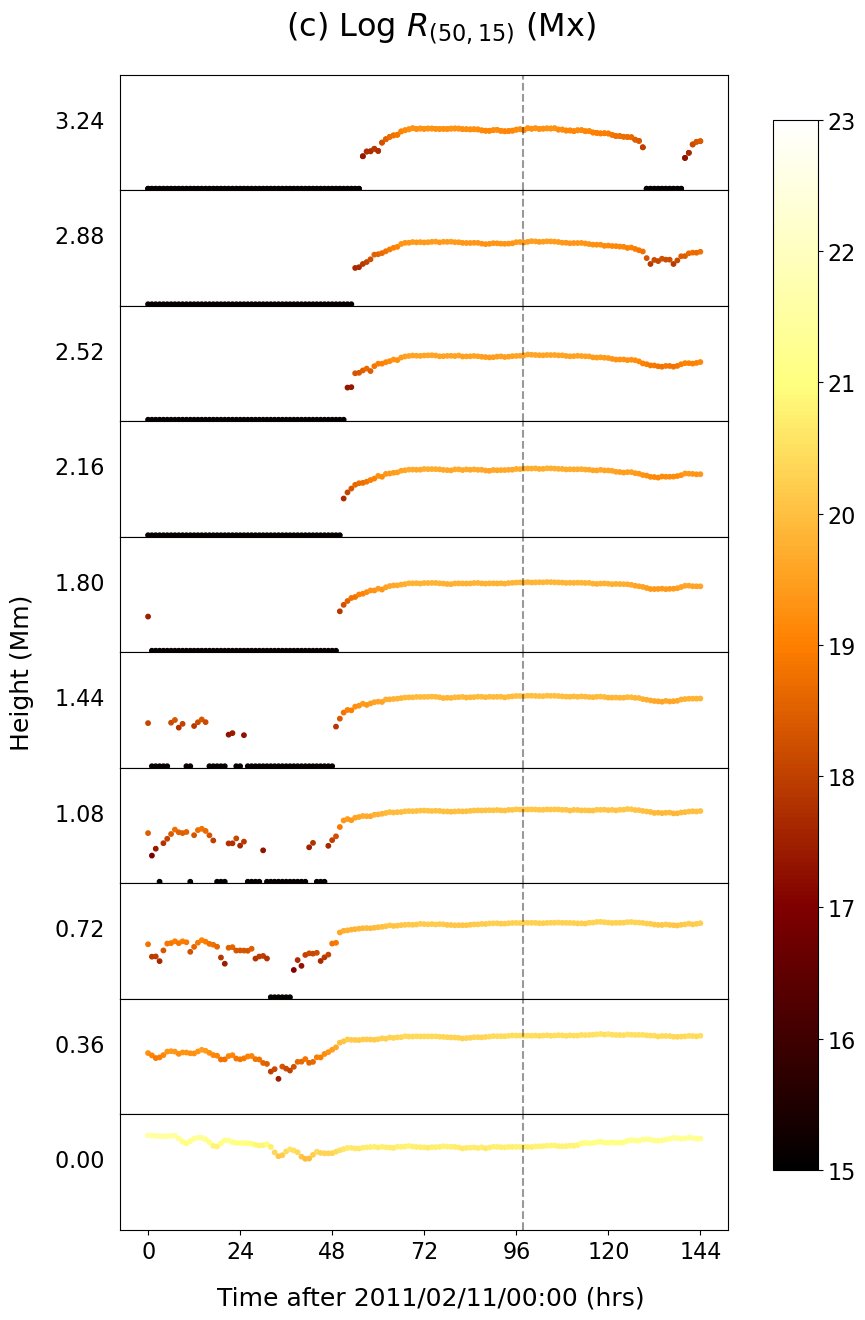}
\end{subfigure}

\caption{Multiple-height stack plots for (a) $R_{(150,15)}$; (b) $R_{(100,15)}$; (c) $R_{(50,15)}$; vertical dashed line indicates the time of occurrence of the X2.2 flare; colourbar indicates the logarithm of R-value (in Mx); any black lines or points indicate null output; The OHRs in this case are as follows: (a) 0.36 - 1.44 Mm, (b) 0.36 - 2.52 Mm, (c) 1.08 - 3.24 Mm; For $R_{(50,15)}$, we do not consider 0.72 Mm within the OHR as the black line is continuous for less than 6 hrs}
\label{fig:11158}
\end{figure}

Before the occurrence of the first X-class flare for ARs 11158, 11166, 12017 and 12673, a significant increase is seen in both the unsigned flux and R-value (see Figure \ref{fig:11166}a in Appendix A for AR 11166 and Figure \ref{fig:fluxclubbed} for other ARs). Qualitatively, therefore, there is some degree of correlation between the temporal variation of the R-value on the photosphere and unsigned flux near PILs for these ARs (compare Figure \ref{fig:fluxclubbed}a with Figures \ref{fig:11158}a, \ref{fig:11158}b and \ref{fig:11158}c, especially at the panel corresponding to the photosphere). It is already known that flux emergence near PILs is linked to X-class flaring activity \citep{Toriumi2022}. However, since we find the jump in R-value at about the same time, we hypothesise that the jump in R-value could be linked to flaring activity, although a statistical study may still be needed to verify this. The exact procedure of identifying a jump in R-value is discussed in the next paragraph. For the purpose of simplicity, we introduce two parameters; (i) $T_{fe}$ denoting the time of flux emergence near PILs on the photosphere and (ii) $T_{fo}$ denoting the latest time-stamp in our dataset just before flare onset. For ARs 11158, 12017 and 12673, $T_{fe}$ indicates the time when a sharp increase in the unsigned flux was observed. For AR 11166, $T_{fe}$ indicates the time when the flux shows a consistently increasing trend following a period of decrease. Table~\ref{table:fluxrval} lists the values of the unsigned flux around PILs and the photospheric R-values at $T_{fe}$ and $T_{fo}$. Using these values we can make a quantitative estimate of the increase of these parameters. For example, in case of AR 11158, the flux at $T_{fo}$ was about five times the value at $T_{fe}$, while the photospheric R-value at $T_{fo}$ was about ten times the value at $T_{fe}$. This suggests that for AR 11158, high-gradient PILs contribute to a higher share in the total flux around PILs immediately before a flare compared to the time when flux emergence is observed. Similar trends are seen for ARs 11166, 12017 and 12673 (refer to Table \ref{table:fluxrval}). 

{\begin{table}[ht!]
\begin{center}
\hskip-3.0cm
\begin{tabular}{| c | c | c | c | c | c | c |}
 \hline
 \multicolumn{7}{|c|}{} \\
 [-0.8em]
 \multicolumn{7}{|c|}{Unsigned Flux $\phi$ near PILs and $R_{(150,15)}$ (both in 10$^{20}$ Mx) at $T_{fe}$ and $T_{fo}$} \\
 [-1em]
 \multicolumn{7}{|c|}{} \\
 \hline
 \phantom{x} & \phantom{x} & \phantom{x} & \phantom{x} & \phantom{x} & \phantom{x} & \phantom{x} \\ [-0.9em]
 AR & $T_{fe}$ & $\phi$ ($T_{fe}$) & $R_{(150,15)}$ (0 Mm) & $T_{fo}$ & $\phi$ ($T_{fo}$) & $R_{(150,15)}$  (0 Mm) \\ 
 [0.3em]
 \hline
 11158 & 2011/02/13 01:00 & 5.63 & 0.24 & 2011/02/15 01:00 & 25.54 & 2.31  \\
 11166 & 2011/03/06 16:00 & 5.93 & 0.18 & 2011/03/09 23:00 & 30.76 & 2.40 \\
 12017 & 2014/03/28 00:00 & 2.94 & 0.05 & 2014/03/29 17:00 & 8.30 & 0.76 \\
 12673 & 2017/09/03 18:00 & 12.70 & 0.87 & 2017/09/06 06:00 & 61.17 & 7.35 \\
\hline
\end{tabular}
\end{center}
\caption{Table comparing changes in $R_{(150,15)}$ and $\phi$ for ARs 11158, 11166, 12017 and 12673 at $T_{fe}$ and $T_{fo}$ in the photosphere}    
\label{table:fluxrval}
\end{table}}   

{\begin{table}[!ht]
\begin{center}
\hskip-1.9cm
\begin{tabular}{| c | c | c | c |}
 \hline
 \multicolumn{4}{|c|}{} \\
 [-0.8em]
 \multicolumn{4}{|c|}{R-value increase times $T_{rv}$ for different R-value models (AR 11158)} \\
 [-1em]
 \multicolumn{4}{|c|}{} \\
 \hline
 \phantom{x} & \phantom{x} & \phantom{x} & \phantom{x} \\ [-0.9em]
 Height (in Mm) & $T_{rv}{[R_{(150,15)}]}$ & $T_{rv}{[R_{(100,15)}]}$ & $T_{rv}{[R_{(50,15)}]}$ \\ 
 [0.3em]
 \hline
 0.00 & - - - & - - - & - - - \\
 0.36 & 2011/02/13 00:00 & 2011/02/12 14:00 & - - - \\
 0.72 & 2011/02/13 03:00 & 2011/02/13 00:00 & - - - \\
 1.08 & 2011/02/13 06:00 & 2011/02/13 03:00 & 2011/02/12 23:00 \\
 1.44 & 2011/02/13 16:00 & 2011/02/13 07:00 & 2011/02/13 01:00 \\
 1.80 & - - - & 2011/02/13 13:00 & 2011/02/13 02:00 \\
 2.16 & - - - & 2011/02/13 14:00 & 2011/02/13 03:00 \\
 2.52 & - - - & 2011/02/13 17:00 & 2011/02/13 04:00 \\
 2.88 & - - - & - - - & 2011/02/13 06:00 \\
 3.24 & - - - & - - - & 2011/02/13 08:00 \\
\hline
\end{tabular}
\end{center}
\caption{R-value increase times within the OHRs for AR 11158; $T_{rv}$ for any given model, indicates the first time-stamp where non-zero R-value output was computed following a continuous period of null values}
\label{table:emergence}

\end{table}}

\begin{figure}[!ht]
\centering
\textbf{R-value plots for AR 12017}\par\medskip

\begin{subfigure}
  \centering
  \includegraphics[width=.32\linewidth]{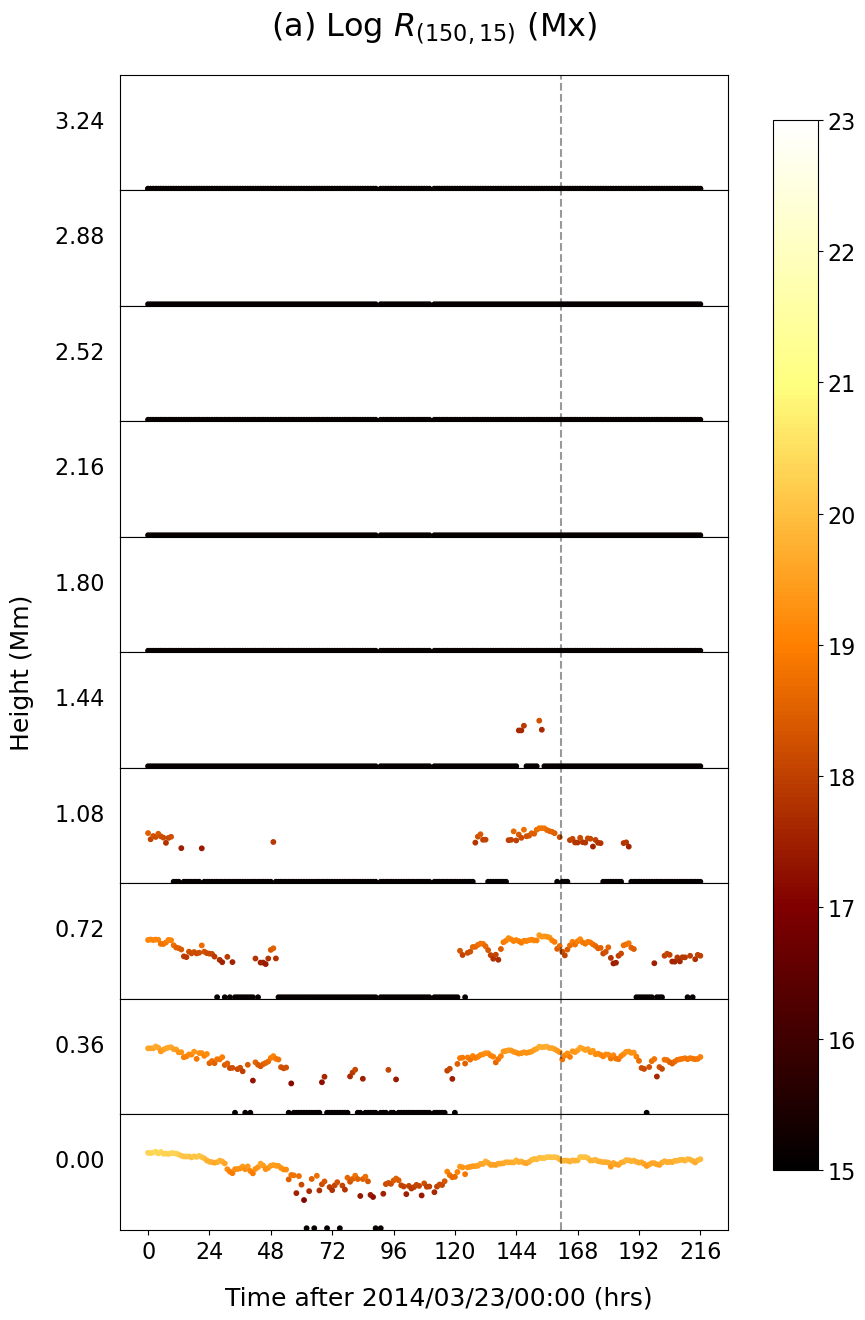}  
\end{subfigure}
\begin{subfigure}
  \centering
  \includegraphics[width=.32\linewidth]{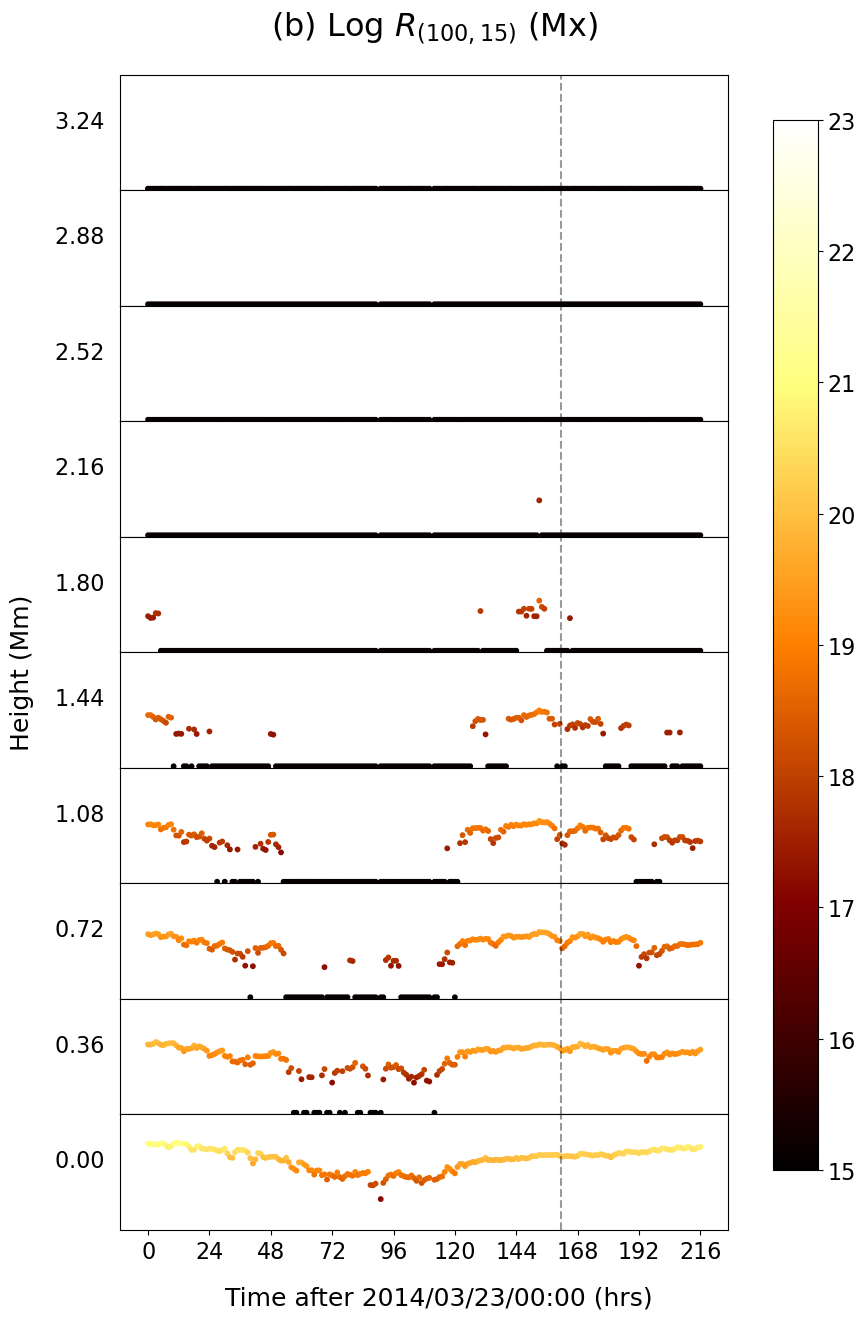}
\end{subfigure}
\begin{subfigure}
  \centering
  \includegraphics[width=.32\linewidth]{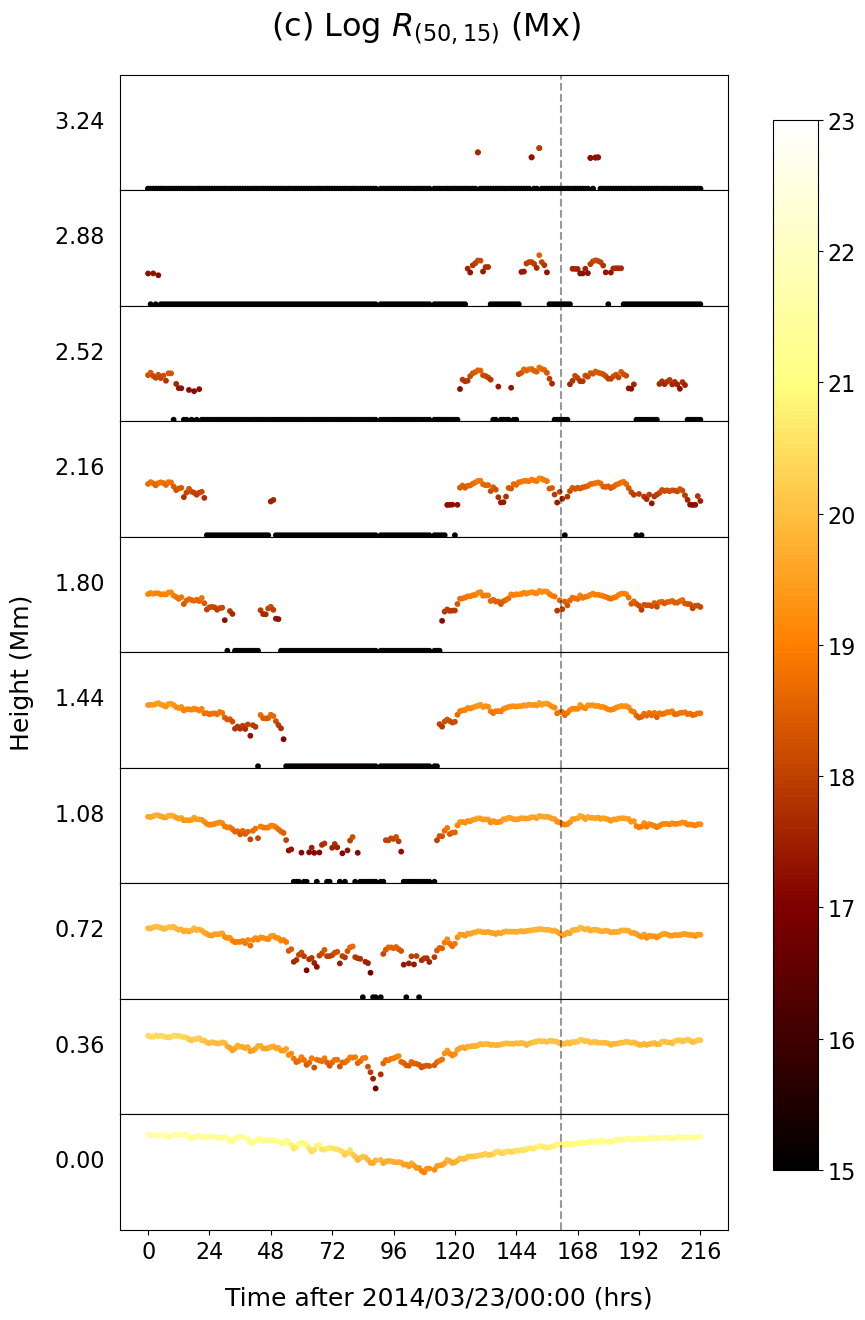}
\end{subfigure}

\caption{Multiple-height stack plots for (a) $R_{(150,15)}$; (b) $R_{(100,15)}$; (c) $R_{(50,15)}$; vertical dashed line indicates the time of occurrence of the X1.0 flare; colourbar indicates the logarithm of R-value (in Mx); any black lines or points indicate null output; The OHRs in this case are as follows: (a) 0.36 - 1.08 Mm, (b) 0.72 - 1.44 Mm, (c) 1.08 - 2.16 Mm
}
\label{fig:12017}

\end{figure}

\phantom{...}

\begin{figure}[h]
\vspace*{-0.58cm}
\centering
\textbf{R-value plots for AR 12673}\par\medskip

\begin{subfigure}
  \centering
  \includegraphics[width=.32\linewidth]{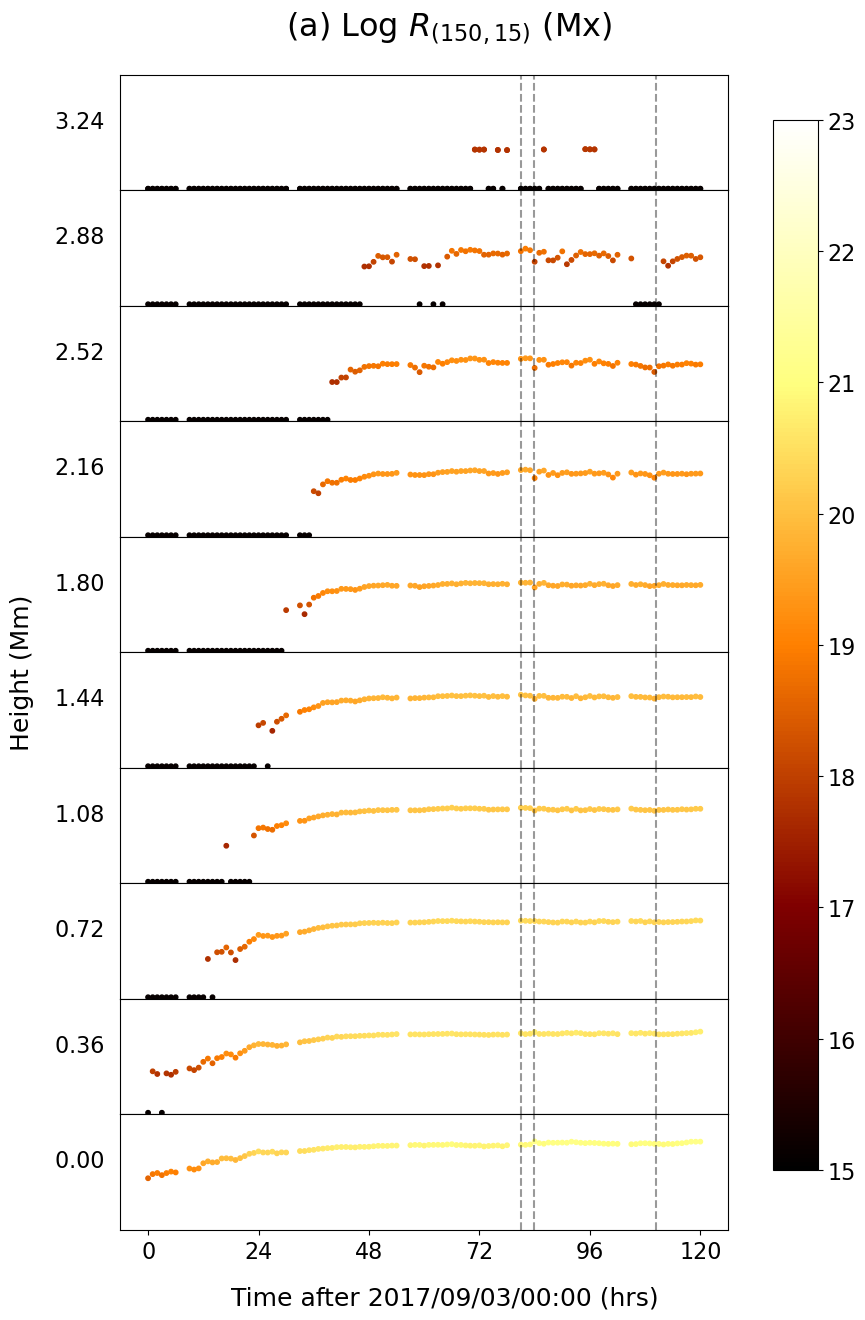}  
\end{subfigure}
\begin{subfigure}
  \centering
  \includegraphics[width=.32\linewidth]{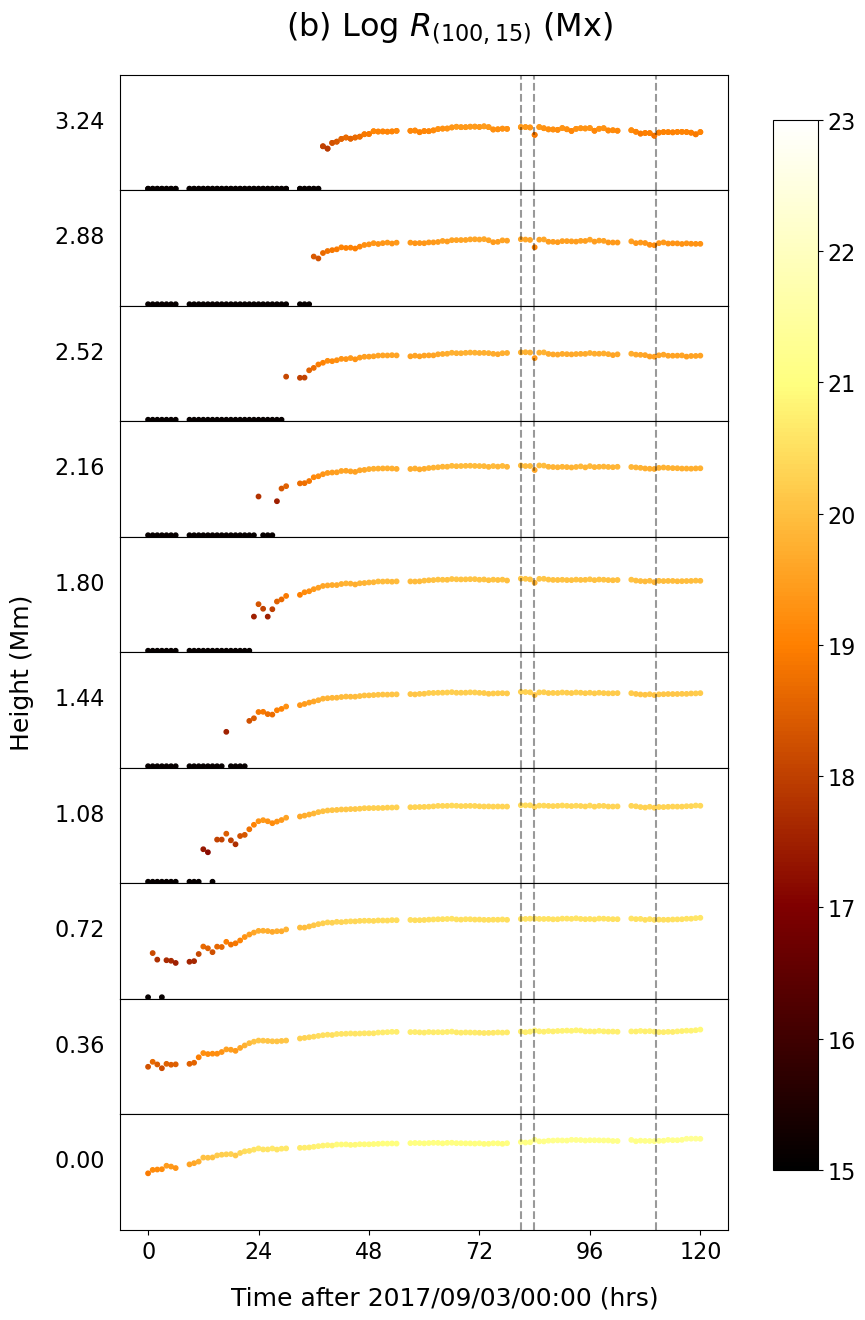}
\end{subfigure}
\begin{subfigure}
  \centering
  \includegraphics[width=.32\linewidth]{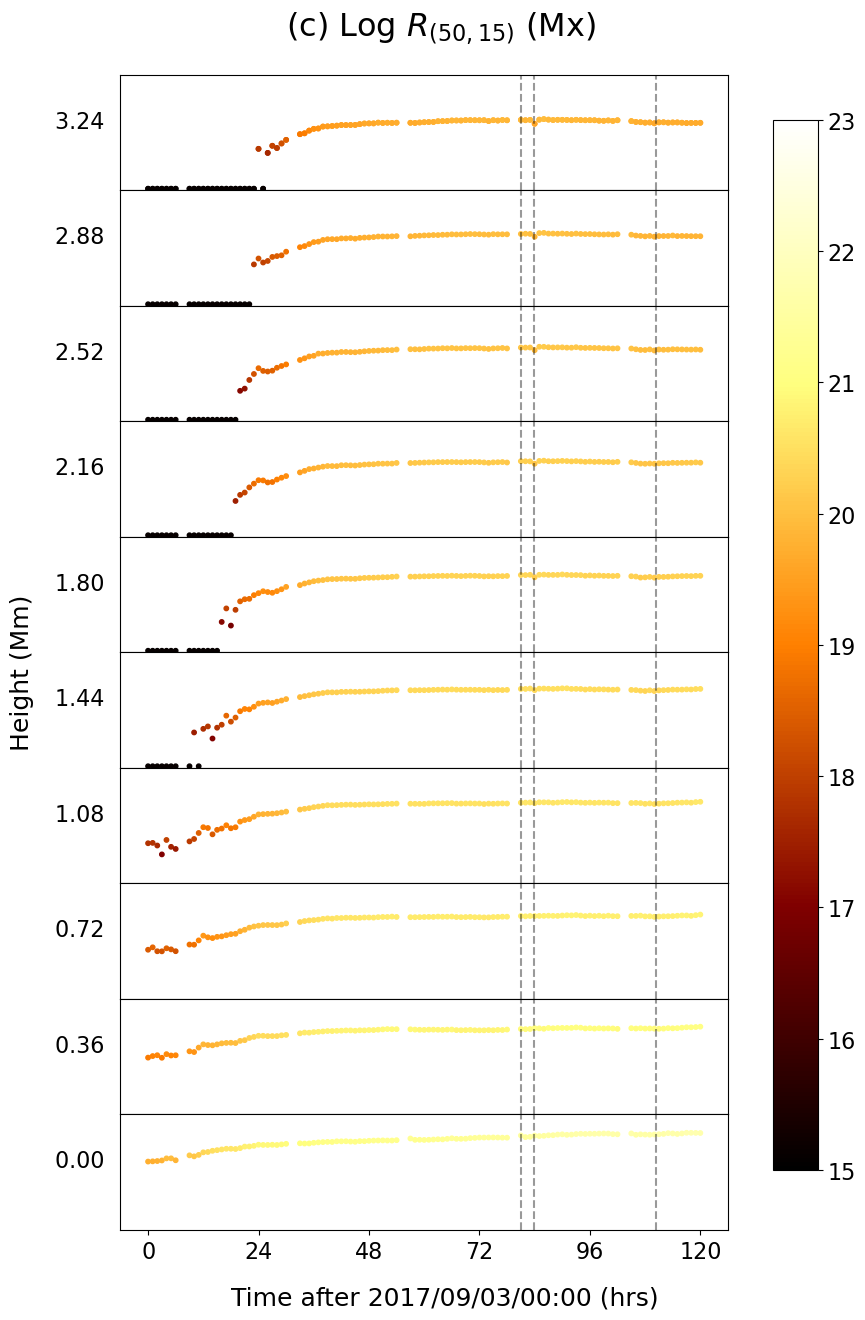}
\end{subfigure}

\caption{Multiple-height stack plots for (a) $R_{(150,15)}$; (b) $R_{(100,15)}$; (c) $R_{(50,15)}$; vertical dashed lines indicate time of occurrence of X-class flares; colourbar indicates the logarithm of R-value (in Mx); any black lines or points indicate null output; The OHRs in this case are as follows: (a) 0.72 - 2.88 Mm, (b) 1.08 - 3.24 Mm, (c) 1.44 - 3.24 Mm}
\label{fig:12673}

\end{figure} 

AR 11158 hosted the first X-class flare of SC24 at 01:44 UTC on Feb 15, 2011. The sunspot group that was $\beta\gamma$-type on Feb 11, 2011, transformed into a $\beta\gamma\delta$-type sunspot on Feb 16, within a day of the eruption of the X2.2 flare. It can be seen from Figure~\ref{fig:fluxclubbed}a that flux levels on Feb 16 were higher compared to the pre-flare levels (Feb 11-12). Let us consider the $R_{(150,15)}$ trends in height for AR 11158 (Figure~\ref{fig:11158}a). It may be seen that at some time around $T_{fe}$ at 0.36 Mm altitude, the black line (denoting null output from the code; owing to weak fields not breaching $B_{th}$) disappears and $R_{(150,15)}$ suddenly jumps. This suggests that strong magnetic flux begins to emerge at 0.36 Mm at around $T_{fe}$. This is also indicative of a high-gradient PIL setup in the extrapolated magnetic field map at 0.36 Mm. At higher altitudes (up to 1.44 Mm), the jump in $R_{(150,15)}$ is observed at later moments in time compared to the temporal variation seen at 0.36 Mm. For $R_{(150,15)}$, we consider the height range of 0.36 - 1.44 Mm as the OHR. So, here, the OHR is to be understood as 'a collection of heights where a clear and sustained jump in R-value is observed'. Please note here that a jump is identified in retrospect when after a period of null R-value output, the R-value begins to show some finite value continuously for minimum duration of 6 hrs. The choice of time interval (i.e. 6 hrs in this case) is somewhat subjective and has been deliberately introduced to distinguish between data and noise, the noise being cases where the R-value fluctuates between zero and finite values at time-scales shorter than 6 hrs. Additional constraints to define the jump may be imposed as more examples (i.e. flare-producing ARs) are studied in future. Within the context of OHR, we define $T_{rv}$ which denotes the time of R-value increase, i.e. the time when the R-value begins to show some finite value after a period of null output. In essence, post the $T_{rv}$, the R-value must not descend back to null values (depicted by black lines in Figure \ref{fig:11158} for example) and it must have been preceded by black lines continuously for a minimum of 6 hrs. Table \ref{table:emergence} lists the times of R-value increase at different heights for AR 11158. \\

Computing R-value with a $B_{th}$ of 150 G, at higher altitudes often leads to continuous flat black lines, indicating null R-values (see Figure \ref{fig:11158}a). This is because the extrapolated $B_z$ is weaker compared to the photospheric $B_z$ and the $B_{th}$ of 150 G is too high to detect high-polarity regions. The results obtained for the magnetic structure model constructed with $R_{(100,15)}$ for AR 11158 are consistent with what has been observed from the $R_{(150,15)}$ model. The R-value increase times are close to $T_{fe}$ just like the $R_{(150,15)}$ model but the OHR is extended further up to 2.52 Mm (refer to Table~\ref{table:emergence}). For the $R_{(100,15)}$ model at 0.36 Mm, the black lines are indicative of the time window when $R_{(100,15)}$ is really low in the photosphere (see panels corresponding to 0 and 0.36 Mm in Figure \ref{fig:11158}b). However, for heights between 0.72 - 1.44 Mm, the time window corresponding to the black lines largely coincides with the time when the unsigned flux is really low before it begins to increase (compare Figures \ref{fig:fluxclubbed}a and \ref{fig:11158}b). In the case of $R_{(50,15)}$, the OHR is between 1.08 - 3.24 Mm. It is important to note here that the time of increase in $R_{(50,15)}$ is more consistent in height compared to that of $R_{(150,15)}$ and $R_{(100,15)}$ (compare Figure~\ref{fig:11158}c with Figures \ref{fig:11158}a and \ref{fig:11158}b). \\

Overall, a progressive increase in the sunspot group complexity was observed either throughout or for a significant part of the observation time window for the 'emergent' ARs 11158, 11166, 12017 and 12673. For example, for AR 11166, the complexity of the sunspot group changed from $\beta$-type on Mar 04, 2011 to $\beta\gamma$-type on Mar 06 and subsequently to $\beta\gamma\delta$-type on Mar 08, 2011. We find that as the sunspot becomes more complex and as we approach the X-class flare in time, the R-value seems to sustain non-zero values at higher altitudes. For example, on Mar 05, the highest altitude where $R_{(150,15)}$ is fairly continuously non-zero is 0.36 Mm but on Mar 09, $R_{(150,15)}$ we find non-zero values at 1.44 Mm (see Figure~\ref{fig:11166}b). This shows that there may be a link between the height-wise variation of the R-value as a function of time and the temporal evolution of the sunspot group complexity. \\

For AR 12017 at first the flux is high (see Figure \ref{fig:fluxclubbed}b around Mar 23-24) and the flux drops to near-zero values and begins to increase significantly again. The R-value also shows a similar trend. Upon inspecting the R-value in the photosphere, we find a 'V' shape trend. This pattern becomes more noticeable at higher altitudes as we can see the plotted black lines corresponding to null outputs in R-value, such as 0.36 Mm for $R_{(150,15)}$ or 1.44 Mm for $R_{(50,15)}$ (see Figure~\ref{fig:12017}). \\

Unlike ARs 11158, 11166 and 12017, an isolated X-class flare was not observed for AR 12673, rather, a series of 4 X-class flare events were observed (3 of which occurred within the $\pm 60^{\circ}$ from the disk center). It is also important to note that in the 72 hrs preceding the first of these X-class flares, 12 M-class flares were reported in AR 12673. For AR 12673, the increase in unsigned flux was more gradual compared to ARs 11158, 11166 and 12017 and $T_{fe}$ coincides with the time when magnetic bipoles emerge around the central PIL \citep{Liu2019}. The first flare with a magnitude X2.2 occurs at a time when a steady increase in the flux is seen (see Figure~\ref{fig:fluxclubbed}). An OHR can be determined from all models; 0.72 - 2.88 Mm for $R_{(150,15)}$,  1.08 - 3.24 Mm for $R_{(100,15)}$ and  1.44 - 3.24 Mm for $R_{(50,15)}$ (see Figure~\ref{fig:12673}). To sum up, all these OHRs have one common feature, by being between 1-3 Mm. In general, at higher altitudes, the computation of R-value loses scientific relevance because of weak field strengths. \\

\begin{figure}[h]
\centering
\textbf{Plots for Unsigned Flux near PILs}\par\medskip

\begin{subfigure}
  \centering
  \includegraphics[width=.32\linewidth]{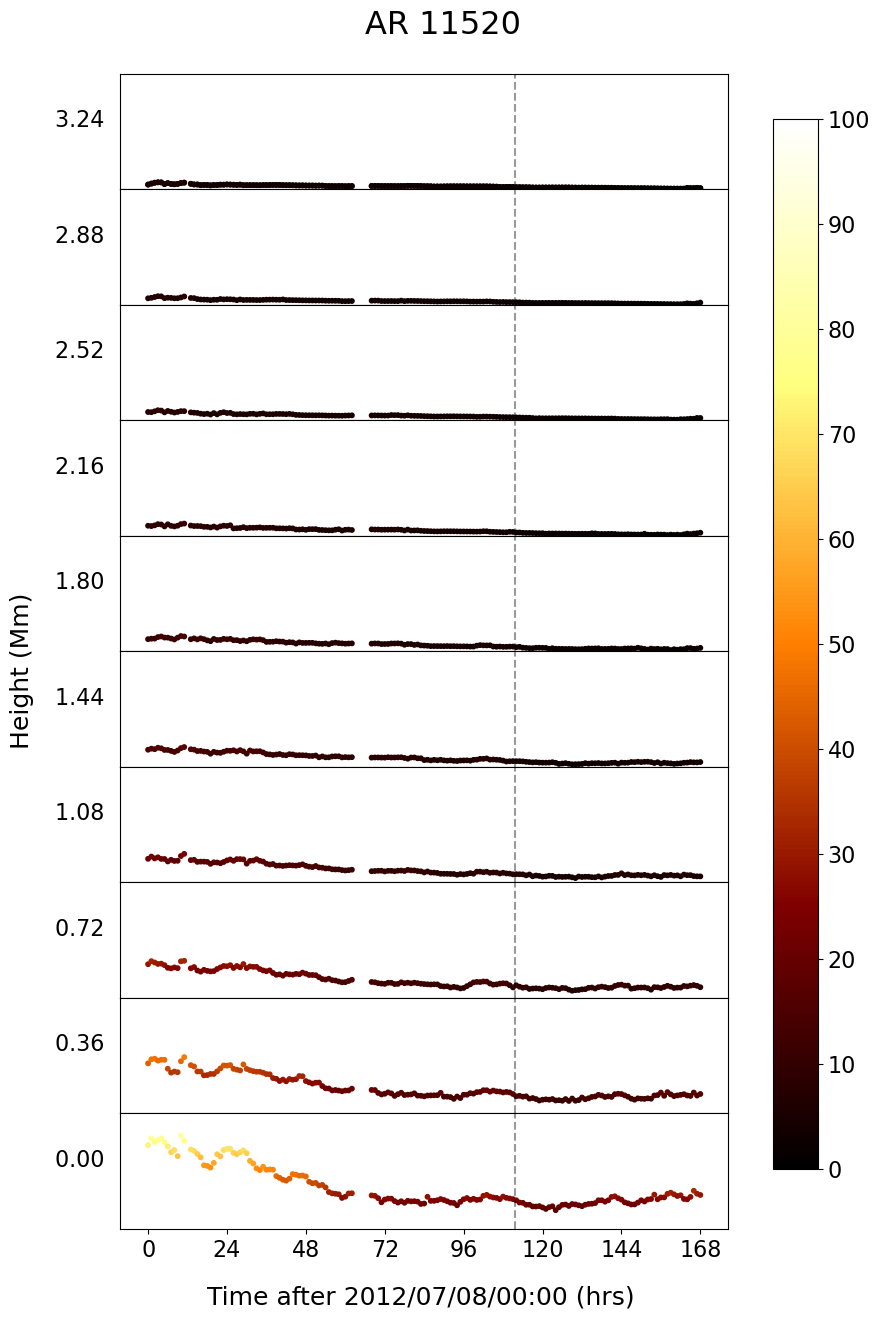}  
\end{subfigure}
\begin{subfigure}
  \centering
  \includegraphics[width=.32\linewidth]{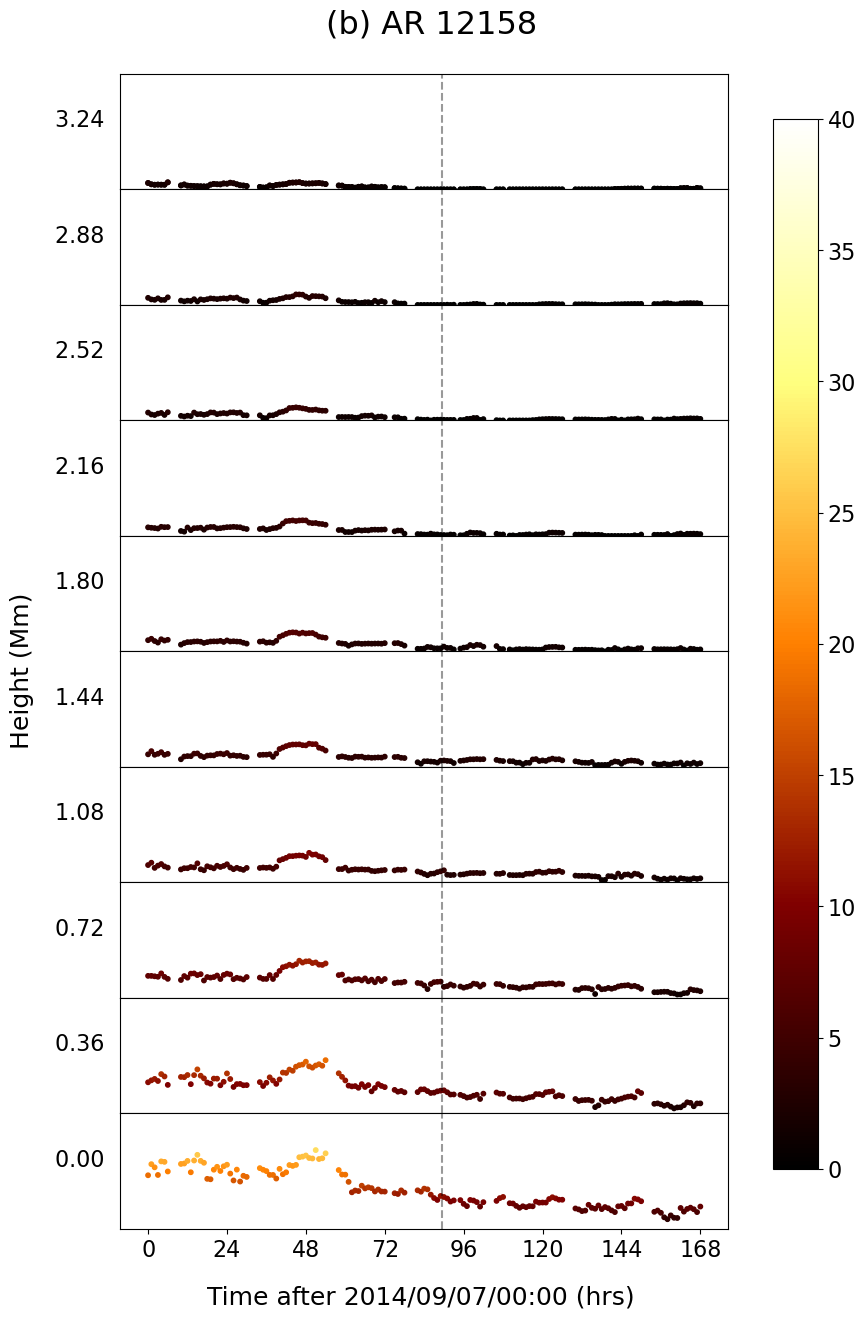}
\end{subfigure}
\begin{subfigure}
  \centering
  \includegraphics[width=.32\linewidth]{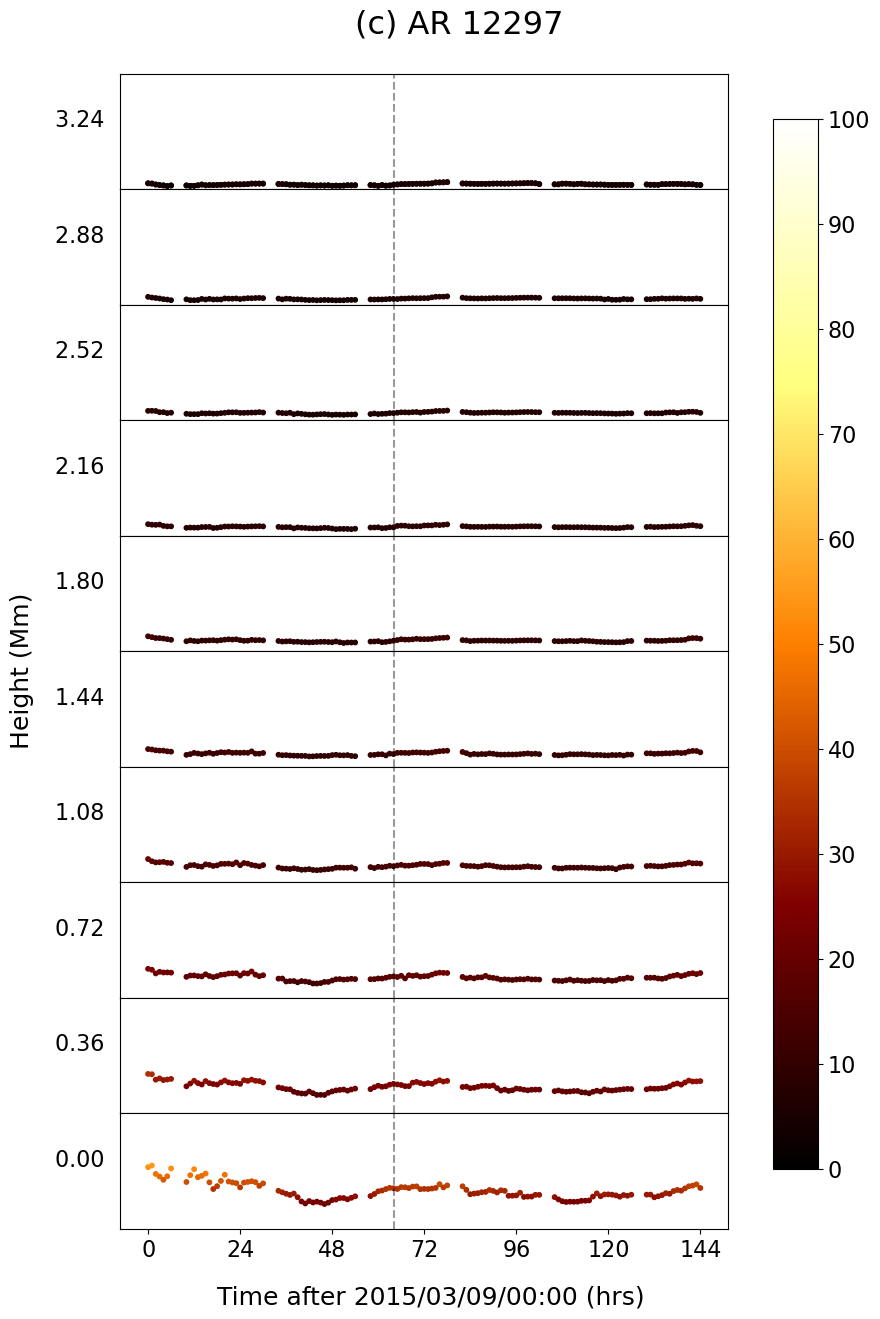}
\end{subfigure}

\caption{Multiple-height stack plots for unsigned flux near PILs for; (a) AR 11520; (b) AR 12158; (c) AR 12297; colourbar indicates the unsigned flux (in $10^{20}$ Mx); vertical dashed line in each plot indicates time of occurrence of the X-class flare}
\label{fig:fluxclubbed2}

\end{figure}

\begin{figure}[h]
\centering
\textbf{R-value plots for AR 11520}\par\medskip

\begin{subfigure}
  \centering
  \includegraphics[width=.32\linewidth]{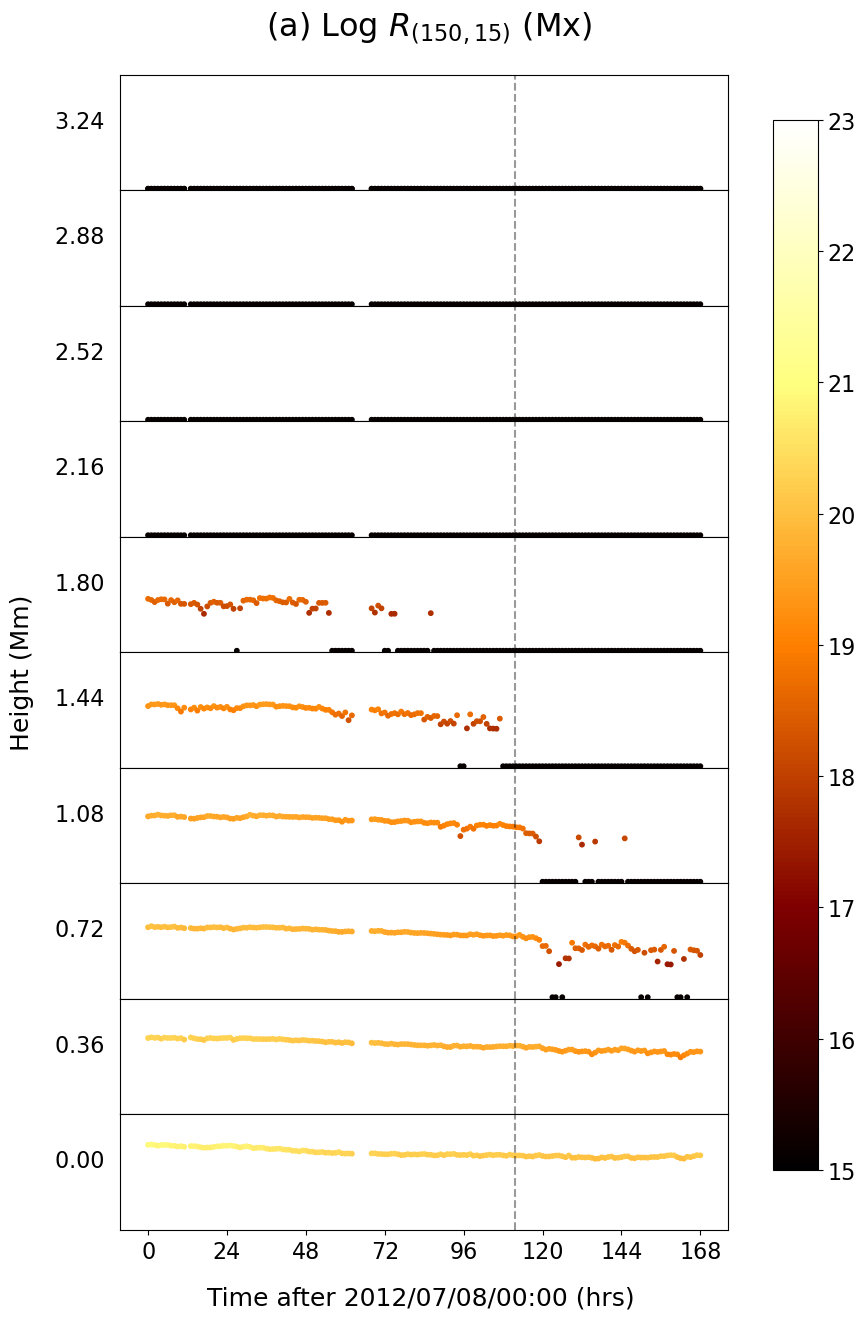}  
\end{subfigure}
\begin{subfigure}
  \centering
  \includegraphics[width=.32\linewidth]{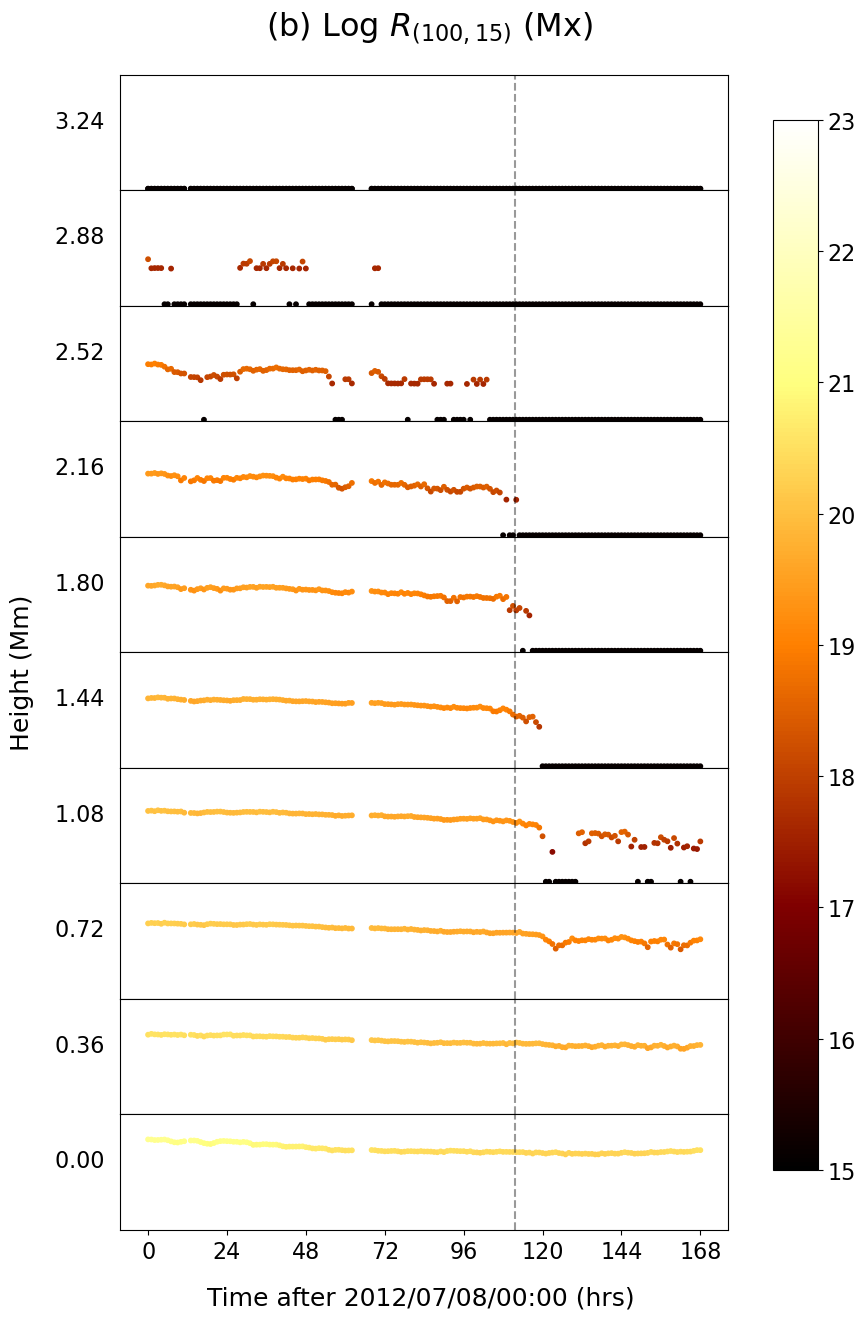}
\end{subfigure}
\begin{subfigure}
  \centering
  \includegraphics[width=.32\linewidth]{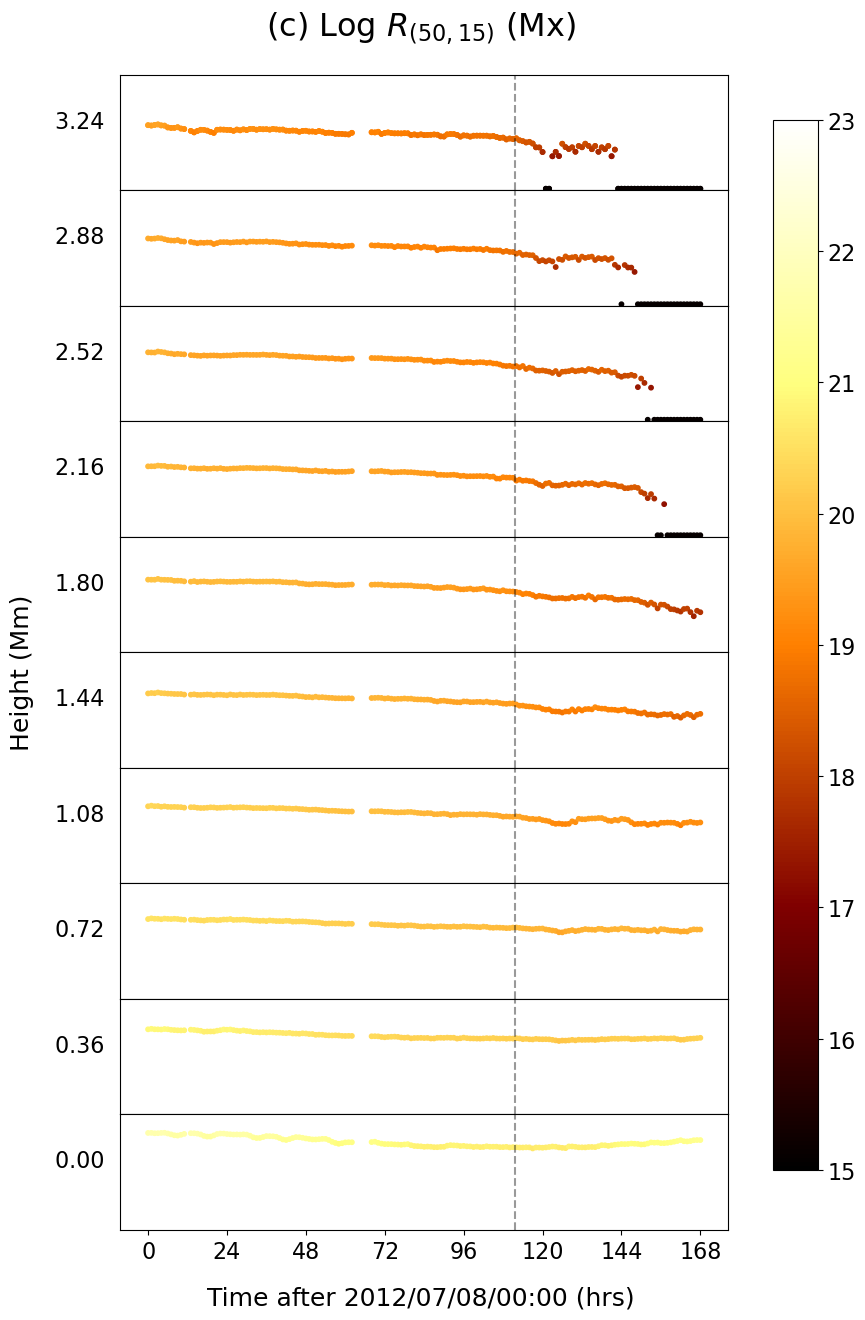}
\end{subfigure}

\caption{Multiple-height stack plots for (a) $R_{(150,15)}$; (b) $R_{(100,15)}$; (c) $R_{(50,15)}$; vertical dashed line indicates time of occurrence of the X1.4 flare; colourbar indicates the logarithm of R-value (in Mx); any black lines or points indicate null output}
\label{fig:11520}

\phantom{...}

\end{figure}

\begin{figure}[ht]
\centering
\textbf{Sample R-value plots for non-flaring cases}\par\medskip

\begin{subfigure}
  \centering
  \includegraphics[width=.32\linewidth]{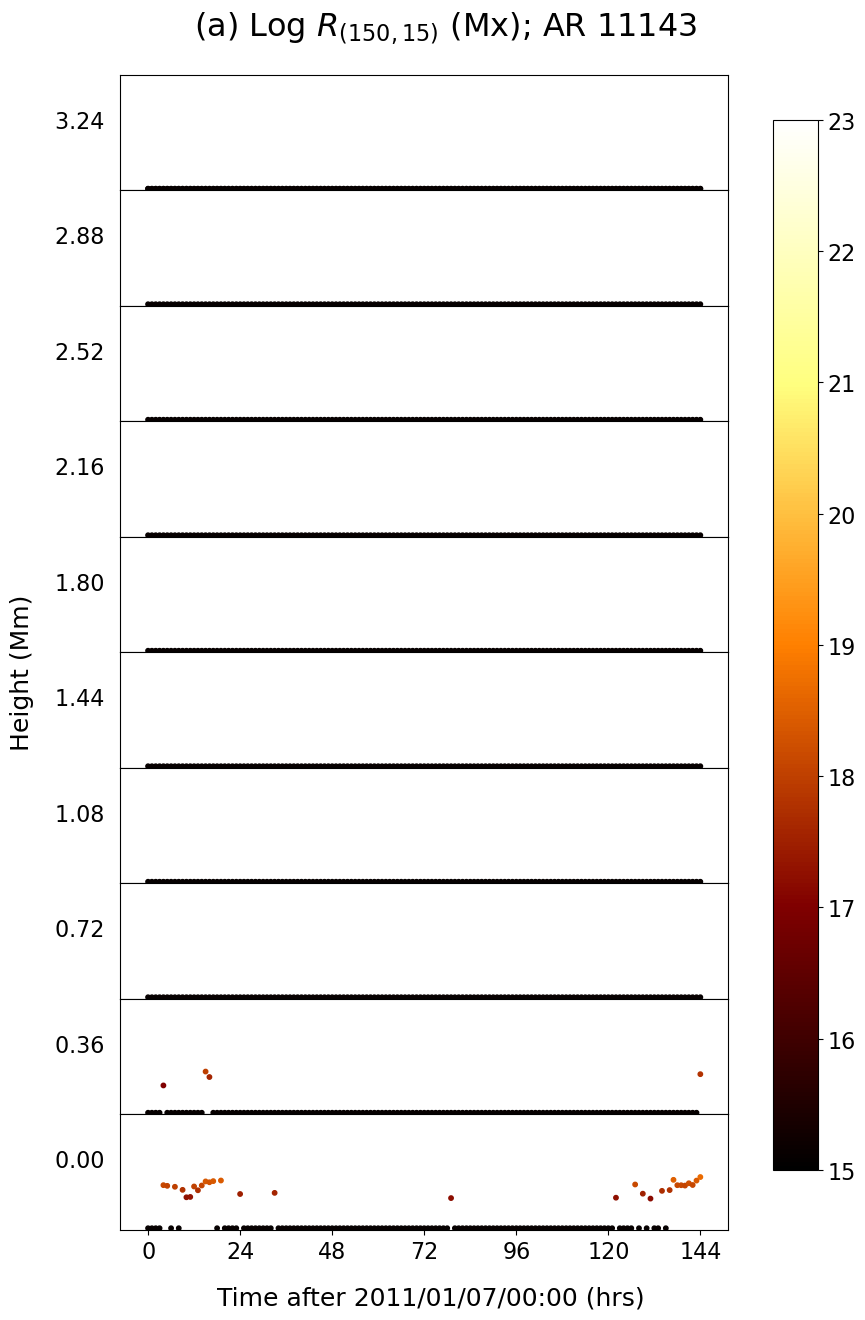}  
\end{subfigure}
\begin{subfigure}
  \centering
  \includegraphics[width=.32\linewidth]{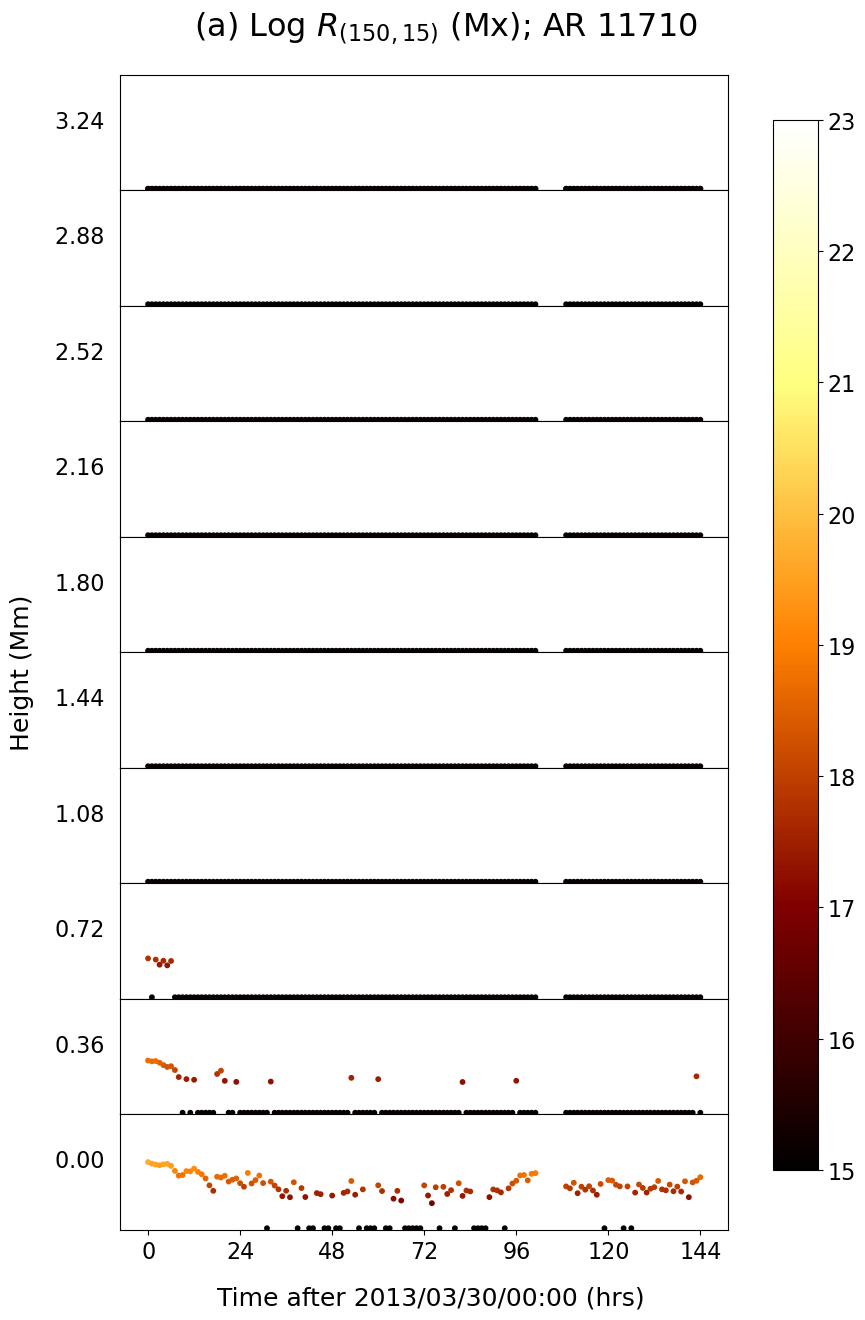}
\end{subfigure}
\begin{subfigure}
  \centering
  \includegraphics[width=.32\linewidth]{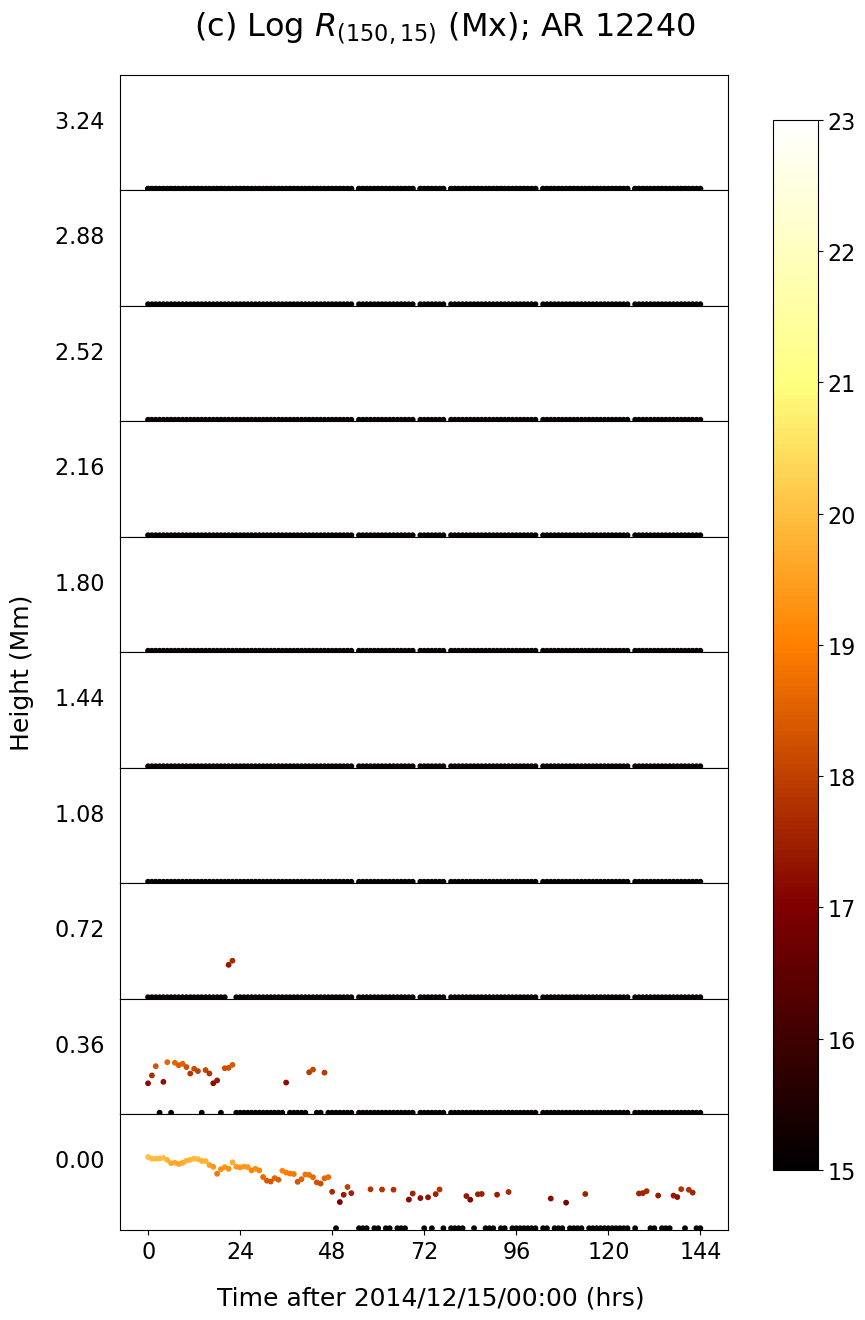}
\end{subfigure}

\caption{Multiple-height $R_{(150,15)}$ stack plots for non-flaring cases; (a) AR 11143, (b) AR 11710 and (c) AR 12240; colourbar indicates the logarithm of R-value (in Mx); any black lines or points indicate null output}
\label{fig:NFcases}

\end{figure}

\subsection{ARs associated with a prior decrease in the unsigned flux near PILs}
\label{ssec:rd_dec}

\phantom{...}

Sometimes, X-class flares may occur during a period of gradual decrease in the unsigned flux near PILs as in the cases of ARs 11520, 12158 and 12297 (see Figure~\ref{fig:fluxclubbed2}). While for ARs 11520 and 12158, decreases in $R_{(50,15)}$, $R_{(100,15)}$ and $R_{(150,15)}$ were observed with the passage of time (see Figure~\ref{fig:11520} as a sample) across different heights, for AR 12297, $R_{(50,15)}$, $R_{(100,15)}$ and $R_{(150,15)}$ remained nearly constant in time across different heights. In summary, the R-value trends from these cases did not yield anything conclusive and an OHR could not be defined and determined. However a common feature for all these ARs is that they were not emergent ARs and were almost always associated with a complex sunspot group throughout the observation time. For example, AR 12158 initially hosted a $\beta$$\delta$-type sunspot that transformed into a $\beta$$\gamma$$\delta$-type before eventually decreasing its complexity to $\gamma$$\delta$-type sunspot towards the end of the observation time. For AR 11520, the black line at 1.80 Mm for the $R_{(150,15)}$ model (about 60 hrs before the X1.4 flare) is almost consistent in time after 09:00 UTC on 10 July 2012 (see Figure~\ref{fig:11520}). It may be an intuitive idea to propose that the R-value disappearance may be linked to X-class flares. However, at this stage, we refrain from proposing such a hypothesis because of the lack of a sufficient number of examples, therefore retaining the idea as a 'conjecture', subject to a more extensive study in the future. It is known that eruptive solar flares at times could be driven by magnetic flux cancellation (\citealp{Zhang2001}; \citealp{Burtseva2013}) and it might be possible that the X-class flares related to ARs 11520, 12158 and 12297 were related to magnetic flux cancellation. \\

\subsection{R-value in height and time for non-flaring ARs}
\label{ssec:rd_nf}

\phantom{...}

The vertical variation of $R_{(150,15)}$ may be used as a discriminant flaring and non-flaring ARs. We consider an AR to be non-flaring if it does not host any flares stronger than C1.0. We studied 3 non-flaring cases: AR 11143 (Jan 7-13, 2011), AR 11710 (Mar 30 - Apr 5, 2013) and AR 12240 (Dec 15-21, 2014) . All these ARs hosted a $\beta$ sunspot during the time-window of study. PF extrapolation followed by computation of R-value in 3D showed that $R_{(150,15)}$ does not consistently sustain non-zero values at 0.36 Mm for AR 11143 or barely does so for ARs 11710 and 12240. This is clearly distinct from what we observed for X-class flares and even M class flares (see Appendix C). For the M-class or X-class flare cases, we find that the maximum height where $R_{(150,15)}$ sustains non-zero values can be as low as 0.72 Mm and as high as 3.24 Mm. Some sample results for the non-flaring ARs are shown in Figure \ref{fig:NFcases}. Although we have restricted the discussion to the variation of $R_{(150,15)}$ as a discriminant at the photosphere (or low altitudes), it is important to note that if we reduce $B_{th}$, the R-value may still sustain non-zero values at altitudes greater than 0.72 Mm (see GitHub project repository for examples). $R_{(50,15)}$ and $R_{(100,15)}$ may still serve as discriminants between flaring and non-flaring ARs but more examples are needed to be studied for further confirmation. \\

\subsection{Physical significance of results}
\label{ssec:inf}

\phantom{...}

\begin{figure}[ht]
\centering

\begin{subfigure}
\centering
\includegraphics[width=.32\linewidth]{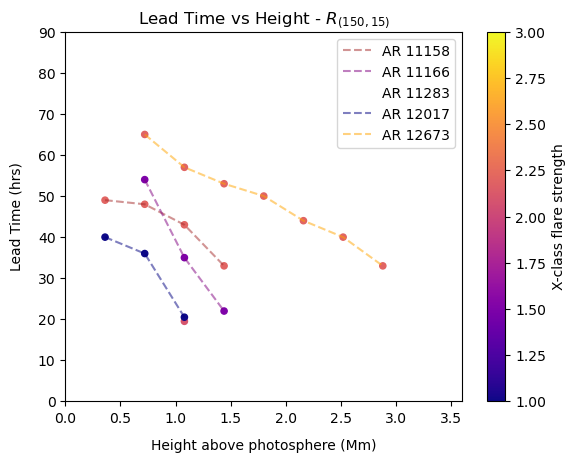}  
\end{subfigure}
\begin{subfigure}
\centering
\includegraphics[width=.32\linewidth]{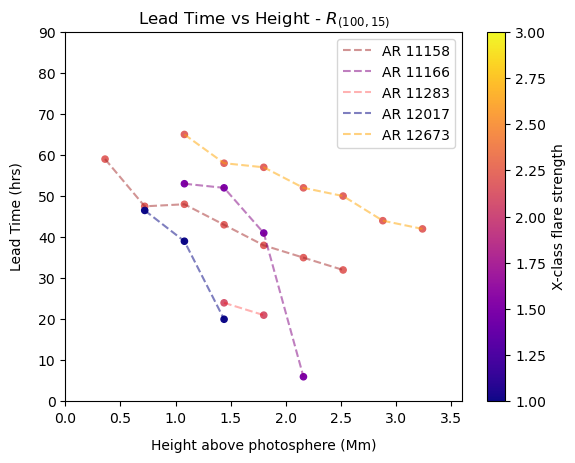}
\end{subfigure}
\begin{subfigure}
\centering
\includegraphics[width=.32\linewidth]
{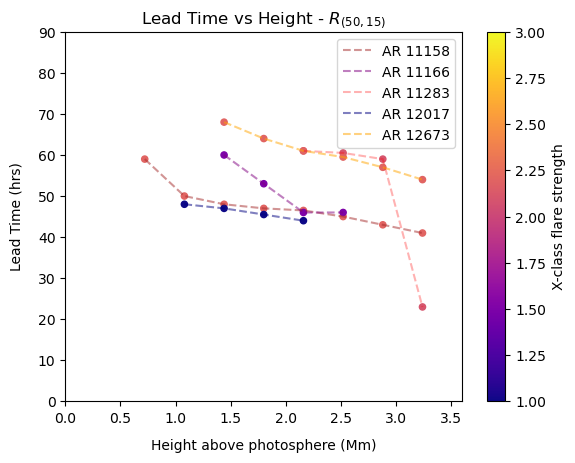}
\end{subfigure}

\caption{Plots between lead time (i.e. $T_{diff}$) and height for 5 different ARs with an OHR; for $R_{(150,15)}$ (left), $R_{(100,15)}$ (middle) and $R_{(50,15)}$ (right). The colourbar in each plot denotes flare strength in terms of peak soft X-ray flux (in 10$^{-4}$ Wm$^{-2}$). For exact numerical values of $T_{diff}$, refer to Tables \ref{table:heighttime150}, \ref{table:heighttime100} and \ref{table:heighttime50} in Appendix D.}
\label{fig:ht}
\end{figure}

We were able to identify OHRs for ARs linked to flux emergence, i.e. ARs 11158, 11166, 12017 and 12673 and the special case of AR 11283 (see Appendix B). For a given AR and its OHR, a $T_{rv}$ (i.e. the time of R-value increase) was determined for all heights included in the OHR. The difference between $T_{rv}$ and the latest time-stamp preceding the flare $T_{pf}$ gives $T_{diff}$, which is an estimate of the lead time. Rather than the using the exact time of flare occurrence, $T_{pf}$ was used so as to keep $T_{diff}$ an integer for convenience. The height-wise variation of the lead time is listed in Tables \ref{table:heighttime150},   \ref{table:heighttime100} and \ref{table:heighttime50}, corresponding to $R_{(150,15)}$, $R_{(100,15)}$ and $R_{(50,15)}$ respectively (see Appendix D). To visualise the data in conjunction with flare strength, we have presented Figure \ref{fig:ht}. \\

A general inference from the height-wise lead time data is that, as we approach the flare in time, the lead time decreases. However, the nature (linear/non-linear) or rate of decrease is different for different ARs. In certain cases this decrease may follow a linear trend. For example, linear regression calculations for the $R_{(150,15)}$ lead time vs height plot for AR 12673 showed a $R^2$ value of 0.97 (see Figure \ref{fig:ht} (left)). The data also suggests that flare strength may not be well correlated with lead time. For example, the lead time-height curves for AR 11158 and AR 11166 intersect between 0.72 Mm and 1.08 Mm, suggesting that there may not be any particular correlation between flare strength and lead time. While at 0.72 Mm the weaker flare (X1.5; AR 11166) is associated with a higher lead time, at 1.08 Mm the stronger flare (X2.2; AR 11158) is associated with a shorter lead time. This aspect needs to be studied further with a larger statistical sample for further conclusions. For $R_{(50,15)}$, the OHR is most likely to occur at heights greater than 1.08 Mm (upto 3.24 Mm) but for $R_{(150,15)}$, it may be expected between 0.36 - 1.44 Mm (see Tables \ref{table:heighttime150} and \ref{table:heighttime50}). The most frequently occurring heights in the OHRs for $R_{(50,15)}$, $R_{(100,15)}$ and $R_{(150,15)}$ are 1.08, 1.44 and 2.16 Mm, respectively. This indicates that if we reduce $B_{th}$, an OHR may be expected at higher altitudes. Furthermore, the height range discussed above, where the jump of the R-value is more pronounced, aligns with the findings of \citep{Korsos2024}.  They have indicated that utilising a variety of precursor parameters is important in the LSA (up to 2 Mm) to enhance the accuracy of eruption predictions. In addition, certain proxies have been observed to reveal their own pre-flare evolution phase earlier within the LSA \citep{Korsos2018, Korsos2020, Korsos2022}. \\

\subsection{Proposed application of results to solar flare prediction}
\label{ssec:swx}

\phantom{...}

In this Section we describe how the R-value may be employed to predict the first X-class flare for an emerging AR in a real time scenario. We emphasise that the ideas proposed here are simply based on the case studies encountered so far and the exact method to be used in case of prediction must be based on the results from a rigorous statistical study, which is beyond the scope of the current paper. To estimate how much time in advance a prediction can be issued in a best-case scenario, we introduce a new term $T_{diff}^{*}$ that denotes the maximum lead time across all heights in an OHR for an AR. Although, $T_{diff}^*$ gives the maximum time available to issue a prediction for a specific R-value model, it is important to note that this time difference has been obtained only in hindsight. In a real-time prediction scenario, a jump in R-value from zero to non-zero values temporarily may not necessarily imply a sudden change in solar activity because it is possible that it might very well be noise. So, in the event of a real-time prediction scenario, we further impose the condition that the R-value must sustain non-zero values for at least 24 hrs after increase so that a prediction warning may be issued with some confidence. This period of 24 hrs can be thought of as a 'confidence interval' (denoted by $C_{int}$) is once again somewhat indicative/conjecture-oriented. Our choice of $C_{int}$ has been decided on the basis of AR examples we have encountered so far and $C_{int}$ may need modification as more examples are studied in future. Based on inputs from the three R-value models, the maximum value of $T_{diff}^{*}$ across different R-value models may be considered as the optimal time $T_{opt}$. The $T_{opt}$ for each case (or AR) is indicated in bold (refer to Table \ref{table:timelist}). The $C_{int}$ was then subtracted from $T_{opt}$ to obtain an estimate of the response time $T_{res}$ (refer to Table \ref{table:timelist}). The idea behind defining $T_{res}$ is to quantify the time available to respond to a flare warning after an alert has been sent out after observing the AR for a certain confidence interval. For example, had the concept of OHR been used to provide a warning for the X2.2 class flare for AR 12673, the exact time of sending out the alert would have been 2017/09/04 12:00 UTC (24 hrs post increase in $R_{(50,15)}$) and $T_{res}$ would have been approximately 44 hrs. \\

{\begin{table}[!ht]
\centering
\begin{tabular}{| c | c | c | c | c | c | c | c | c |}
 \hline
 \phantom{x} & \phantom{x} & \phantom{x} & \phantom{x} & \phantom{x} & \phantom{x} & \phantom{x} & \phantom{x} & \phantom{x} \\ [-0.9em]
 No. & AR & Flare & $T_{pf}$ & $T_{diff}^*$ [$R_{(150,15)}$] & $T_{diff}^*$ [$R_{(100,15)}$] & $T_{diff}^*$ [$R_{(50,15)}$] &  \phantom{..} $T_{opt}$ \phantom{..} & $T_{res}$ = $T_{opt}$ - $C_{int}$ \\ [0.3em]
\hline
 01 & 11158 & X2.2 & 2011/02/15 01:00 & 49 hrs & {\textbf{59 hrs}} & 50 hrs & 59 hrs & 35 hrs \\
 02 & 11166 & X1.5 & 2011/03/09 23:00 & 54 hrs & 53 hrs & {\textbf{60 hrs}} & 60 hrs & 36 hrs \\
 03 & 11283 & X2.1 & 2011/09/06 22:00 & 20 hrs & 24 hrs & {\textbf{61 hrs}} & 61 hrs & 37 hrs \\
 04 & 11520 & X1.4 & 2012/07/12 15:00 & - - - & - - - & - - - & - - - & - - - \\
 05 & 12017 & X1.0 & 2014/03/29 17:00 & 40 hrs & 47 hrs & {\textbf{48 hrs}} & 48 hrs & 24 hrs \\
 06 & 12158 & X1.6 & 2014/09/10 17:00 & - - - & - - - & - - - & - - - & - - - \\
 07 & 12297 & X2.1 & 2015/03/11 16:00 & - - - & - - - & - - - & - - - & - - - \\
 08 & 12673 & X2.2 & 2017/09/06 08:00 & 65 hrs & 65 hrs & {\textbf{68 hrs}} & 68 hrs & 44 hrs \\
\hline
\end{tabular}
\caption{A list of optimal times (indicated in bold) and response times for each OHR for X-class flare cases; column 4 lists the latest time-stamp preceding the flare $T_{pf}$; Columns 5-7 list the maximum lead time across all heights $T_{diff}^*$ for $R_{(150,15)}$, $R_{(100,15)}$ and $R_{(50,15)}$ respectively. The maximum $T_{diff}^*$ across multiple R-value models is $T_{opt}$ and is listed in column 8. Column 9 lists the response time $T_{res}$ after subtracting the confidence interval $C_{int}$ from $T_{opt}$. For exact information on $T_{diff}$ and $T_{diff}^*$, refer to Tables \ref{table:heighttime150}, \ref{table:heighttime100} and \ref{table:heighttime50}.}
\label{table:timelist}
\end{table}}

\newpage

\section{SUMMARY AND CONCLUSIONS}
\label{sec:cs}

\phantom{...}

Detailed information on measuring the pre-eruptive conditions in the solar atmosphere is important to obtain more accurate future solar flare prediction methods. \cite{Korsos2020} and \cite{Korsos2022} proposed and elaborated that the prediction of major solar eruptions could be improved by incorporating data from the LSA, which extends to approximately 4 Mm above the photosphere. They noted that using PF extrapolation data allows for earlier identification of the pre-flare evolution phase of predictor parameters, particularly in the region above the photosphere within the LSA (up to 2 Mm). The PF offers a simplified yet insightful representation of the 3D magnetic field of an AR, capturing its essential large-scale structure without the complexities of currents. It is important to stress here that free magnetic energy or the dynamics of flares cannot be obtained from PF extrapolation. However, PF extrapolation can provide a meaningful insight into the topology of the field, and that's where its value is in the current context (\citealp{Wiegelmann2012, Korsos2024}).  Therefore to further explore the idea of studying the pre-eruptive conditions of ARs, we made use of the concept developed by \cite{Korsos2020} and \cite{Korsos2022} and analysed the evolution of the  R-value as a function height for the selected ARs. We specifically investigated whether there is an OHR where the R-value provides certain hints about the occurrence of an upcoming large flare. Based on our case study, we conclude the following: \\

\begin{itemize}
    \item[$\bullet$\hspace{0.3cm}] Since the R-value is a filtered version of the unsigned flux and is calculated based on the magnetic field values around high-gradient PILs, it only reinforces the argument that there may be a strong correlation between flux emergence and R-value increase.
    \item[$\bullet$\hspace{0.3cm}] The variation of R-value before the first X-class flare is quite different from that of 'non-flaring' ARs i.e. ARs that do not host flares stronger or equal to C1.0. It is seen that the $R_{(150,15)}$ decays to zero output faster in height for non-flaring ARs compared to flaring ARs.  $R_{(150,15)}$ serves as a good discriminant to distinguish between flaring and non-flaring ARs. 
    \item[$\bullet$\hspace{0.3cm}] Having tested five different models of R-value with fixed values of $B_{th}$ and $D_{sep}$, we found that the R-value is more sensitive to $B_{th}$ in comparison to $D_{sep}$ (representative example in Appendix A, for more examples see the GitHub project repository). We found that the OHR for R-value is impacted by the choice of $B_{th}$ and there is no specific choice for $B_{th}$ that works best for all cases. For example, considering a target height range of 0.00 - 3.24 Mm, a $B_{th}$ of 100 G is a good choice to study AR 11158 but a $B_{th}$ of 50 G works better for AR 11166 (refer to Table \ref{table:timelist}). Here, a good choice for the threshold is the one that best optimises the response time. Overall, $R_{(50,15)}$ may be adjudged as the best performing R-value model for having maximised $T_{diff}$ in four out of five cases. For the purpose of real time prediction, $R_{(50,15)}$ may be primarily used for prediction while $R_{(150,15)}$ and $R_{(100,15)}$ may be used for purposes of correlating and validating the information received slightly in advance from $R_{(50,15)}$. We have also seen that it is not necessary for an OHR to exist but provided it exists for multiple models, it is shifted to lower heights upon increasing $B_{th}$.
    \item[$\bullet$\hspace{0.3cm}] Previous studies on R-value have mostly focused on its evolution on the photosphere. Instead, we studied it in the LSA and we were able to define the OHR at heights where a definitive jump in the R-value was observed. Thus, the concept of OHR was extended beyond its definition based on the $WG_M$ morphological parameter and magnetic helicity to a new parameter such as R-value. Based on the current case study, we find that the concept of OHR when defined in terms of R-value may work best for 'predicting' X-class flares which are linked to 'flux emergence near PILs', as compared to cases which do not exhibit the characteristics of flux emergence and are already associated to complex sunspots to start with (statement in general terms; statistical significance is beyond the scope of this paper). ARs 11158, 11166, 12017 and 12673 are the best examples of such cases. In these cases, we also found that the sunspot group became progressively more complex (for example $\beta$-type at the start of the observation time window and $\beta\gamma\delta$-type towards the end). However, for cases like ARs 12297 or 12158, for which an OHR could not be determined, a $\delta$ sunspot could be found throughout/or at the start of the observation time window. 
    \item[$\bullet$\hspace{0.3cm}] Based on our calculations of the OHR for five ARs linked to X-class flares, we found that if the R-value is studied for a confidence interval of 24 hrs, it may be possible to have an optimal time of 48-68 hrs and a response time of 24-44 hrs.  
    
\end{itemize}

\phantom{...}

The biggest limitation we encountered in our study was the absence of a statistically large dataset for X-class flares that satisfied the criteria discussed in Section \ref{sec:ardataset}. We hope that in the future, as more examples are studied, we may be able to improve the definition of a jump in R-value by reviewing the continuity interval (6 hrs as defined currently). In the future, building on the approach of this study, we will also address the question of whether the joint application of different predictor parameters enhances prediction skills by applying them throughout the LSA. We plan to extend these calculations to more cases of X-class flares and weaker flares such as M-class flares and other methods of extrapolations. \\

\section{Acknowledgement}

\phantom{...}

This research work is supported by Space Weather Awareness Training Network (SWATNet) and it comes under the aegis of the  project: 'Three-Dimensional Solar Flare predicting' (\href{https://swatnet.eu/}{https://swatnet.eu/}). SWATNet has received funding from the European Union’s Horizon 2020 research and innovation programme under the Marie Sklodowska-Curie Grant Agreement No. 955620. MBK, RE and MG acknowledge support from ISSI-BJ ("Step forward in solar flare and coronal mass ejection (CME) predicting"). 
 RE is grateful to STFC (UK, grant number ST/M000826/1) and PIFI (China, grant number No. 2024PVA0043). MBK is grateful for the Leverhulme Trust Found ECF-2023-271. MBK acknowledges support by UNKP-22-4-II-ELTE-186, ELTE Hungary. RE and MBK  also thank for the support received from NKFIH OTKA (Hungary, grant No. K142987). This work was also supported by the NKFIH Excellence Grant TKP2021-NKTA-64.  AN and SP acknowledge support by the ERC Synergy Grant ‘Whole Sun’ (GAN: 810218). \\
  
\newpage
\appendix
\section{Qualitative comparison of input parameter sensitivities: Case study of AR 11166}
\label{sec:app_a}

\phantom{...}

\begin{figure}
\centering
\textbf{Plots for AR 11166}\par\medskip

\begin{subfigure}
  \centering
  \includegraphics[width=.32\linewidth]{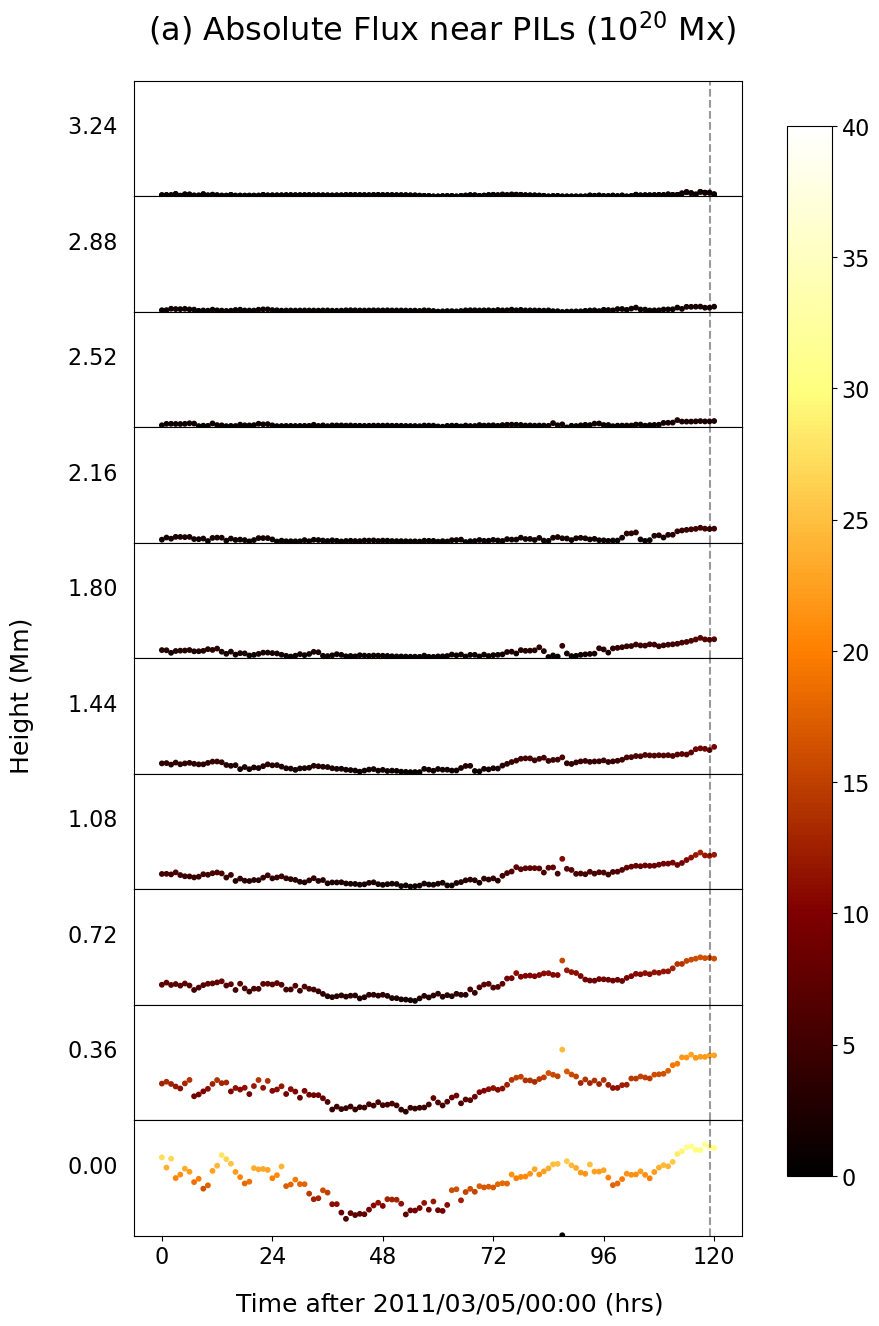}  
\end{subfigure}
\begin{subfigure}
  \centering
  \includegraphics[width=.32\linewidth]{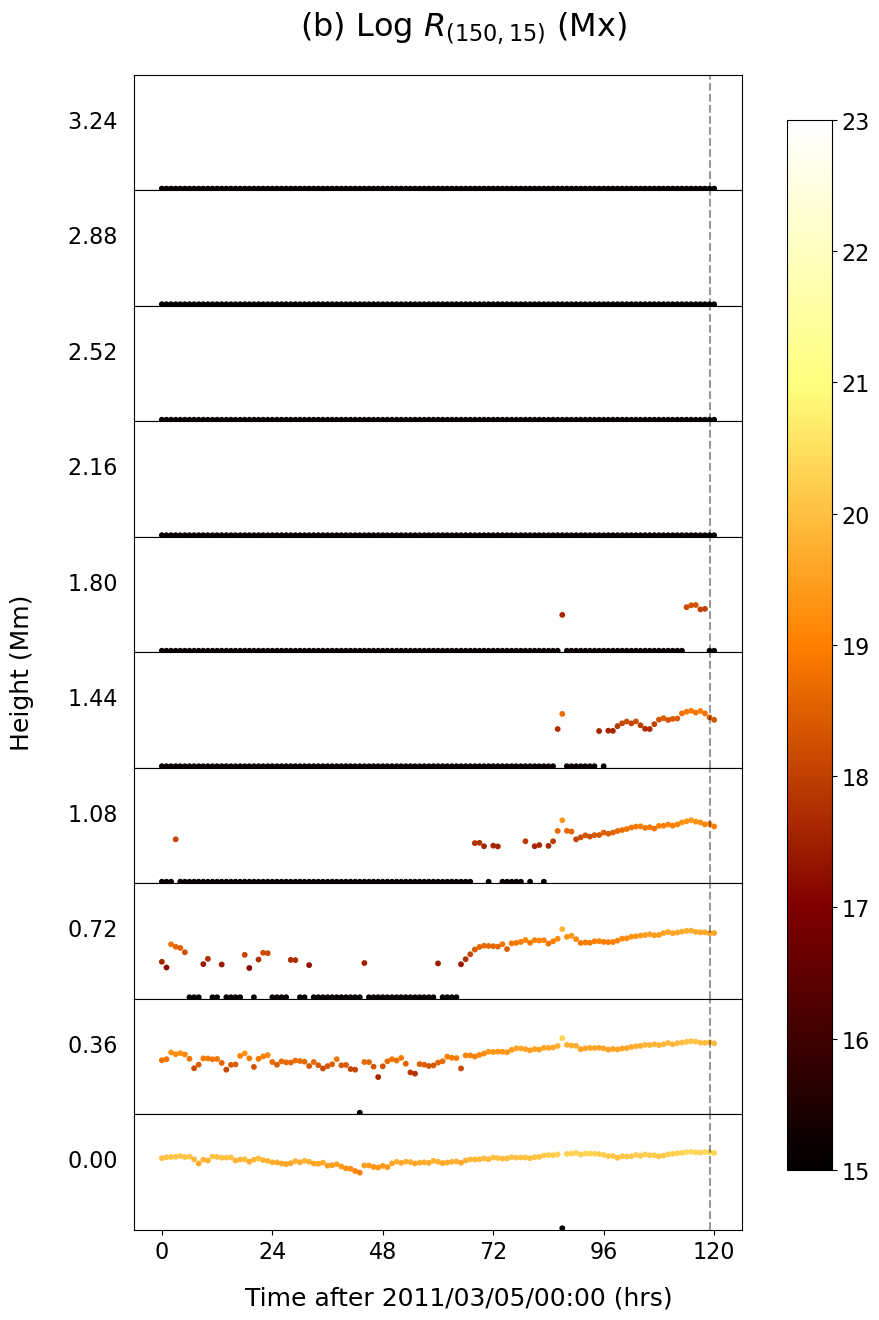}
\end{subfigure}
\begin{subfigure}
  \centering
  \includegraphics[width=.32\linewidth]{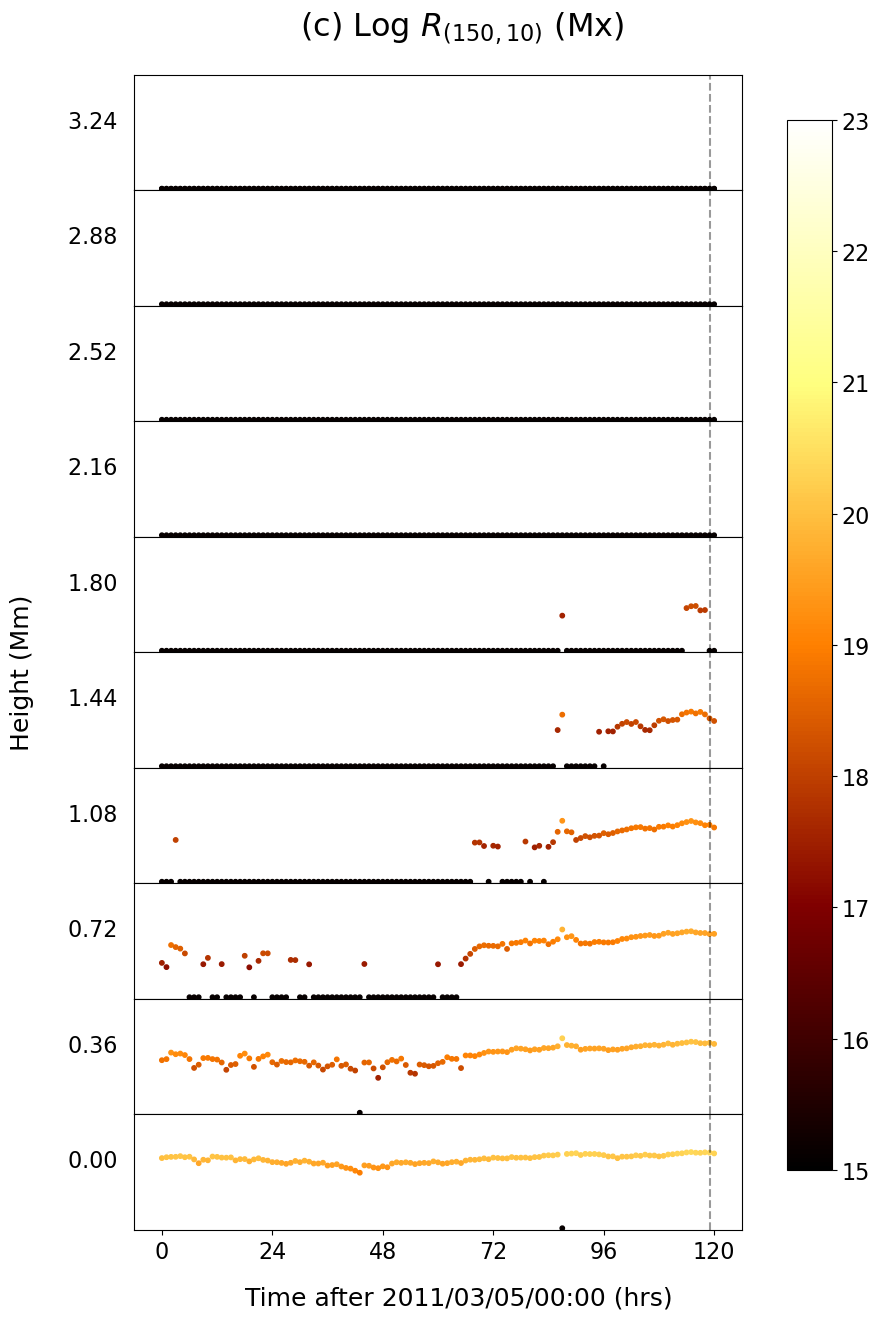}
\end{subfigure}


\begin{subfigure}
  \centering
  \includegraphics[width=.32\linewidth]{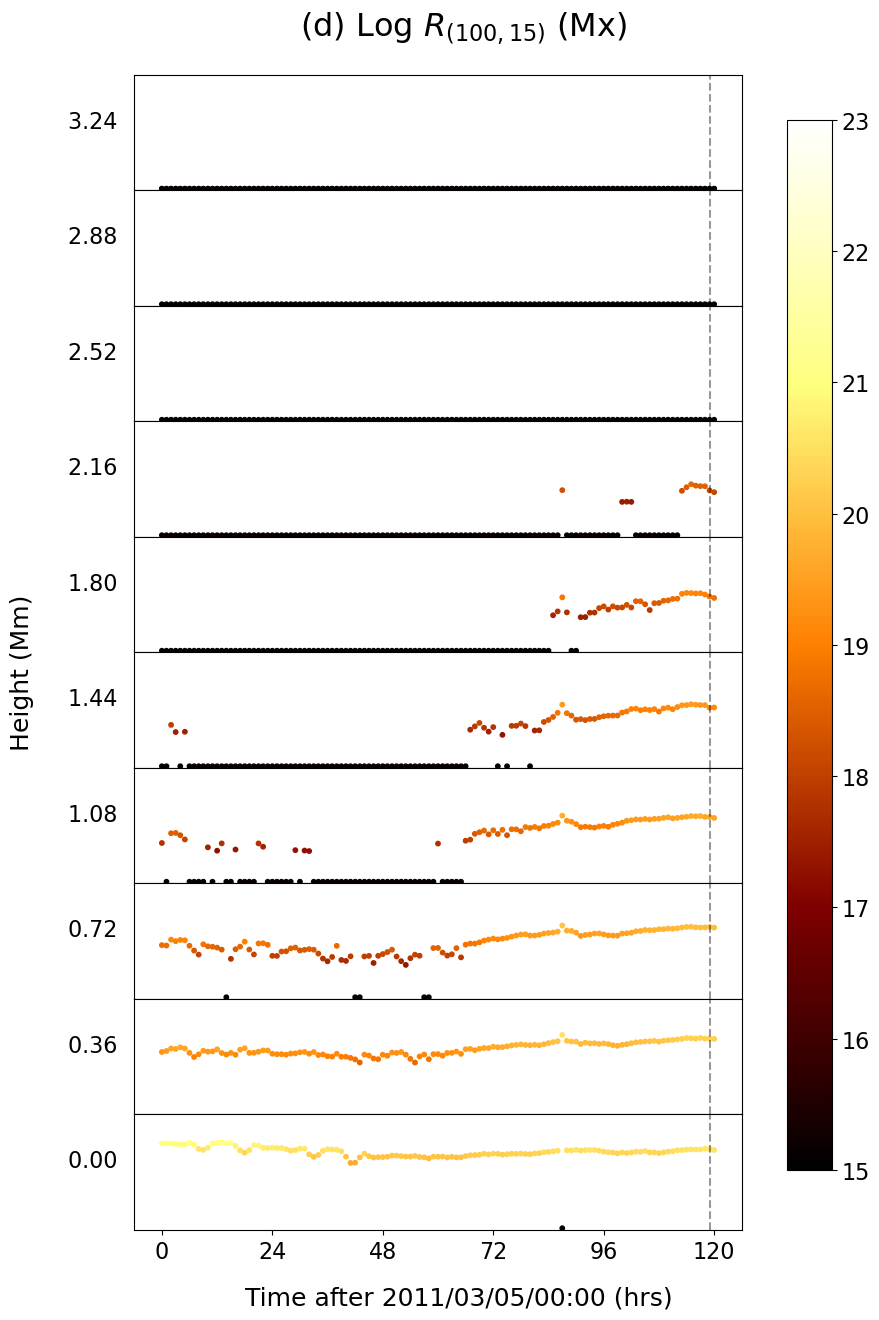}  
\end{subfigure}
\begin{subfigure}
  \centering
  \includegraphics[width=.32\linewidth]{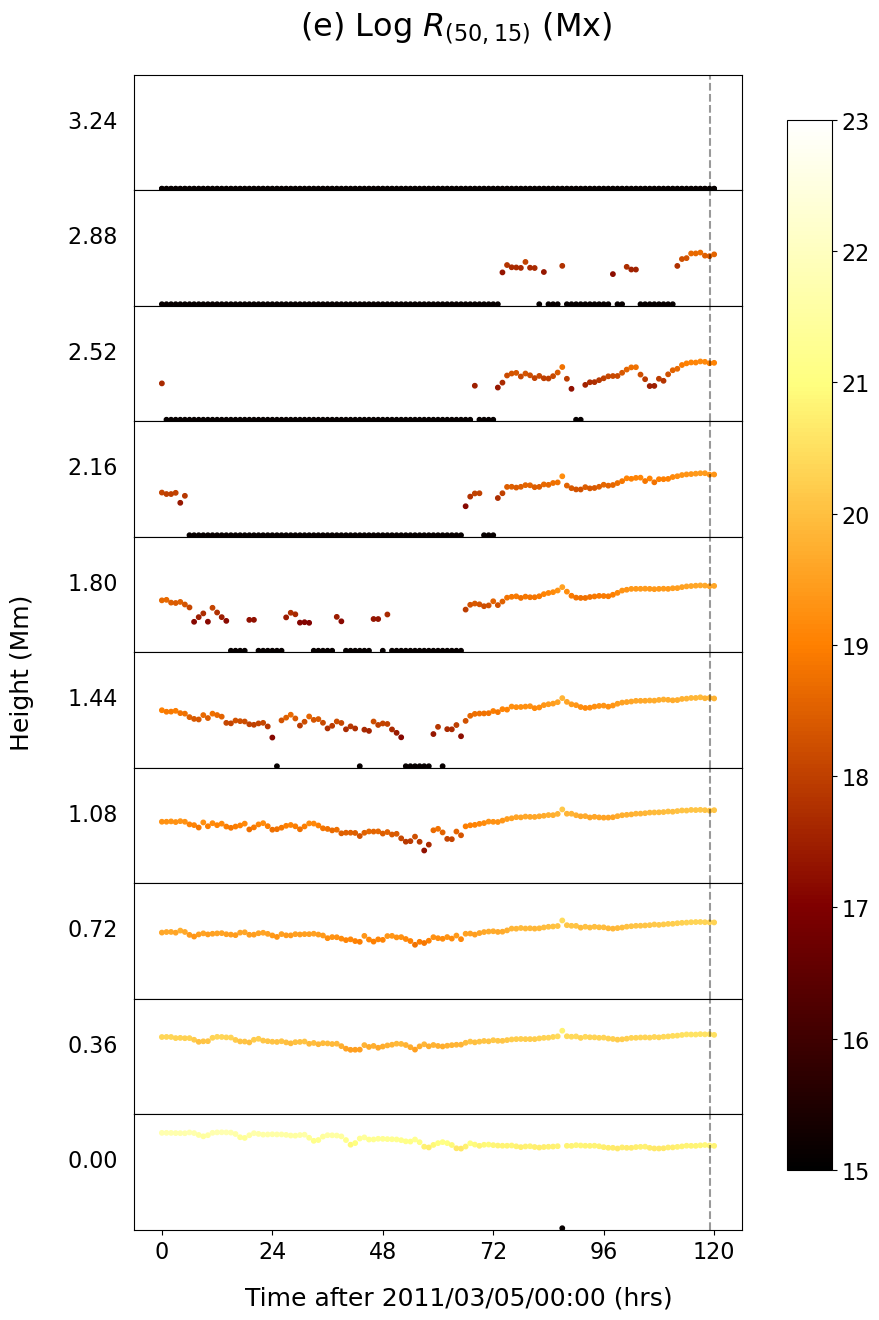} 
\end{subfigure}
\begin{subfigure}
  \centering
  \includegraphics[width=.32\linewidth]{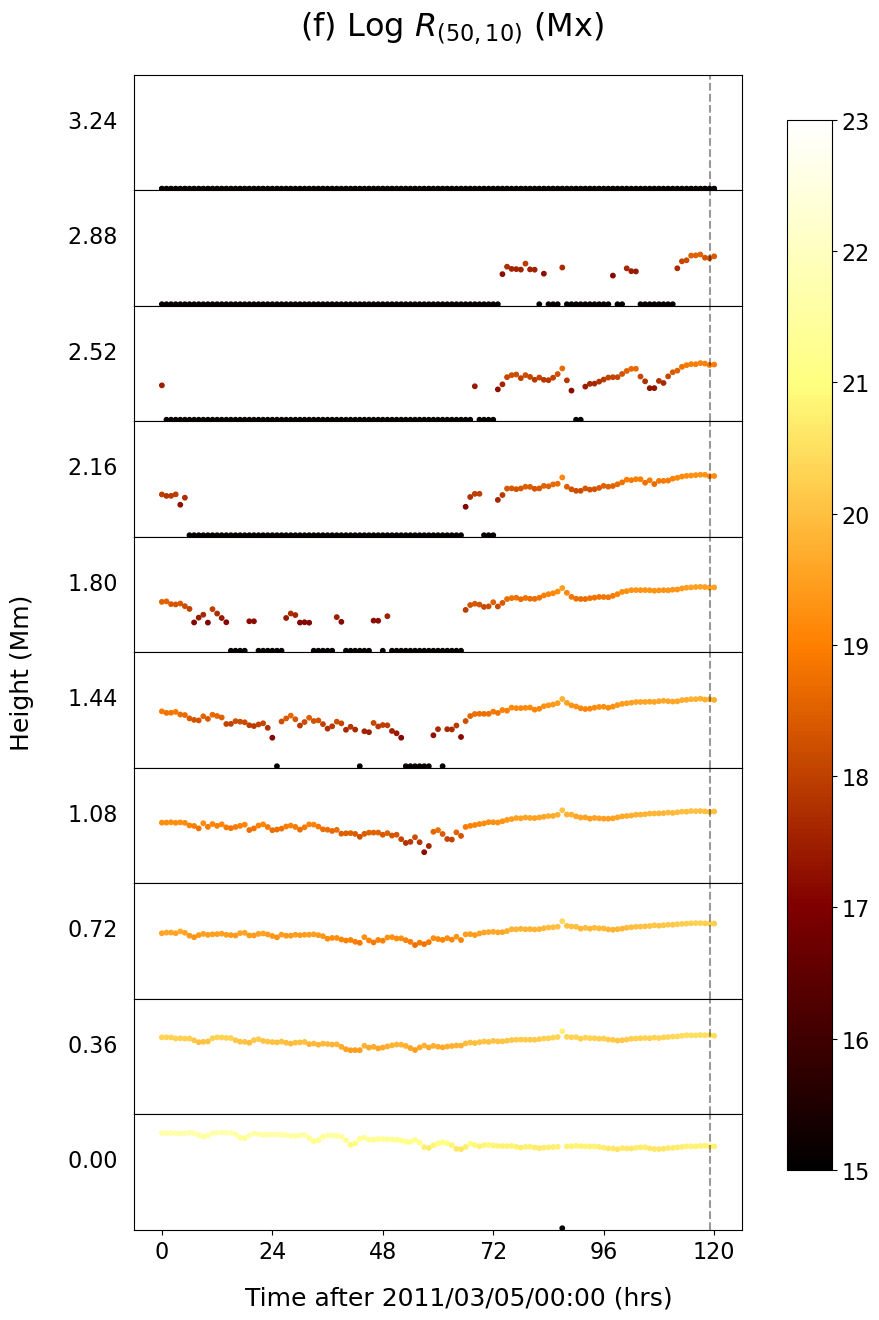} 
\end{subfigure}

\caption{Multiple-height stack plots for AR 11166; (a) unsigned flux around PILs, (b) $R_{(150,15)}$, (c) $R_{(150,10)}$, (d) $R_{(100,10)}$, (e) $R_{(50,15)}$, (f) $R_{(50,10)}$; vertical dashed line indicates the time of occurrence of the X1.5 flare at 23:13 UTC on Mar 9, 2011; note that plots b \& c do not exhibit any significant difference (same goes for plots e \& f); implying that R-value is not too sensitive to $D_{sep}$; the colourbar in plot a indicates unsigned flux (in $10^{20}$ Mx); the colourbars in plots b-f indicate the logarithm of R-value (in Mx); any black lines or points in plots b-f indicate null output}
\label{fig:11166}
\end{figure}

Changing $D_{sep}$ does not impact the R-value in height and time as much as changing $B_{th}$ (see Figure~\ref{fig:11166}). For a quantitative example, consider Tables \ref{table:dsep-sensitivity} and \ref{table:bth-sensitivity}. It may be observed from Table~\ref{table:dsep-sensitivity} that if $B_{th}$ is kept fixed at 150 G, reducing $D_{sep}$ from 15 to 10 Mm reduces the computed R-value by approximately 5-10\%, while for $B_{th}$ fixed at 50 G, the reduction is approximately 10-20\%. However, Table~\ref{table:bth-sensitivity} shows that reducing $B_{th}$ from 150 to 100 G (with $D_{sep}$ fixed at 15 Mm) can cause R-value to increase by an order of magnitude (check height 1.08 Mm). 


{\begin{table}[!h]
\begin{center}
\begin{tabular}{| c | c | c | c | c | c | c | c |}
 \hline
 \multicolumn{8}{|c|}{} \\
 [-0.8em]
 \multicolumn{8}{|c|}{Unsigned Flux near PILs and R-values (both in 10$^{20}$ Mx) for AR 11166 at 18:00 UTC, Mar 08, 2011} \\
 [-1em]
 \multicolumn{8}{|c|}{} \\
 \hline
 \phantom{x} & \phantom{x} & \phantom{x} & \phantom{x} & \phantom{x} & \phantom{x} & \phantom{x} & \phantom{x} \\ [-0.9em]
 Height (Mm) & Unsigned Flux & $R_{(150,15)}$ & $R_{(150,10)}$ & $R_{(150,10)}$/$R_{(150,15)}$ & $R_{(50,15)}$ & $R_{(50,10)}$ & $R_{(50,10)}$/$R_{(50,15)}$ \\ [0.3em]
\hline
 0.00 & 23.54 & 2.1758 & 2.0698 & 0.95 & 7.4204 & 6.7488 & 0.91 \\
 0.36 & 15.21 & 0.5619 & 0.5240 & 0.93 & 2.5417 & 2.2674 & 0.89 \\
 0.72 & 11.06 & 0.1382 & 0.1228 & 0.89 & 1.1530 & 0.9991 & 0.87 \\
 1.08 & 5.39 & 0.0114 & 0.0103 & 0.90 & 0.5310 & 0.4435 & 0.84 \\
 1.44 & 3.99 & N.V. & N.V. & N.V. & 0.2230 & 0.1812 & 0.81 \\
 1.80 & 0.90 & N.V. & N.V. & N.V. & 0.0626 & 0.0529 & 0.85 \\
 2.16 & 1.61 & N.V. & N.V. & N.V. & 0.0196 & 0.0157 & 0.80 \\
 2.52 & 0.48 & N.V. & N.V. & N.V. & N.V. & N.V. & N.V. \\
 2.88 & 0.18 & N.V. & N.V. & N.V. & N.V. & N.V. & N.V. \\
 3.24 & 0.00 & N.V. & N.V. & N.V. & N.V. & N.V. & N.V. \\
\hline
\end{tabular}
\end{center}
\caption{A comparison of different models of R-values to explore the sensitivity to $D_{sep}$ (refer to columns 5 and 8); N.V.: No value, indicates null output. In columns 3 and 4, $B_{th}$ is fixed at 150 G but $D_{sep}$ is 15 and 10 Mm, respectively. In columns 6 and 7, $B_{th}$ is fixed at 50 G but $D_{sep}$ is 15 and 10 Mm, respectively. It is seen that $R_{(150,15)}$ is about 10\% of the value of the unsigned flux near PILs on the photosphere.}
\label{table:dsep-sensitivity}
\end{table}}
{\begin{table}[!h]
\centering
\begin{tabular}{| c | c | c | c | c | c | c | c | c |}
 \hline
 \multicolumn{9}{|c|}{} \\
 [-0.8em]
 \multicolumn{9}{|c|}{R-values (in 10$^{20}$ Mx) for AR 11166 at 18:00 UTC, Mar 08, 2011} \\
 [-1em]
 \multicolumn{9}{|c|}{} \\
 \hline
 \phantom{x} & \phantom{x} & \phantom{x} & \phantom{x} & \phantom{x} & \phantom{x} & \phantom{x} & \phantom{x} & \phantom{x} \\ [-0.9em]
 Height (Mm) & $R_{(150,15)}$ & $R_{(100,15)}$ & $R_{(50,15)}$ & $R_{(100,15)}$/$R_{(150,15)}$ &  $R_{(50,15)}$/$R_{(150,15)}$ & $R_{(150,10)}$ & $R_{(50,10)}$ & $R_{(50,10)}$/$R_{(150,10)}$ \\ [0.3em]
\hline
 0.00 & 2.1758 & 3.5746 & 7.4204 & 1.64 & 3.41 & 2.0698 & 6.7488 & 3.26 \\
 0.36 & 0.5619 & 1.1095 & 2.5417 & 1.97 & 4.52 & 0.5240 & 2.2674 & 4.33 \\
 0.72 & 0.1382 & 0.3935 & 1.1530 & 2.85 & 8.34 & 0.1228 & 0.9991 & 8.13 \\
 1.08 & 0.0114 & 0.1291 & 0.5310 & 11.36 & 46.74 & 0.0103 & 0.4435 & 43.18 \\
 1.44 & N.V. & 0.0218 & 0.2230 & N.V. & N.V. & N.V. & 0.1812 & N.V. \\
 1.80 & N.V. & N.V. & 0.0626 & N.V. & N.V. & N.V. & 0.0529 & N.V. \\
 2.16 & N.V. & N.V. & 0.0196 & N.V. & N.V. & N.V. & 0.0157 & N.V. \\
 2.52 & N.V. & N.V. & N.V. & N.V. & N.V. & N.V. & N.V. & N.V. \\
 2.88 & N.V. & N.V. & N.V. & N.V. & N.V. & N.V. & N.V. & N.V. \\
 3.24 & N.V. & N.V. & N.V. & N.V. & N.V. & N.V. & N.V. & N.V. \\
\hline
\end{tabular}
\caption{A comparison of different models of R-value to explore the sensitivity to $B_{th}$ (refer to columns 5, 6 and 9); N.V.: No value, indicates null output. In columns 2, 3 and 4, $D_{sep}$ is fixed at 15 Mm but $B_{th}$ is 150, 100 and 50 G, respectively. In columns 7 and 8, $D_{sep}$ is fixed at 10 Mm but $B_{th}$ is 150 and 50 G, respectively.}
\label{table:bth-sensitivity}
\end{table}}


\newpage

\clearpage
\newpage


\section{Outlier case of AR 11283}
\label{sec:app_b}

\phantom{...}

AR 11283 defied classification into either of the two categories as described in Section \ref{ssec:rd_inc}. In this AR, the two X-class flares occur between Sep 6-8, when the unsigned flux near PILs is not too high and is somewhat stable in time. However high levels of unsigned flux were reported around Sep 1 and after Sep 8, 2011. The complexity of the sunspot group gradually increased with time, $\beta$-type between Sep 1-5, $\beta\gamma$-type on Sep 6 and $\beta\gamma\delta$-type between Sep 7-8. \\

\begin{figure}[ht]
\centering
\textbf{Plots for AR 11283}\par\medskip

\begin{subfigure}
  \centering
  \includegraphics[width=.32\linewidth]{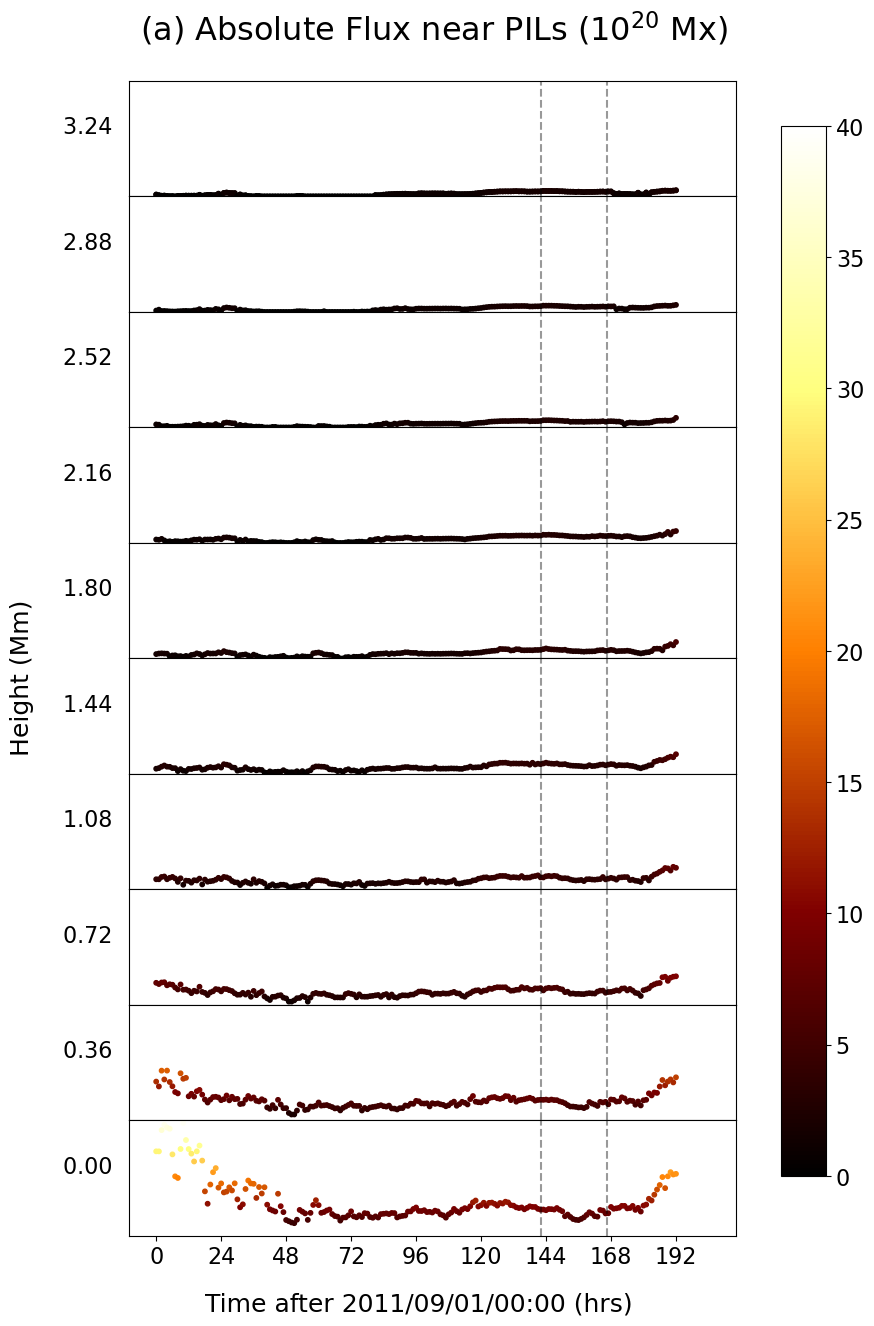}  
\end{subfigure}
\begin{subfigure}
  \centering
  \includegraphics[width=.32\linewidth]{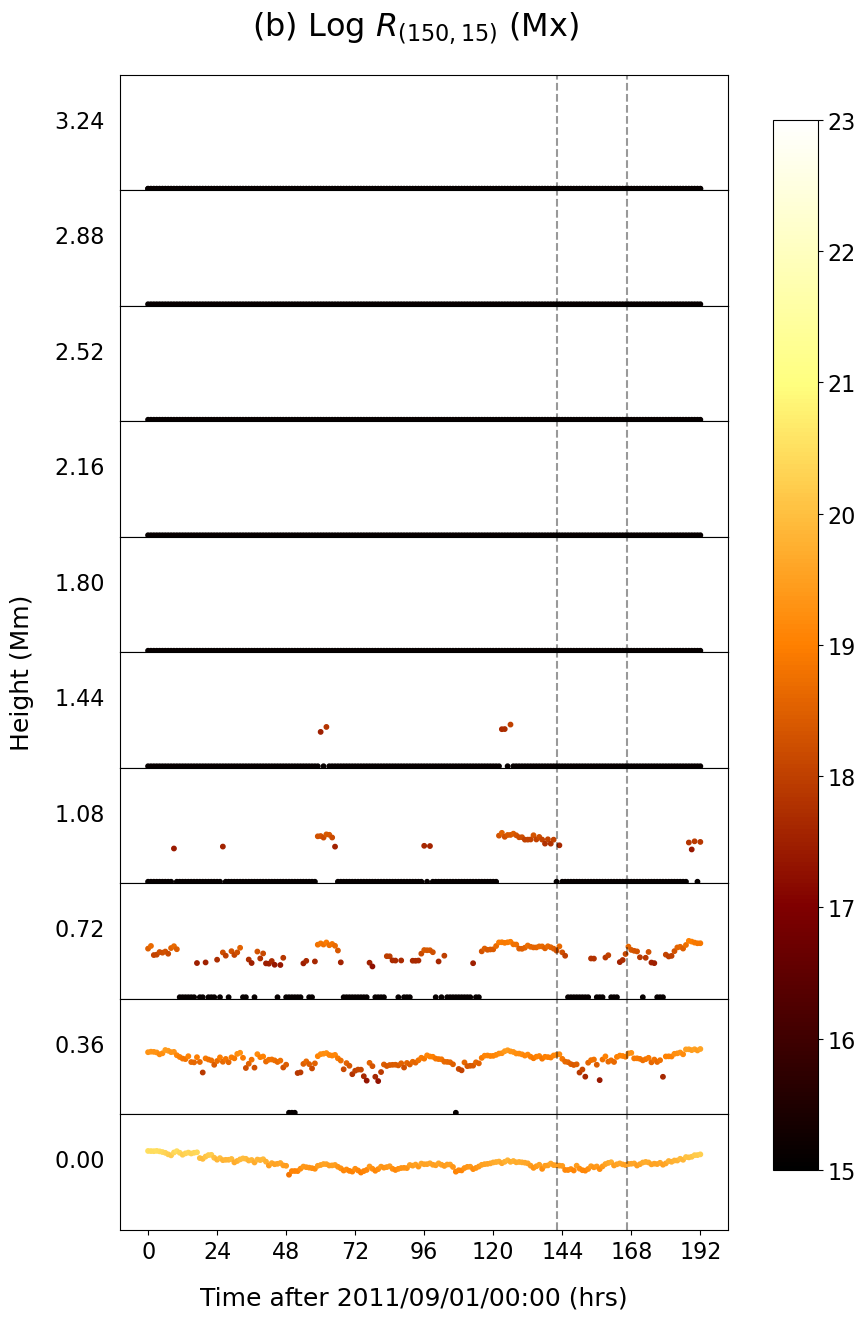}
\end{subfigure}
\begin{subfigure}
  \centering
  \includegraphics[width=.32\linewidth]{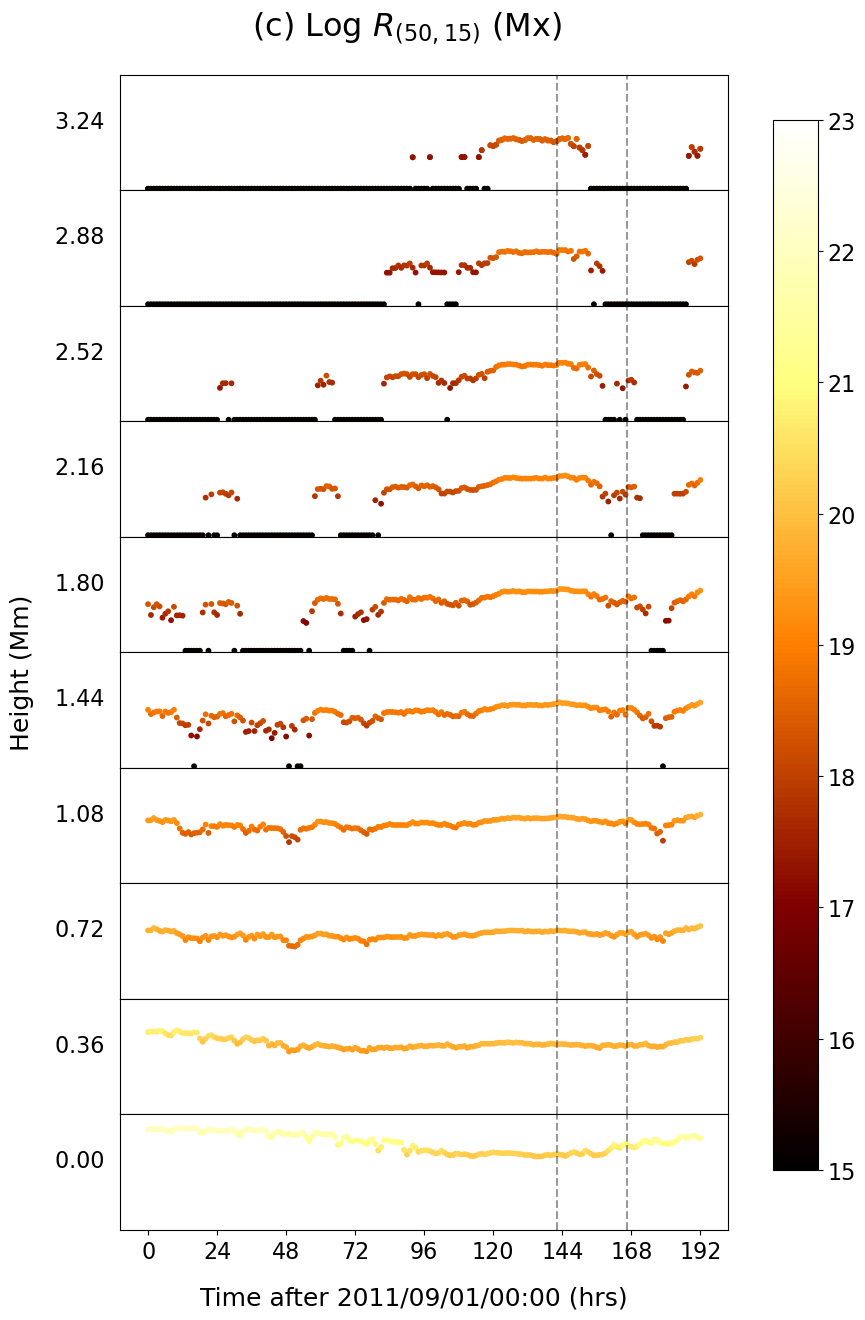}
\end{subfigure}

\caption{Multiple-height stack plots for AR 11283; (a) unsigned flux around PILs; (b) $R_{(150,15)}$; (c) $R_{(50,15)}$; vertical dashed lines indicate X-class flare occurrence times on Sep 6 and Sep 7, 2011; the colourbar in plot a indicates unsigned flux (in $10^{20}$ Mx); the colourbars in plots b-f indicate the logarithm of R-value (in Mx); black lines or points in plots b and c indicate null output}
\label{fig:11283}

\end{figure}

Although this AR is somewhat similar to the cases discussed in Section \ref{ssec:rd_inc} (as in a significant increase in the unsigned flux is seen after Sep 8), it is rather distinctive when it comes to results obtained from the R-value models. At first glance, it is difficult to determine an OHR from the $R_{(150,15)}$ and $R_{(100,15)}$ models because the black lines are not continuous over a long time (48 hrs) and they are not followed by a period of consistently high R-value as we have seen for cases described in 4.1 (see Figure \ref{fig:11283}). For example, $R_{(150,15)}$ at 0.72 Mm fluctuates between null and finite values before a temporary increase in R-value is seen at 22:00 UTC on Sep 5. However, determining an OHR for $R_{(50,15)}$ is not that difficult and the height range of 2.16 - 3.24 Mm is considered as the OHR. This is because a similar pattern, consistent in height, can be found for heights 2.16 - 2.52 Mm. \\

At a time of about 24 hrs before the time of occurrence of the X2.1 flare, $R_{(50,15)}$ has a relatively lower value in the photosphere compared to what it is at around Sep 1-2. Interestingly, at the heights in the OHR, the opposite is true. $R_{(50,15)}$ sustains a finite output 24 hrs before the flare but null-values are seen at around Sep 1-2. \\


\clearpage
\newpage


\section{Case studies for weaker flare classes}
\label{sec:app_c}

\phantom{...}

\subsection{M-class flares}

\phantom{...}

In this section, an attempt is made to check if the pre-flare trends before M-class flares are similar to what we have seen previously in case of X-class flares. This shall help us understand if pre-flare variations of R-value (in height and time) could possibly distinguish between an impending X-class flare and a M-class flare. Since we have only studied 4 cases here, our motivation here is not to make explicit statistical conclusions but to gain an insight into the R-value configuration in height and time before the first M-class flare hosted by an AR associated with a $\delta$ sunspot. The selection criteria of these cases remain the same as we had for X-class flares. The flares under focus are the first M-class flares for ARs 11620, 11719, 11818 and 12497 (refer to Table \ref{table:dataset2}). We followed the same procedures as before for X-class flares to determine the OHR, $T_{rv}$, $T_{diff}$, $T_{diff}^*$, $T_{opt}$ and $T_{res}$ for M-class flares (refer to Tables \ref{table:ohrlistM} and \ref{table:timelistM}). \\

\phantom{...}

{\begin{table}[ht!]
\centering
\begin{tabular}{| c | c | c | c | c | c | c | c |}
 \hline
 \multicolumn{8}{|c|}{} \\
 [-0.9 em]
 \multicolumn{8}{|c|}{Observed Data - M Class Flares} \\[0.4 em]
 \hline
 \phantom{x} & \phantom{x} & \phantom{x} & \phantom{x} & \phantom{x} & \phantom{x} & \phantom{x} & \phantom{x} \\ [-0.9em]
No. & AR & Class & $T_{start}$ &  $T_{end}$ & $T_{flare{\phantom{.}}onset}$ 
 & $T_{fo}$ - $T_{start}$ (hrs) &  Dimensions (Mm$^2$) \\
[0.3em]
\hline
 01 & 11620 & M2.2 & 2012/11/24 00:00 & 2012/11/30 00:00 & 2012/11/28 21:20 & 117.33 & 318.96 x 109.80 \\
 02 & 11719 & M6.5 & 2013/04/08 00:00 & 2013/04/13 00:00 & 2013/04/11 06:55 & 78.92 & 388.80 x 247.68 \\
 03 & 11818 & M3.3 & 2013/08/12 00:00 & 2013/08/19 00:00 & 2013/08/17 18:16 & 138.27 & 225.72 x 124.20 \\
 04 & 12497 & M1.0 & 2016/02/08 00:00 & 2016/02/15 00:00 & 2016/02/12 10:36 & 106.60 & 306.00 x 228.60 \\
 \hline
\end{tabular}

\phantom{...}
\caption{Table listing the details of studied ARs, flares, corresponding time windows and AR dimensions}
\label{table:dataset2}
\end{table}}

{\begin{table}[!ht]
\hskip-2.0cm
\centering
\begin{tabular}{| c | c | c | c | c |}
 \hline
 \phantom{x} & \phantom{x} & \phantom{x} & \phantom{x} & \phantom{x} \\ [-0.9em]
 No. & AR & OHR [$R_{(150,15)}]$ & OHR [$R_{(100,15)}$] & OHR [$R_{(50,15)}$] \\ [0.3em]
\hline
 01 & 11620 & 0.00 - 1.80 Mm & 0.36 - 2.52 Mm & 0.72 - 3.24 Mm \\
 02 & 11719 & - - - & - - - & - - - \\
 03 & 11818 & 0.72 - 1.80 Mm & 1.08 - 2.52 Mm & 1.44 - 3.24 Mm \\
 04 & 12497 & 0.72 - 1.08 Mm & 1.08 - 2.16 Mm & 1.80 - 3.24 Mm \\
\hline
\end{tabular}
\caption{A list of OHRs for different models of R-value for M-class flare cases}
\label{table:ohrlistM}
\end{table}}

{\begin{table}[!ht]
\centering
\begin{tabular}{| c | c | c | c | c | c | c | c | c |}
 \hline
 \phantom{x} & \phantom{x} & \phantom{x} & \phantom{x} & \phantom{x} & \phantom{x} & \phantom{x} & \phantom{x} & \phantom{x} \\ [-0.9em]
 No. & AR & Flare & $T_{pf}$ & $T_{diff}^*$ [$R_{(150,15)}$] & $T_{diff}^*$ [$R_{(100,15)}$] & $T_{diff}^*$ [$R_{(50,15)}$] &  \phantom{..} $T_{opt}$ \phantom{..} & $T_{res}$ = $T_{opt}$ - $C_{int}$ \\ [0.3em]
\hline
 01 & 11620 & M2.2 & 2012/11/28 21:00 & \textbf{85 hrs} & 82 hrs & 83 hrs & 85 hrs & 61 hrs \\
 02 & 11719 & M6.5 & 2013/04/11 06:00 & - - - & - - - & - - - & - - - & - - - \\
 03 & 11818 & M3.3 & 2013/08/17 18:00 & 54 hrs & 59 hrs & \textbf{97 hrs} & 97 hrs & 73 hrs \\
 04 & 12497 & M1.0 & 2016/02/12 10:00 & \textbf{18 hrs} & 7 hrs & \textbf{18 hrs} & 18 hrs & -6 hrs \\
\hline
\end{tabular}
\caption{A list of optimal times (indicated in bold) and response times for each OHR for X-class flare cases; column 4 lists the latest time-stamp preceding the flare '$T_{pf}$'; columns 5-7 list the maximum lead time across all heights '$T_{diff}^*$' for $R_{(150,15)}$, $R_{(100,15)}$ and $R_{(50,15)}$ respectively. The maximum '$T_{diff}^*$' across multiple R-value models is '$T_{opt}$' and is listed in 
column 8; column 9 lists the response time '$T_{res}$' after subtracting the confidence interval '$C_{int}$' from $T_{opt}$. In case of AR 12497, $T_{opt}$ was less than the confidence interval '$C_{int}$' as the jump in R-value was immediately followed by a M-class flare and hence, we have a negative value for $T_{res}$.}
\label{table:timelistM}
\end{table}}


\newpage
\clearpage
\newpage

\begin{figure}[ht]
\centering
\textbf{R-value plots for AR 11620}\par\medskip

\begin{subfigure}
  \centering
  \includegraphics[width=.32\linewidth]{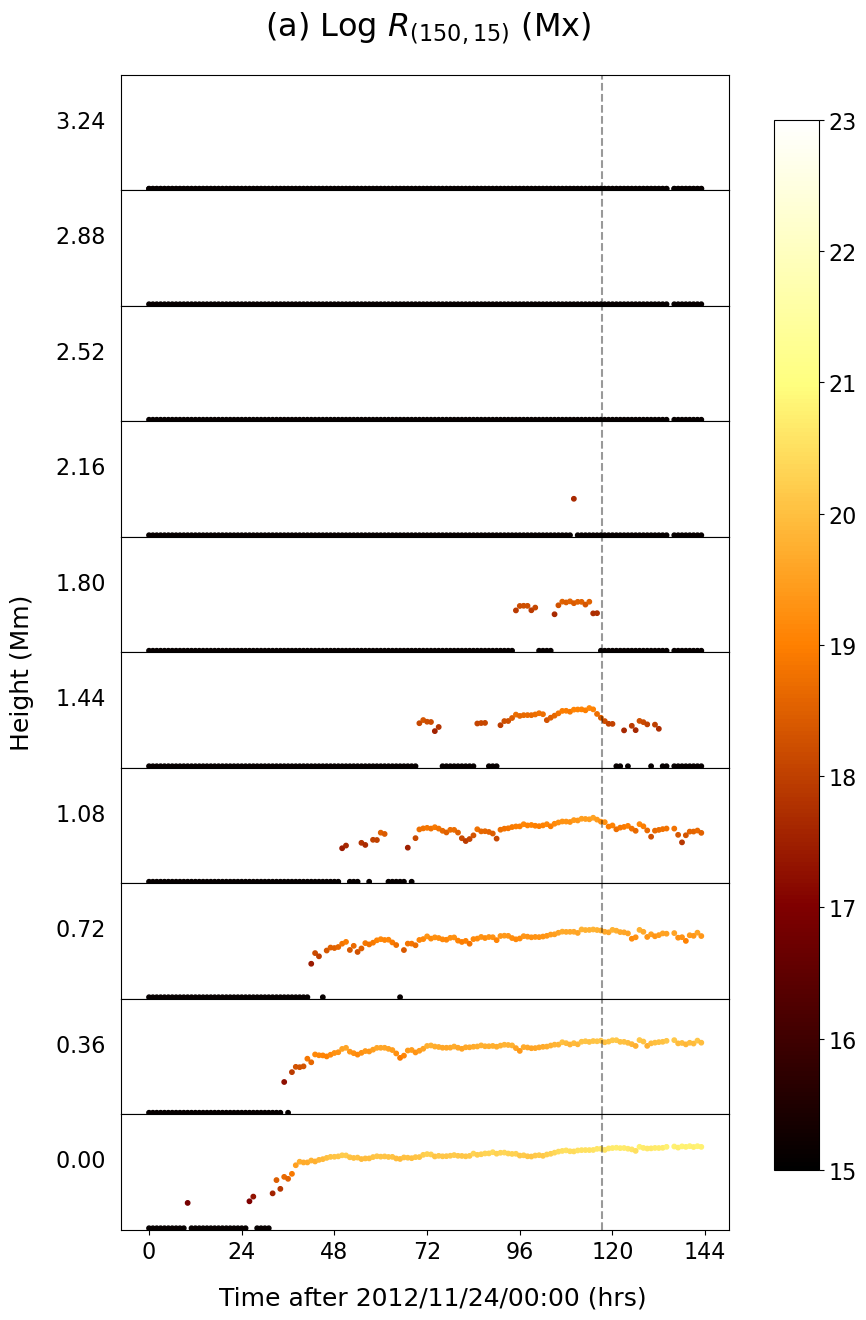}  
\end{subfigure}
\begin{subfigure}
  \centering
  \includegraphics[width=.32\linewidth]{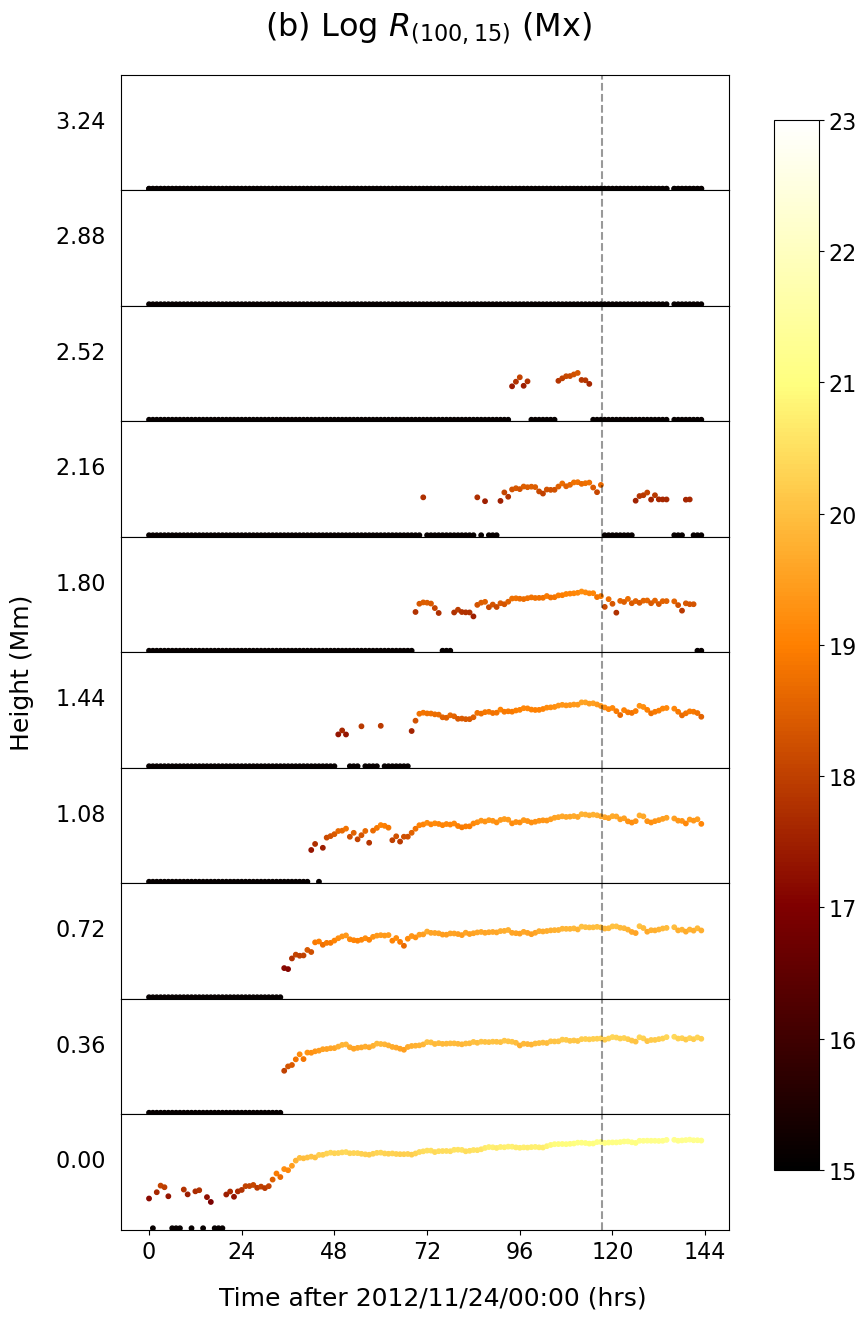}
\end{subfigure}
\begin{subfigure}
  \centering
  \includegraphics[width=.32\linewidth]{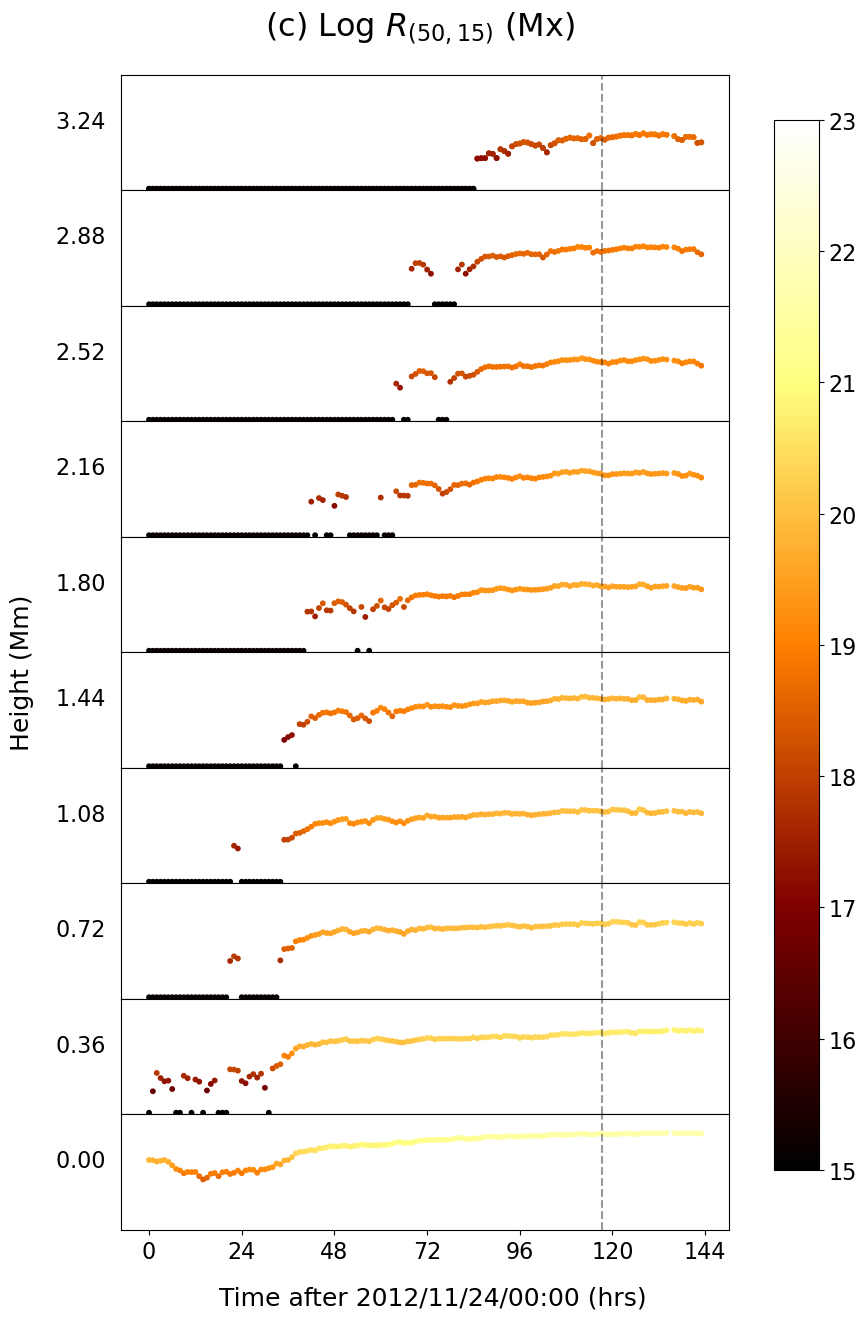}
\end{subfigure}

\caption{Multiple-height stack plots for (a) $R_{(150,15)}$; (b) $R_{(100,15)}$; (c) $R_{(50,15)}$; 
vertical dashed line indicates time of occurrence of the M2.2 flare; colourbar indicates the logarithm of R-value (in Mx); any black lines or points indicate null output}
\label{fig:11620}

\end{figure}
\begin{figure}[ht]
\centering
\textbf{R-value plots for AR 11719}\par\medskip

\begin{subfigure}
  \centering
  \includegraphics[width=.32\linewidth]{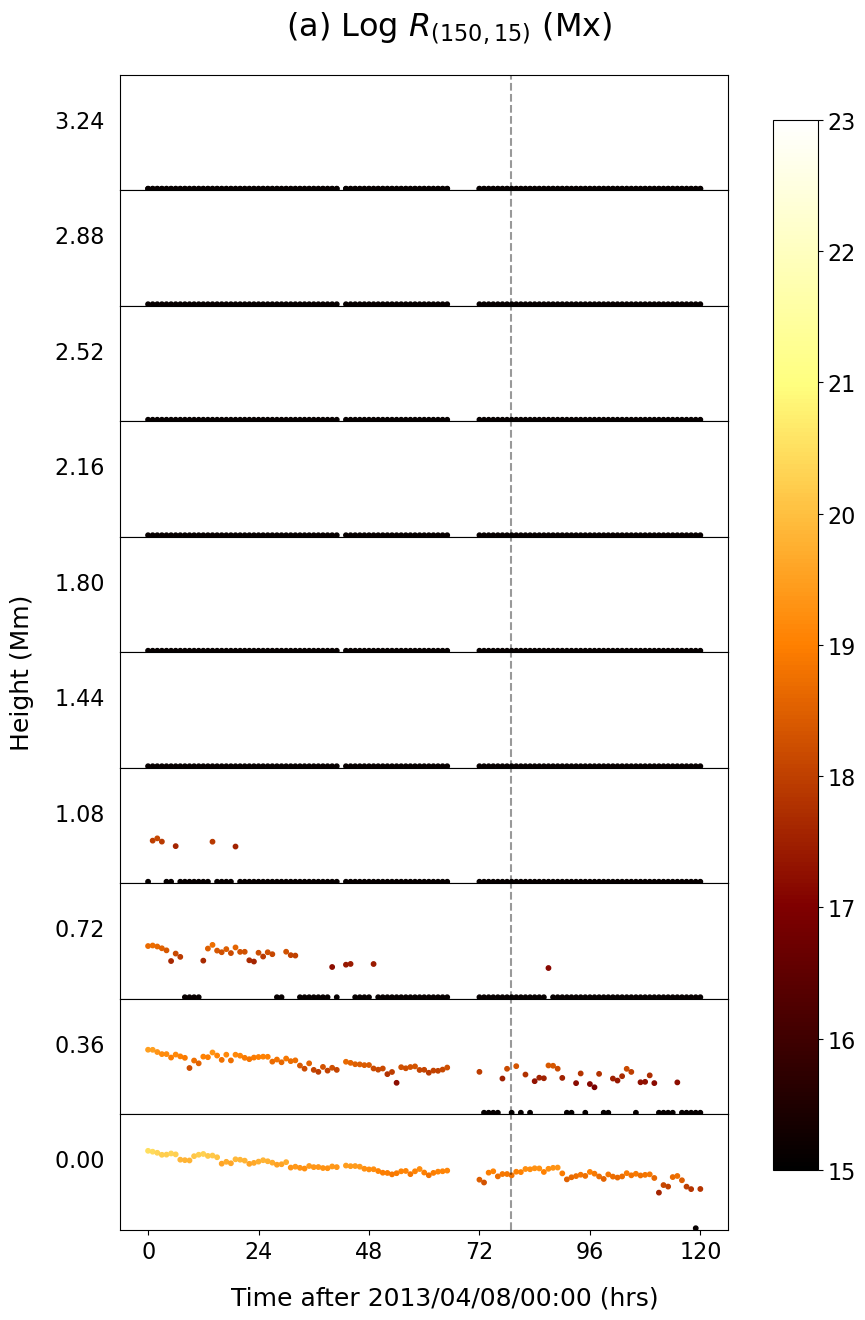}  
\end{subfigure}
\begin{subfigure}
  \centering
  \includegraphics[width=.32\linewidth]{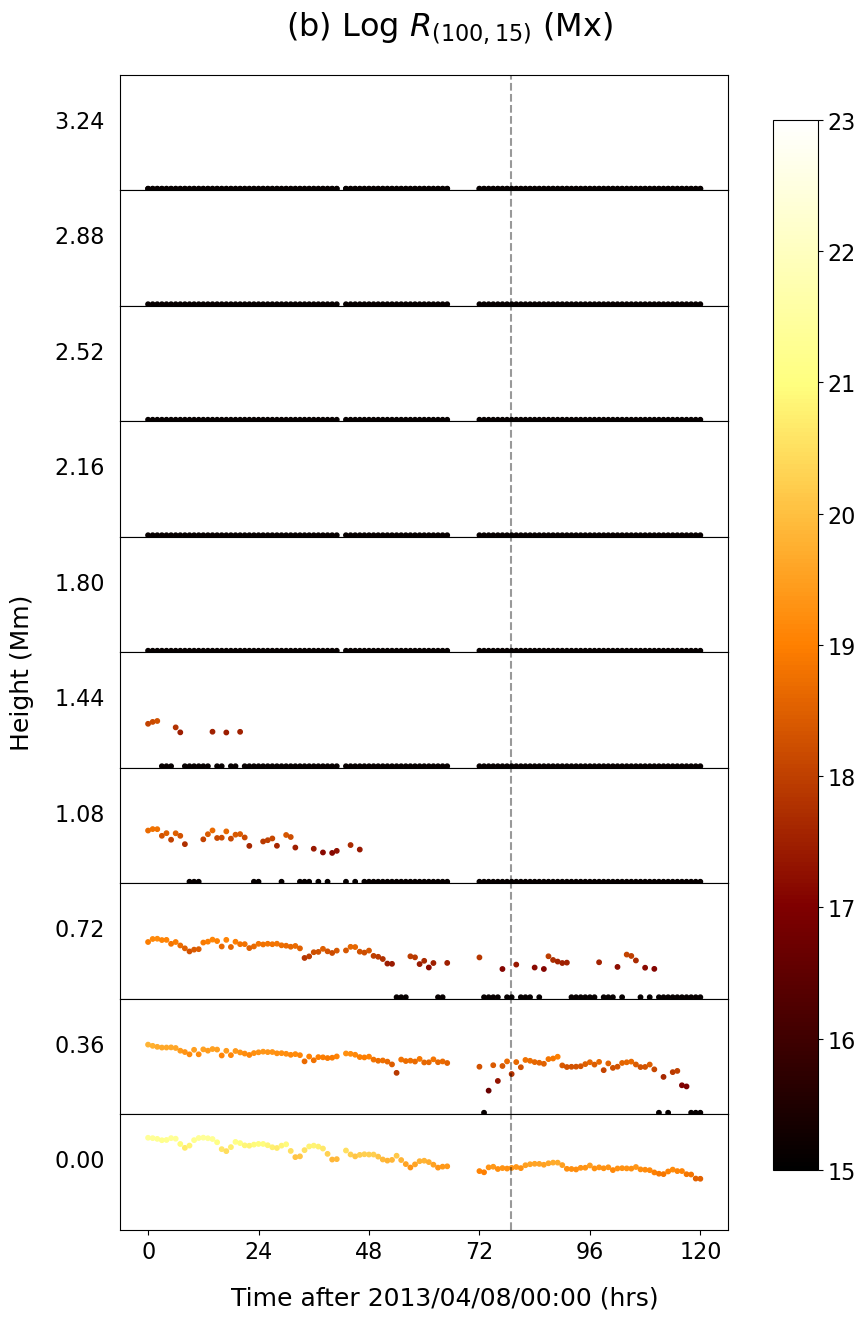}
\end{subfigure}
\begin{subfigure}
  \centering
  \includegraphics[width=.32\linewidth]{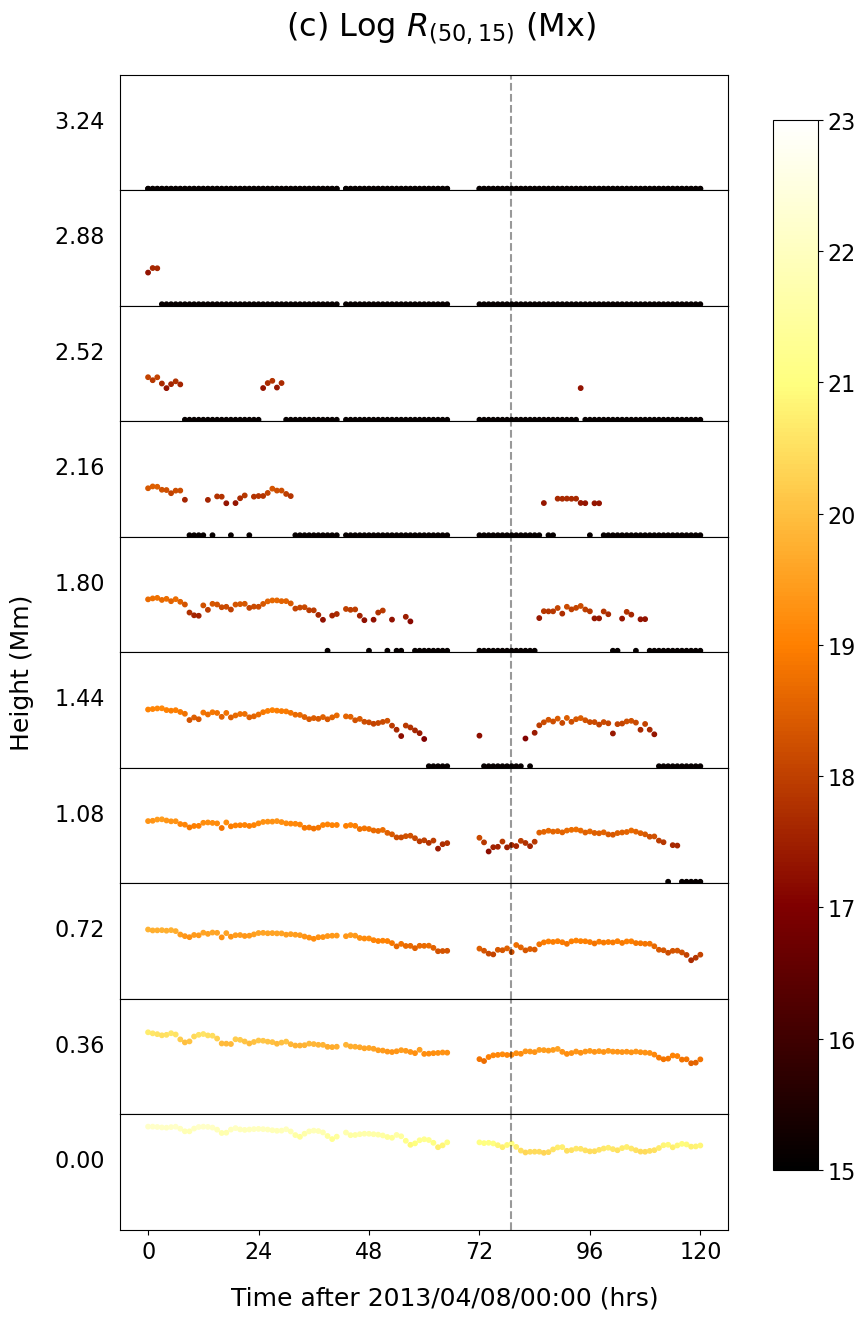}
\end{subfigure}

\caption{Multiple-height stack plots for (a) $R_{(150,15)}$; (b) $R_{(100,15)}$; (c) $R_{(50,15)}$; 
vertical dashed line indicates time of occurrence of the M6.5 flare; colourbar indicates the logarithm of R-value (in Mx); any black lines or points indicate null output}
\label{fig:11719}

\end{figure}

An OHR could be determined in 3 out of the 4 cases. While the evolution of R-value for AR 11719 (see Figure \ref{fig:11719}) appears similar to the cases explained in Section \ref{ssec:rd_dec}, the evolution of R-value for ARs 11620, 11818 and 12497 closely resembles the X-class flare cases for emerging ARs (refer to Section \ref{ssec:rd_inc}; as an example see Figure \ref{fig:11620}). This suggests that studying the R-value along with the concept of OHR may not be sufficient in distinguishing between an impending X-class flare and a M-class flare qualitatively. \\

\subsection{C-class flare case: AR 12353}

\phantom{...}

\begin{figure}[ht]
\centering
\textbf{R-value plots for AR 12353}\par\medskip

\begin{subfigure}
  \centering
  \includegraphics[width=.32\linewidth]{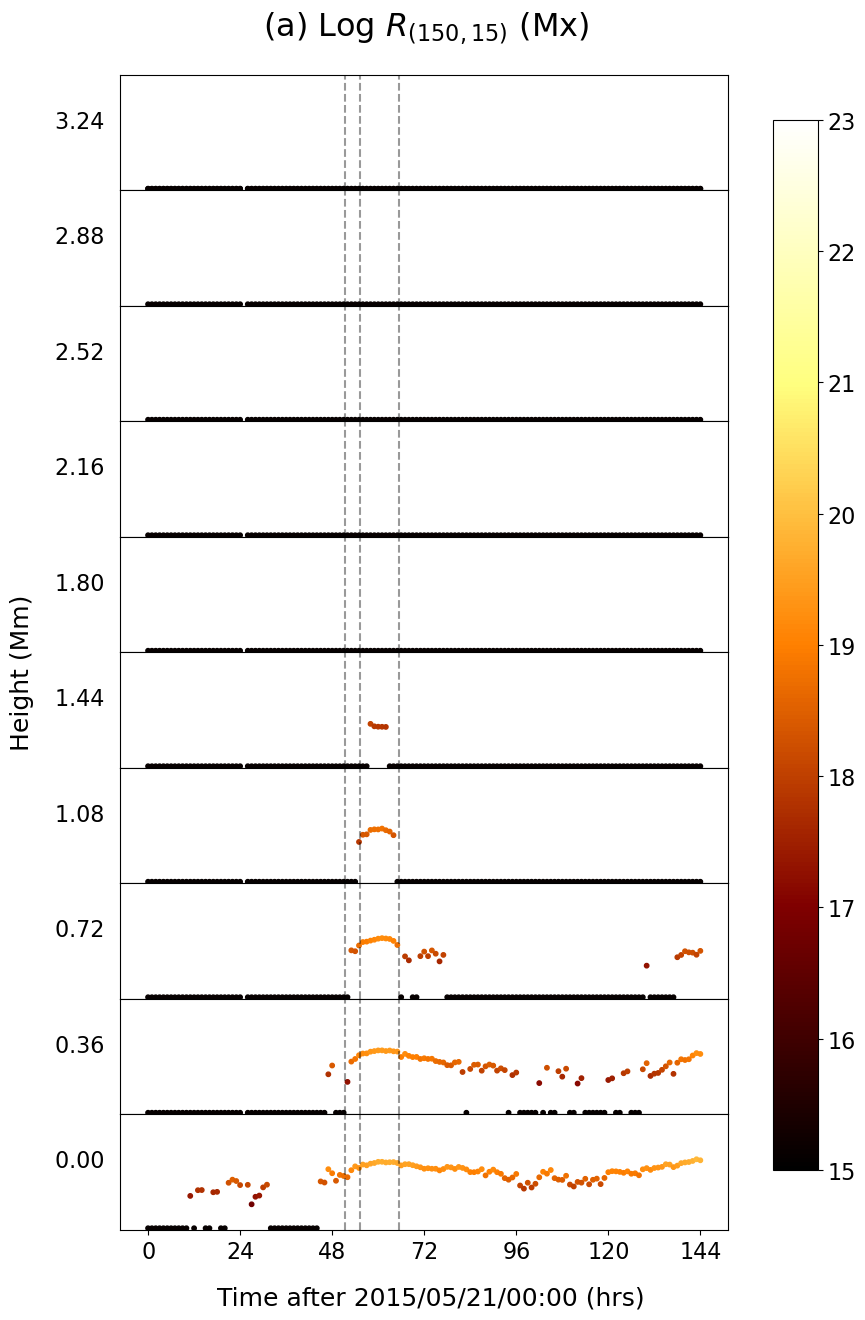}  
\end{subfigure}
\begin{subfigure}
  \centering
  \includegraphics[width=.32\linewidth]{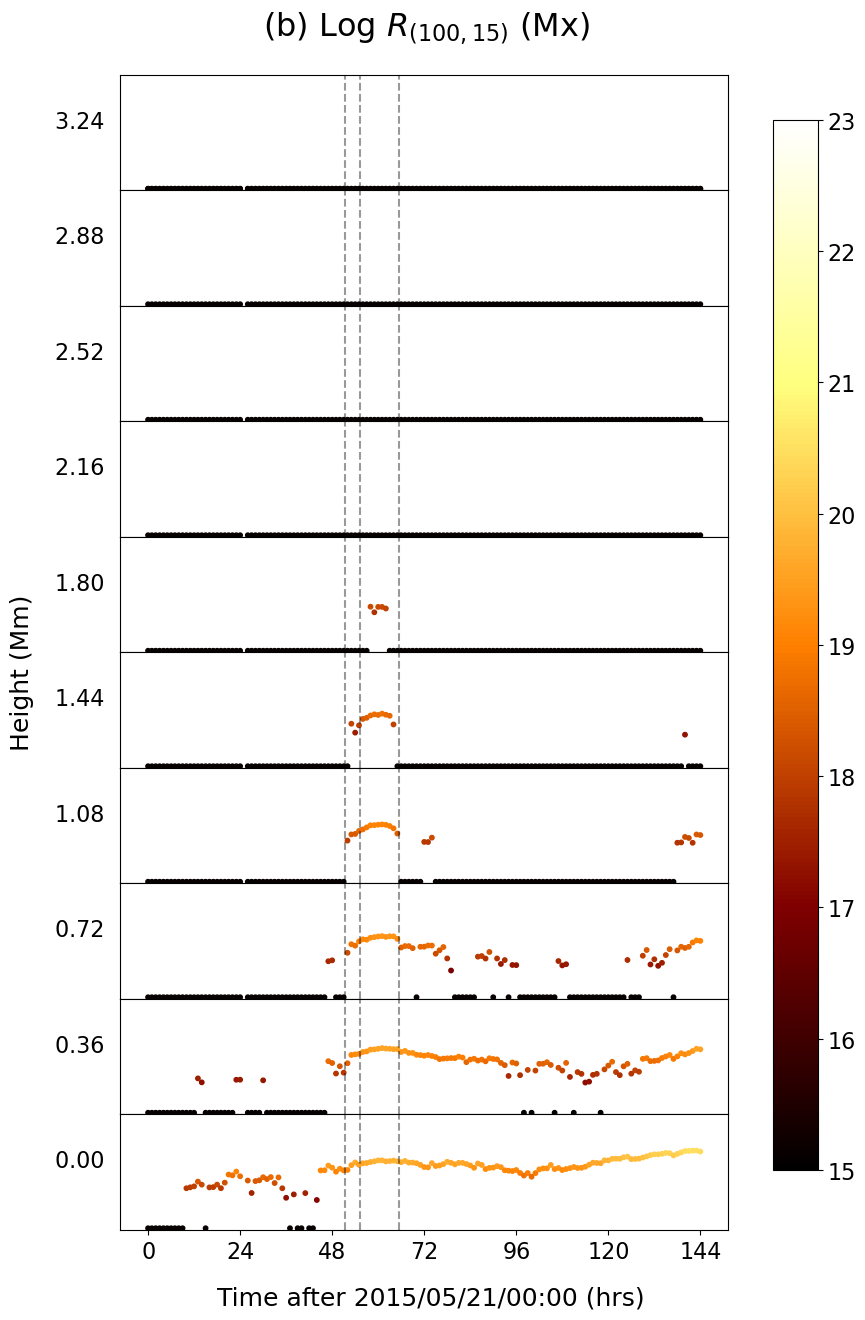}
\end{subfigure}
\begin{subfigure}
  \centering
  \includegraphics[width=.32\linewidth]{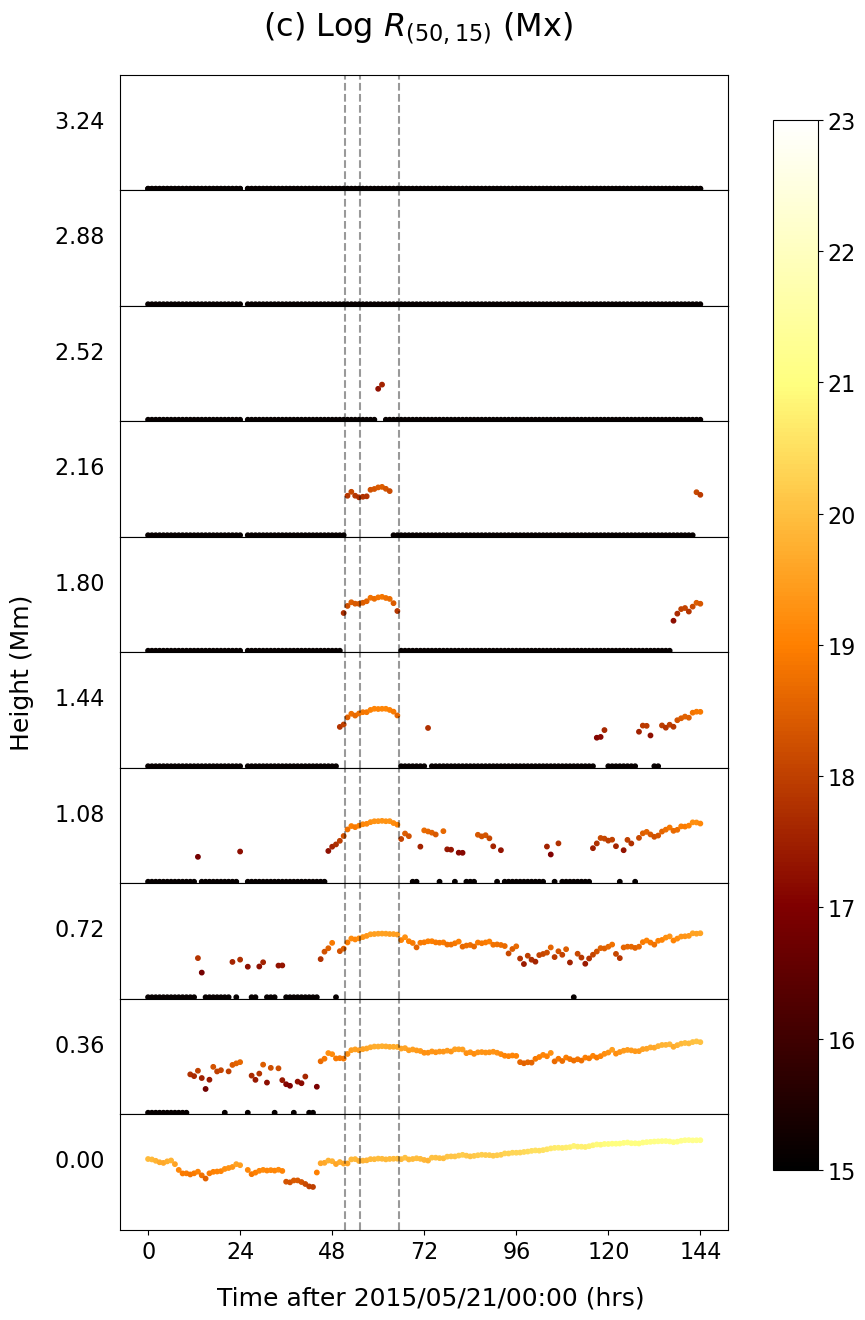}
\end{subfigure}

\caption{Multiple-height stack plots for (a) $R_{(150,15)}$; (b) $R_{(100,15)}$; (c) $R_{(50,15)}$; vertical dashed lines indicate C-class flare occurrence times on May 23, 2015; colourbar indicates the logarithm of R-value (in Mx); any black lines or points indicate null output}
\label{fig:12353}

\end{figure}

ARs associated with a $\delta$ sunspot are normally flaring in nature and are associated with M-class or X-class flares. It is rare for ARs hosting a $\delta$ sunspot to be solely associated with C-class flares or remain non-flaring. One such example is AR 12353. AR 12353 produced 3 C-class flares on May 23, 2015; C1.0 (03:27 UTC); C1.1 (07:18 UTC) and C2.3 (17:30 UTC) in chronological order. In the SHARP data repository, SHARP number 5596 corresponds to not just AR 12353 but also AR 12352. Since it was not possible to isolate AR 12353, PF extrapolation and subsequent computation of $R_{(150,15)}$, $R_{(100,15)}$ and $R_{(50,15)}$ were carried out on the entire SHARP data. A jump in $R_{(150,15)}$ to non-zero values (in the photosphere) is seen at about 6 hrs before the first flare. Interestingly, at heights of 0.36 Mm and 0.72 Mm, all the 3 C-class flares happen when $R_{(150,15)}$ shows a clear jump and is non-zero. The jump in $R_{(150,15)}$ is seen at all heights up to 1.44 Mm (see Figure \ref{fig:12353}a). This suggests that it could be possible that a jump in R-value may be linked to a C-class flare, hosted by an AR with a $\delta$ sunspot. The maximum value of $R_{(150,15)}$ in the photosphere observed for AR 12353 (in the time window of study) was 0.78 x 10$^{20}$ Mx. It is considerably lesser (by orders) than the threshold of 20 x 10$^{20}$ Mx which when breached guarantees the occurrence of an X-class flare \citep{Schrijver2007}. 

\phantom{...}

\newpage

\section{Additional Tables}
\label{sec:app_d}

{\begin{table}[!ht]
\centering
\begin{tabular}{| c | c | c | c | c | c | c | c | c | c | c | c |}
 \hline
 \multicolumn{12}{|c|}{} \\
 [-0.8 em]
 \multicolumn{12}{|c|}{Approximate time between jump in $R_{(150,15)}$ and X-class flare occurence time, calculated height-wise} \\ [-1em]
 \multicolumn{12}{|c|}{} \\
 \hline
 \phantom{x} & \phantom{x} & \phantom{x} & \phantom{x} & \phantom{x} & \phantom{x} & \phantom{x} & \phantom{x} & \phantom{x} & \phantom{x} &
 \phantom{x} & \phantom{x} \\ [-0.9em]
 No. & AR & 0.00 Mm & 0.36 Mm & 0.72 Mm & 1.08 Mm & 1.44 Mm & 1.80 Mm & 2.16 Mm & 2.52 Mm & 2.88 Mm & 3.24 Mm \\ [0.3em]
 \hline
 01 & 11158 & - - - & {\textbf{49 hrs}} & 48 hrs & 43 hrs & 33 hrs & - - - & - - - & - - - & - - - & - - - \\
 02 & 11166 & - - - & - - - & {\textbf{54 hrs}} & 35 hrs & 22 hrs & - - - & - - - & - - - & - - - & - - - \\
 03 & 11283 & - - - & - - - & - - - & {\textbf{20 hrs}} & - - - & - - - & - - - & - - - & - - - & - - - \\
 04 & 11520 & - - - & - - - & - - - & - - - & - - - & - - - & - - - & - - - & - - - & - - - \\
 05 & 12017 & - - - & {\textbf{40 hrs}} & 36 hrs & 20 hrs & - - - & - - - & - - - & - - - & - - - & - - - \\
 06 & 12158 & - - - & - - - & - - - & - - - & - - - & - - - & - - - & - - - & - - - & - - - \\
 07 & 12297 & - - - & - - - & - - - & - - - & - - - & - - - & - - - & - - - & - - - & - - - \\
 08 & 12673 & - - - & - - - & {\textbf{65 hrs}} & 57 hrs & 53 hrs & 50 hrs & 44 hrs & 40 hrs & 33 hrs & - - - \\ [0.3 em]
\hline
\end{tabular}
\caption{Lead times $T_{diff}$ listed at all heights for different ARs (computed using $R_{(150,15)}$); text in bold indicates $T_{diff}^*$ i.e. the maximum lead time across all heights for a given active region.}
\label{table:heighttime150}
\end{table}}

{\begin{table}[!ht]
\centering
\begin{tabular}{| c | c | c | c | c | c | c | c | c | c | c | c |}
 \hline
 \multicolumn{12}{|c|}{} \\
 [-0.8 em]
 \multicolumn{12}{|c|}{Approximate time between jump in $R_{(100,15)}$ and X-class flare occurence time, calculated height-wise} \\ [-1em]
 \multicolumn{12}{|c|}{} \\
 \hline
 \phantom{x} & \phantom{x} & \phantom{x} & \phantom{x} & \phantom{x} & \phantom{x} & \phantom{x} & \phantom{x} & \phantom{x} & \phantom{x} &
 \phantom{x} & \phantom{x} \\ [-0.9em]
 No. & AR & 0.00 Mm & 0.36 Mm & 0.72 Mm & 1.08 Mm & 1.44 Mm & 1.80 Mm & 2.16 Mm & 2.52 Mm & 2.88 Mm & 3.24 Mm \\ [0.3em]
 \hline
 01 & 11158 & - - - & {\textbf{59 hrs}} & 47 hrs & 48 hrs & 43 hrs & 38 hrs & 35 hrs & 32 hrs & - - - & - - - \\
 02 & 11166 & - - - & - - - & - - - & {\textbf{53 hrs}} & 52 hrs & 41 hrs & 6 hrs & - - - & - - - & - - - \\
 03 & 11283 & - - - & - - - & - - - & - - - & {\textbf{24 hrs}} & 21 hrs & - - - & - - - & - - - & - - - \\
 04 & 11520 & - - - & - - - & - - - & - - - & - - - & - - - & - - - & - - - & - - - & - - - \\
 05 & 12017 & - - - & - - - & {\textbf{47 hrs}} & 39 hrs & 20 hrs & - - - & - - - & - - - & - - - & - - - \\
 06 & 12158 & - - - & - - - & - - - & - - - & - - - & - - - & - - - & - - - & - - - & - - - \\
 07 & 12297 & - - - & - - - & - - - & - - - & - - - & - - - & - - - & - - - & - - - & - - - \\
 08 & 12673 & - - - & - - - & - - - & {\textbf{65 hrs}} & 58 hrs & 57 hrs & 52 hrs & 50 hrs & 44 hrs & 42 hrs \\ [0.3 em]
\hline
\end{tabular}
\caption{Lead times $T_{diff}$ listed at all heights for different ARs (computed using $R_{(100,15)}$); text in bold indicates $T_{diff}^*$ i.e. the maximum lead time across all heights}
\label{table:heighttime100}
\end{table}}

{\begin{table}[!ht]
\centering
\begin{tabular}{| c | c | c | c | c | c | c | c | c | c | c | c |}
 \hline
 \multicolumn{12}{|c|}{} \\
 [-0.8 em]
 \multicolumn{12}{|c|}{Approximate time between jump in $R_{(50,15)}$ and X-class flare occurence time, calculated height-wise} \\ [-1em]
 \multicolumn{12}{|c|}{} \\
 \hline
 \phantom{x} & \phantom{x} & \phantom{x} & \phantom{x} & \phantom{x} & \phantom{x} & \phantom{x} & \phantom{x} & \phantom{x} & \phantom{x} &
 \phantom{x} & \phantom{x} \\ [-0.9em]
 No. & AR & 0.00 Mm & 0.36 Mm & 0.72 Mm & 1.08 Mm & 1.44 Mm & 1.80 Mm & 2.16 Mm & 2.52 Mm & 2.88 Mm & 3.24 Mm \\ [0.3em]
 \hline
 01 & 11158 & - - -  & - - - & - - - & {\textbf{50 hrs}} & 48 hrs & 47 hrs & 46 hrs & 45 hrs & 43 hrs & 41 hrs \\
 02 & 11166 & - - - & - - - & - - - & - - - & {\textbf{60 hrs}} & 53 hrs & 46 hrs & 46 hrs & - - - & - - - \\
 03 & 11283 & - - - & - - - & - - - & - - - & - - - & - - - & {\textbf{61 hrs}} & 60 hrs & 59 hrs & 23 hrs\\
 04 & 11520 & - - - & - - - & - - - & - - - & - - - & - - - & - - - & - - - & - - - & - - - \\
 05 & 12017 & - - - & - - - & - - - & {\textbf{48 hrs}} & 47 hrs & 46 hrs & 44 hrs & - - - & - - - & - - - \\
 06 & 12158 & - - - & - - - & - - - & - - - & - - - & - - - & - - - & - - - & - - - & - - - \\
 07 & 12297 & - - - & - - - & - - - & - - - & - - - & - - - & - - - & - - - & - - - & - - - \\
 08 & 12673 & - - - & - - - & - - - & - - - & {\textbf{68 hrs}} & 64 hrs & 61 hrs & 60 hrs & 57 hrs & 54 hrs \\ [0.3 em]
\hline
\end{tabular}
\caption{Lead times $T_{diff}$ listed at all heights for different ARs (computed using $R_{(50,15)}$); text in bold indicates $T_{diff}^*$ i.e. the maximum lead time across all heights}
\label{table:heighttime50}
\end{table}}

\bibliography{sample631}{}

\phantom{...}

\bibliographystyle{aasjournal}

\end{document}